\documentclass[12pt]{article}
\usepackage[utf8]{inputenc}
\usepackage{amsmath,amsfonts,amssymb}
\usepackage{psfrag}
\usepackage{enumerate}
\usepackage{mathrsfs}
\usepackage{graphicx}
\usepackage{wrapfig}
\usepackage{xcolor}
\usepackage{caption}
\usepackage{amsthm}
\usepackage{subfig}
\usepackage{xcolor}

\usepackage{comment}

\usepackage{jheppub}
\allowdisplaybreaks
 \newcommand{\be}{\begin{equation}}
\newcommand{\beq}{\begin{equation}}
 \newcommand{\ee}{\end{equation}}
 \newcommand{\bea}{\begin{align}}
 
 \newcommand{\eea}{\end{align}}

\def\nn{\nonumber}

\global\long\def\mA{\mathcal{A}}%

\global\long\def\mF{\mathcal{F}}%

\global\long\def\mH{\mathcal{H}}%

\global\long\def\mJ{\mathcal{J}}%

\global\long\def\mT{\mathcal{T}}%

\global\long\def\mZ{\mathcal{Z}}%

\global\long\def\e{\epsilon}%
 
\global\long\def\ra{\rightarrow}%

\global\long\def\avg#1{\left\langle #1\right\rangle }%

\global\long\def\f#1#2{\frac{#1}{#2}}%
 
\global\long\def\del{\partial}%
 
\global\long\def\t{\theta}%
 
\global\long\def\a{\alpha}%
 
\global\long\def\b{\beta}%
 
\global\long\def\g{\gamma}%

\global\long\def\s{\sigma}%
 
\global\long\def\r{\rho}%
 
\global\long\def\d{\delta}%
 
\global\long\def\Tr{\text{Tr}}%

\global\long\def\ket#1{\left\langle #1\right|}%
 
\global\long\def\bra#1{\left|#1\right\rangle }%
 
\global\long\def\N{\mathbb{N}}%

\global\long\def\Z{\mathbb{Z}}%
\global\long\def\E{\mathbb{E}}%
 
\global\long\def\R{\mathbb{R}}%

\global\long\def\T{\text{T}}%
 
\global\long\def\w{\omega}%
 
\global\long\def\D{\Delta}%

\global\long\def\S{\Sigma}%

\global\long\def\l{\ell}%

\global\long\def\app{\approx}%
\global\long\def\sgn{\text{sgn}}%
\global\long\def\arctanh{\text{arctanh}}%
\global\long\def\ad{\text{ad}}%
\global\long\def\tfd{\text{tfd}}%
\global\long\def\eff{\text{eff}}%

\title{Seeing behind black hole horizons in SYK}

\author[a]{Ping Gao}
\author[b]{and Lampros Lamprou}%

\affiliation[a]{Center for Theoretical Physics,\\ Massachusetts Institute of Technology,
Cambridge, MA 02139, USA}
\affiliation[b]{University of British Columbia,\\
Vancouver, BC V6T 1Z1, Canada}

\emailAdd{pgao@mit.edu}
\emailAdd{llamprou@phas.ubc.ca}

\abstract{We present an explicit reconstruction of the interior of an AdS$_2$ black hole in Jackiw-Teitelboim gravity, that is entirely formulated in the dual SYK model and makes no direct reference to the gravitational bulk. We do this by introducing a probe ``observer'' in the right wormhole exterior and using the prescription of [arXiv:2009.04476] to transport SYK operators along the probe's infalling worldline and into the black hole interior, using an appropriate SYK modular Hamiltonian. Our SYK computation recovers the precise proper time at which signals sent from the left boundary are registered by our observer's apparatus inside the wormhole. The success of the computation relies on the universal properties of SYK and we outline a promising avenue for extending it to higher dimensions and applying it to the computation of scattering amplitudes behind the horizon.}

\begin{document}

\maketitle

\section{Introduction and summary}
In this work, we perform an explicit computation demonstrating the ability of the recent proposal \cite{Jafferis:2020ora} to holographically reconstruct operators behind black hole horizons, while relying entirely on boundary data. 

The framework of \cite{Jafferis:2020ora} outlines an intrinsically holographic method for \emph{transporting} local operators along the trajectory of a selected bulk ``observer'' or probe, which propagates in some ambient geometry.\footnote{See \cite{Yoshida:2019qqw,Yoshida:2019kyp} for a conceptual similar approach to interior reconstruction} The central idea is that upon tracing out the probe's internal degrees of freedom, the rest of the Universe, which we call the \emph{system}, is endowed with a reduced density matrix, $\rho$, as a consequence of its initial entanglement with the probe. The key observation was that, in certain states, the unitary flow $\rho^{is}$, called \emph{modular flow}, propagates bulk operators, initially localized near the probe, along the probe's worldline by \emph{translating them in proper time} by an amount equal to
\begin{equation}
\tau_{proper}=\frac{\beta_{probe}}{2\pi}\,s \label{eq:conversionintro}
\end{equation}
while keeping their location relative to the worldline fixed. The parameter $\beta_{probe}$ is an effective inverse temperature associated with the probe's mixed state which we will make precise in the main text. 

Practically, the introduction of the observer is achieved by entangling our holographic system with an external reference, representing the observer's internal degrees of freedom; the system's modular flow $\rho^{is}$ is then obtained by tracing out that reference. The reader is encouraged to consult \cite{Jafferis:2020ora} for an in-depth exposition to the method and the arguments for it. The modular time/proper time correspondence, in the form stated here, has a limited regime of validity but it becomes the seed for a general holographic construction of an observer's local proper time Hamiltonian, which is explained in an upcoming paper \cite{LamprouJafferisdeBoer}. The most exciting possibility created by this proposal is obtaining holographic access to the local operators in the interior of black holes, by propagating bulk fields in the exterior\footnote{where reconstruction is well understood} with the modular flow of an infalling probe for the appropriate (finite) amount of modular time (Fig.~\ref{fig:HKLL}).

In this paper, we explicitly apply this method, within its expected regime of validity, in order to test this interior reconstruction. The setup of our computation is the $AdS_2$/SYK correspondence \cite{Sachdev:2010um}, where an eternal $AdS_2$ wormhole solution of Jackiw-Teitelboim gravity is described microscopically by a pair of dynamically decoupled SYK systems (which we call SYK$_l$ and SYK$_r$) in the thermofield double state. Each SYK model \cite{sachdev1993gapless,kitaev2015simple} is a quantum mechanical system that consists of $N$ Majorana fermions $\psi_{l,r}^j$ and has a $q$-local Hamiltonian with random couplings drawn from a Gaussian ensemble \cite{maldacena2016remarks}. The infalling probe we wish to co-move with is a configuration of Majorana fermions introduced near the right asymptotic boundary, entangled with an external reference system of Dirac fermions. The probe is introduced by inserting in the Euclidean path integral that prepares the thermofield double state dual to the empty wormhole (Fig.~\ref{fig:EucTFD}), an operator $U_{sys+ref}$ that entangles our system with the reference. 

Following the proposal of \cite{Jafferis:2020ora}, we proceed by analyzing, directly in the pair of SYK models, the evolution of a fermion $\psi_r$ of SYK$_r$ with the unitary $\rho^{is}$, where $\rho$ is the reduced density matrix of SYK$_l\times$SYK$_r$ after tracing out the reference. To test the success of our reconstruction beyond the horizon, we study the \emph{causal influence} of an excitation $\psi_l(t)$ inserted in the \emph{left} asymptotic boundary at time $t$, on the modular flow of the \emph{right} exterior operator $\rho^{-is} \psi_r \rho^{is}$, as a function of modular time $s$, by evaluating the anticommutator:
 \begin{equation}
     W(s,t) = \Tr\left(\r\{\r^{-is}\psi_r\r^{is},\psi_l(t)\}\right) \label{Wintro}
 \end{equation}
The bulk expectation for $W$ is the following: When the backreaction of the probe is small, the semiclassical geometry of the wormhole implies that the causal propagator $W$  vanishes for the range of proper times the flowed operator remains spacelike separated from the left insertion, and transitions to an $O(1)$ value at timelike separations, with a sharp spike occurring at the proper time when the former crosses the bulk lightcone of the latter. 

Our SYK computation \emph{exactly} reproduces this expected bulk propagator together with the \emph{precise} proper time of lightcone crossing, in the large $q,N$ and low temperature limit, after the determination of the conversion factor $\beta_{probe}$ in (\ref{eq:conversionintro}). Our results, therefore, establish that the method proposed in \cite{Jafferis:2020ora} constitutes a practically useful tool for the holographic reconstruction of black hole interior operators. 

\subsection*{Summary of our results} 
\begin{figure}
    \begin{centering}
\subfloat[\label{fig:necklaceBC}]{\begin{centering}
 \includegraphics[width=5cm]{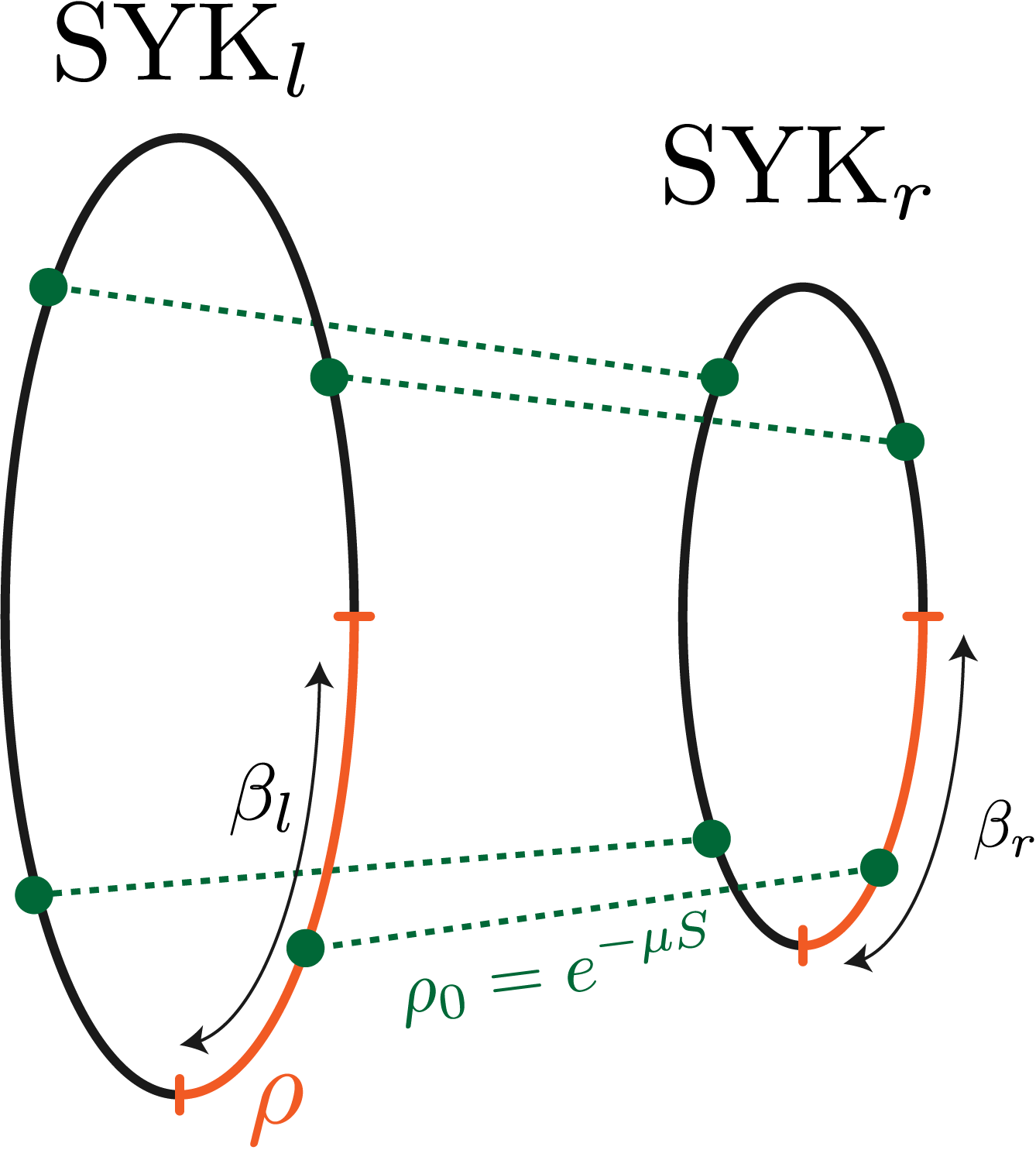}
\par\end{centering}
}$\quad\quad\quad$\subfloat[\label{fig:disconnected}]{\begin{centering}
\includegraphics[width=5cm]{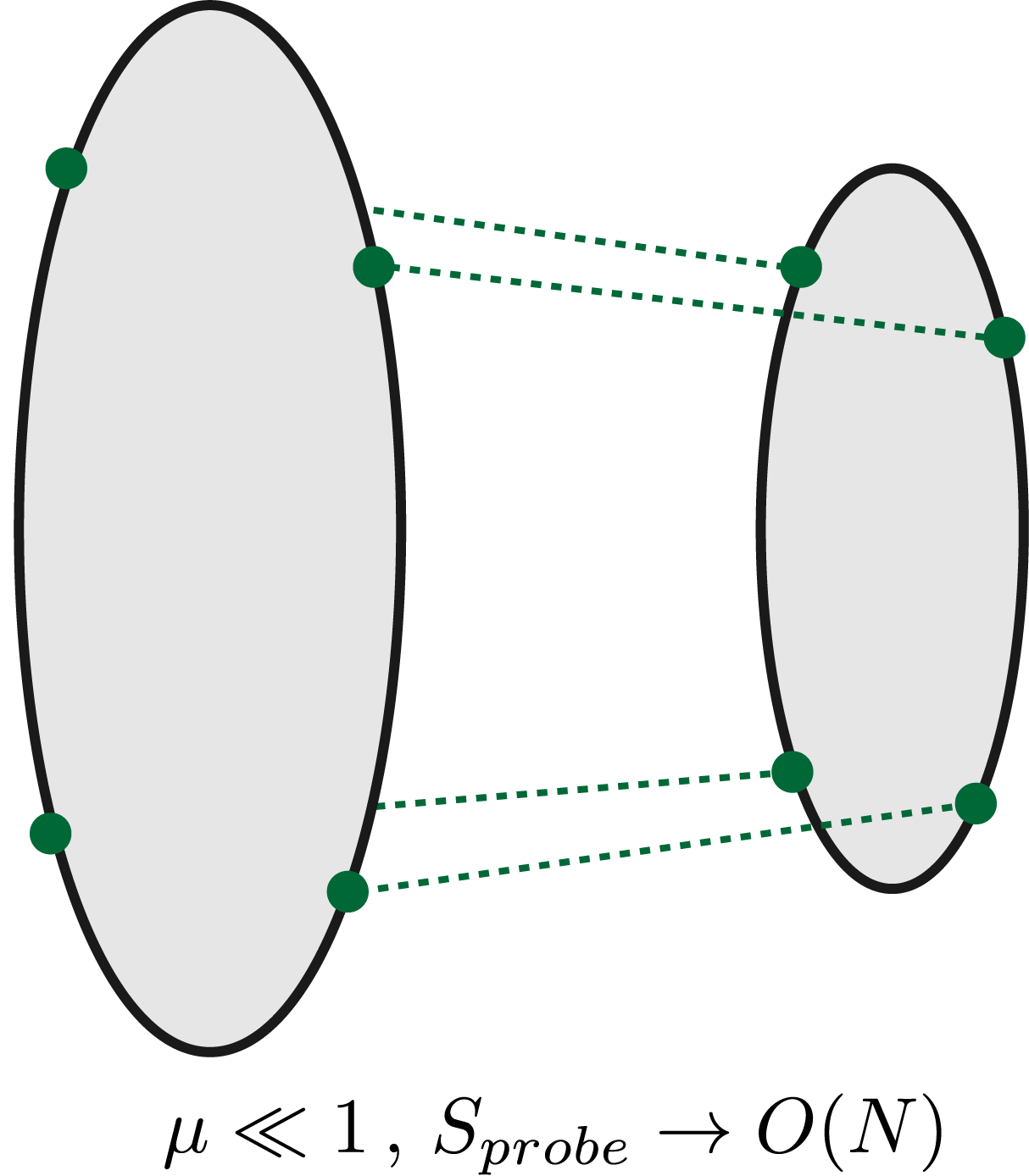}
\par\end{centering}
}\\
\subfloat[\label{fig:connected}]{\begin{centering}
\includegraphics[width=4.5cm]{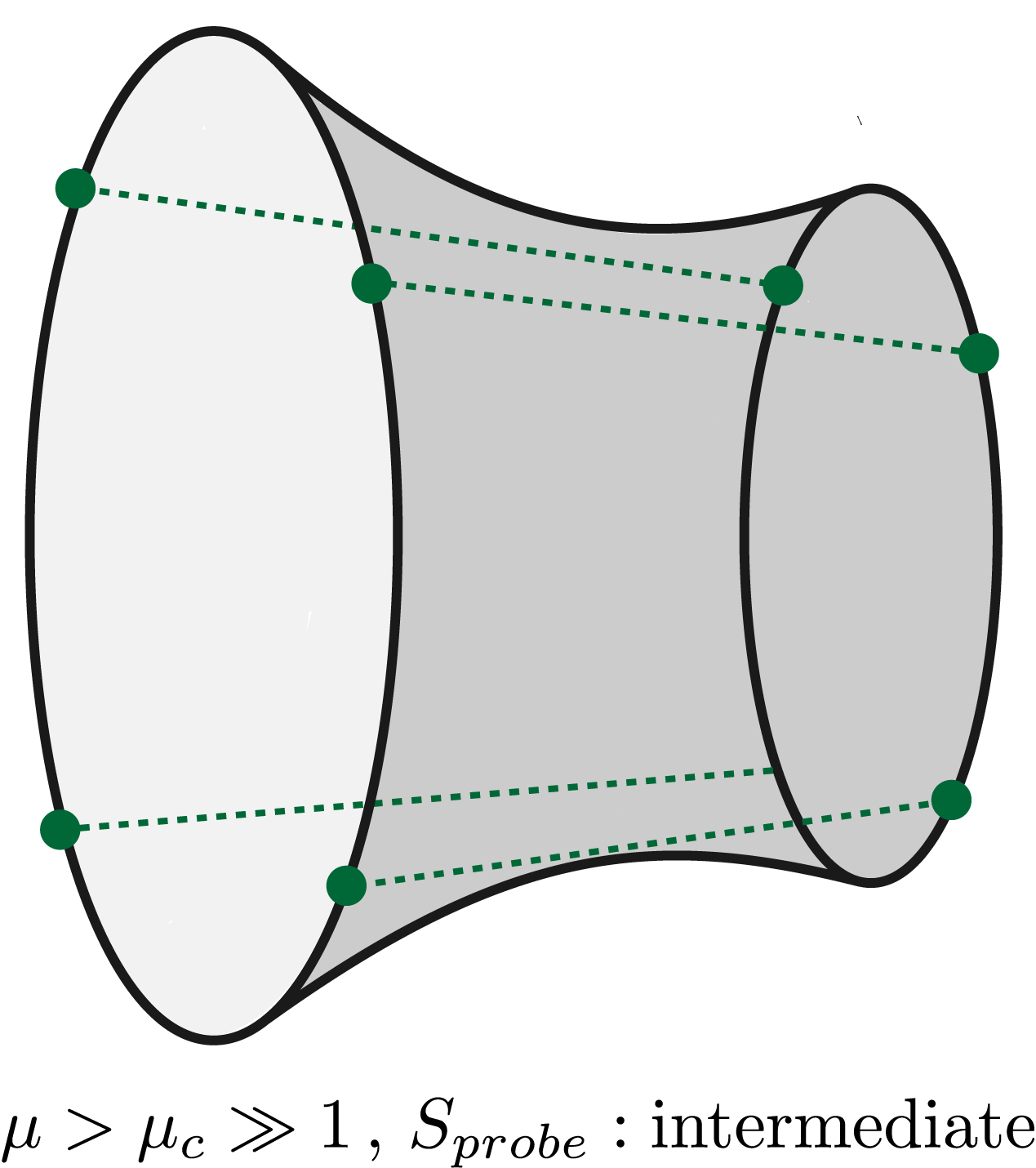}
\par\end{centering}
}
$\quad\quad\quad$\subfloat[\label{fig:degenerate}]{\begin{centering}
\includegraphics[width=4.5cm]{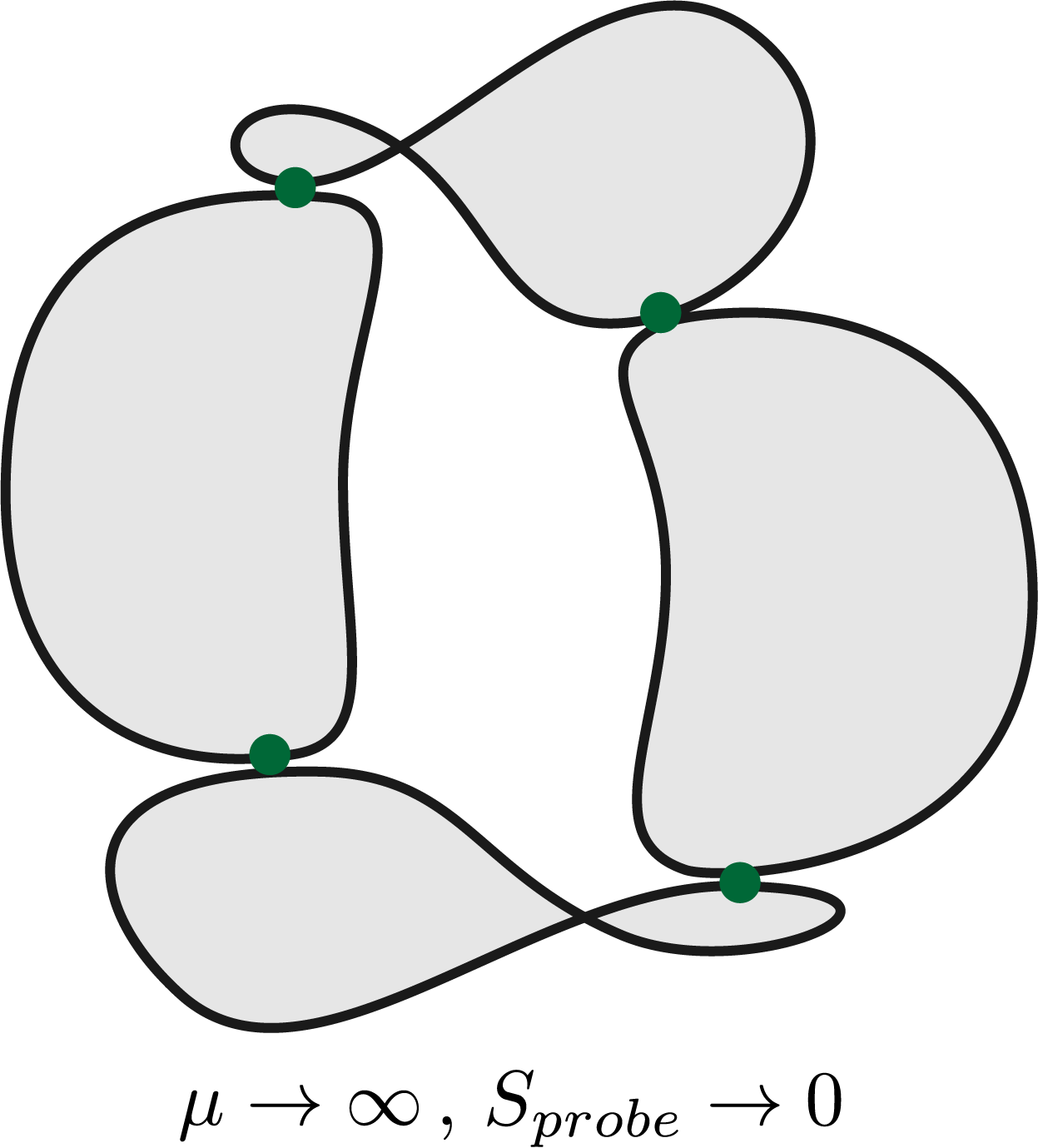}
\par\end{centering}
}
\par\end{centering}
\caption{(a) The ``necklace'' SYK diagram, summarizing the replica manifold for $k=4$ replicas. The green dotted lines connecting SYK$_l$ and SYK$_r$ correspond to local insertions of $\r_0=e^{-\mu S}$ where $S$ is the ``size'' operator (\ref{size}) and $\mu$ a parameter related to the entropy of the probe $S_{probe}$ and defined in Section~\ref{sec:prepare}. This coupling between the two boundary quantum systems appears after we trace out the reference and is a consequence of the entanglement between the probe and the reference. The modular flowed anticommutator (\ref{Wintro}) is obtained by an analytic continuation of the SYK propagator on this ``necklace" diagram. (b) The SYK ``necklace" diagram serves as the boundary condition for the Euclidean path integral of the dual JT gravity. In the limit of probe entropy the dominant saddle is a pair of disconnected geometries with disk topology, leading to trivial modular flow. (c) At intermediate values of the probe entropy for $\mu$ greater than a critical value $\mu_{cr}$, the Euclidean wormhole saddle with cylindrical topology dominates, supported by the $\r_0$ path integral insertions. The modular flowed commutator $W$ becomes non-trivial in this regime, allowing us to propagate into the black hole interior and detect signals sent from the other side. (d) At very small probe entropies, the backreaction of the $\r_0$ becomes large, squeezing the wormhole at the insertion points, and causing it to ``pinch off'' into a product of $k=4$ disconnected disks with perimeter $\beta_l+\beta_r$. Modular flow becomes trivial in this limit.  }
\end{figure}

We setup the SYK computation in Section \ref{sec:2}. We first prepare the SYK state dual to an AdS wormhole that contains a probe entangled with a reference, in Section \ref{sec:prepare} and \ref{sec:2.2}. We devote Section \ref{sec: bulkexpect} to the detailed discussion of the bulk trajectory followed by this infalling probe and the behavior of the bulk-to-boundary causal propagator as a function of the probe's proper time ---the object we aim to compute holographically. In order to perform the dual SYK computation of $W$ and test its agreement with this bulk expectation, we introduce a replica trick, explained in Section~\ref{sec:replica}, which translates the computation of $W$ to the evaluation of the SYK propagator on the Euclidean ``necklace'' diagram shown in Fig.~\ref{fig:necklaceBC}. In the rest of the paper, we present this computation from two different perspectives, using the microscopic SYK dynamics (Section~\ref{sec:3}) and the bulk JT path integral (Section~\ref{sec:4}), in an attempt to clarify the physics that underlies its success.

\begin{figure}
\begin{centering}
\subfloat[\label{fig:EucTFD}]{\begin{centering}
\includegraphics[height=4cm]{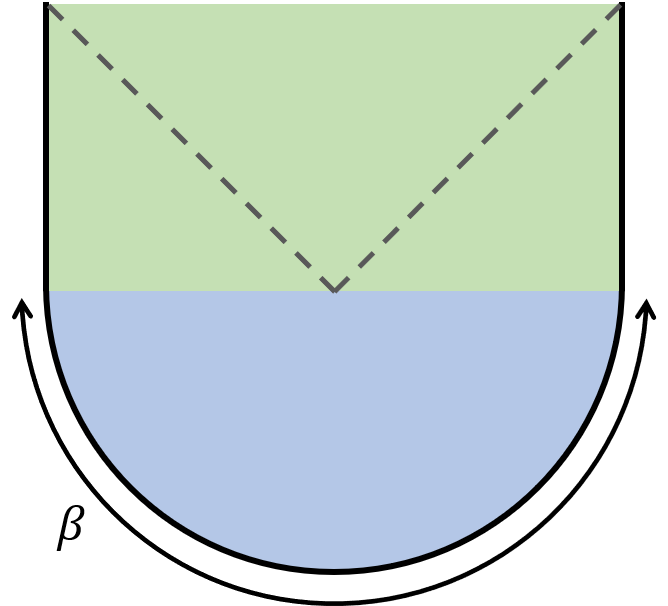}
\par\end{centering}
}\subfloat[\label{fig:Euclidean-path-integral}]{\begin{centering}
\includegraphics[height=4cm]{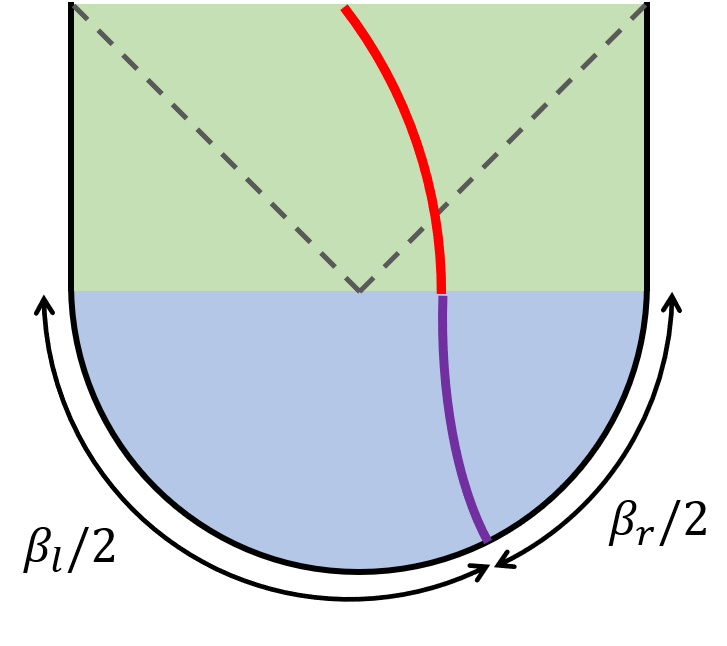}
\par\end{centering}
}$\quad$\subfloat[\label{fig:HKLL}]{\begin{centering}
\includegraphics[height=4cm]{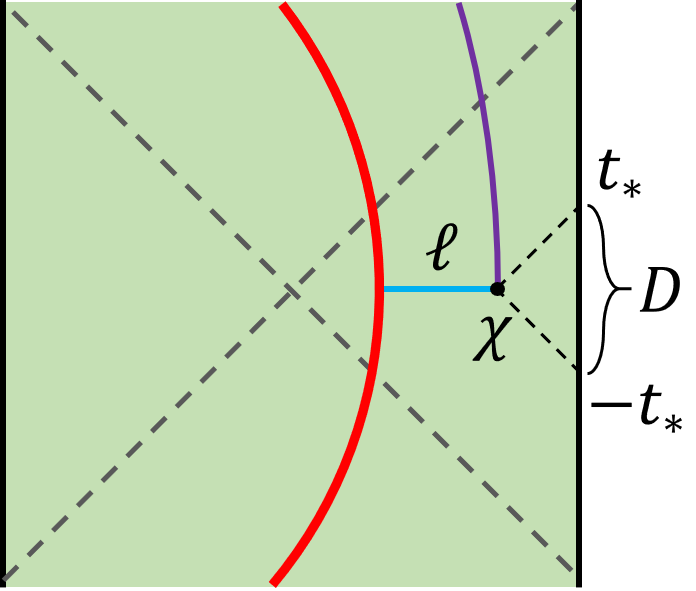}
\par\end{centering}
}
\par\end{centering}
\caption{(a) Euclidean path integral preparation of the thermofield double state. The blue half disk is Euclidean path integral and
the green strip is the Lorentzian continuation. (b) Euclidean path integral preparation of the thermofield double state
with a probe following geodesic (\ref{eq:geod}), which is plotted
as the red curve. The purple curve is
the Euclidean geodesic of the probe. (c) HKLL reconstruction of
a bulk spinor field $\chi$ (black dot) with $\protect\l$ distance
from the probe (red curve). Its boundary representation involves an integral of the HKLL kernel over the
boundary region $D(t_*)=[-t_{*},t_{*}]$ which is spacelike separated from
$\chi$. Translating the bulk field $\chi$, originally located outside the horizon, along the proper time of the red geodesic, while keeping its geodesic distance from this geodesic fixed (purple curve), allows us to probe the $AdS_2$ wormhole interior. In the dual SYK model, this proper time translation is generated by the modular flow $\rho^{is}$ of the red probe, after tracing out the reference system it is entangled with.}
\end{figure}

In order to pave the way for the subsequent technical analysis, Section \ref{sec:replica} offers some intuition for the behavior of the replica correlator in the limits of very large and very small probe entropy $S_{probe}$, showing that both lead to a trivial anticommutator $W(s,t)\ra 0$, for all  $s$, albeit for different reasons, and highlighting the importance of the intermediate $S_{probe}$ regime for getting interesting physics. In particular, $S_{probe}$ serves as an order parameter for the different phases of the dual Euclidean gravity path integral with the ``necklace'' diagram boundary conditions (Fig.~\ref{fig:necklaceBC}). At $S_{probe}\sim O(N)$ the dominant replica saddle consists of two disconnected disks associated with the left and right SYK boundary conditions, respectively ---a factorization that yields a modular flow that does not mix SYK$_l$ and SYK$_r$ hence $W(s,t)\ra 0$, for all  $s$. The bulk interpretation of this behavior comes from the large backreaction of our probe which elongates the ambient wormhole and destroys the shared interior region, rendering the infalling observer incapable of receiving causal signals from the other side. As we decrease $S_{probe}$, a new dominant JT saddle appears describing a Euclidean wormhole with cylindrical topology (Fig.~\ref{fig:connected}) which, however, degenerates again as we take the limit $S_{probe}\to 0$ (Fig.~\ref{fig:degenerate}). It is precisely this Euclidean wormhole phase in the intermediate $S_{probe}$ regime that generates an interesting anticommutator $W$ which reflects the reception of signals sent from the left exterior by the observer falling in from the right. The critical point of this phase transition is studied in Appendix \ref{app:e}. The remainder of our discussion is, thus, focused on studying this phase.

In Section~\ref{sec:3}, we perform the detailed computation working directly with the SYK dynamics, in a $1/q$ perturbative expansion. The computation amounts to obtaining the SYK propagator on the ``necklace'' diagram in Fig.~\ref{fig:necklaceBC}, with the different circles of the ``necklace" glued together via conditions determined by the unitary $U_{sys+ref}$ used to insert the probe as explained in Section~\ref{sec:3.1}. While an exact solution to the equations of motion cannot be obtained due to the strong symmetry constraints discussed in Section \ref{sec:symm} and further in Appendix \ref{app:a}, we find a consistent approximation in Section~\ref{sec:approxsoln} (with more technical details in Appendix \ref{app:slvrec}) that allows us to solve them in a wide parametric regime of interest that is specified in Appendix \ref{app:b}.  

The central ingredients of the computation are: (a) the quenched ensemble average over the random SYK couplings which connects dynamically the different replicas (circles of the ``necklace"), (b) the entanglement with the reference generated by $U_{sys+ref}$ which, after tracing out the latter, results in an explicit coupling between left and right SYKs in the replica diagram, and (c) the emergent $SL(2,R)$ symmetry controlling the maximally chaotic dynamics of the IR sector which captures the universal effect of this coupling on the SYK solution. The replica propagator can be approximately computed when the entropy of the probe is not too large, and after an appropriate analytic continuation discussed in Section~\ref{sec:3.4} it yields the expected bulk answer for $W$. This result can be combined with the standard HKLL reconstruction of bulk operators in the exterior of the black hole, in order to study the modular flow of a bulk field located at a finite distance from the infalling probe (Fig. \ref{fig:HKLL}). From this pure SYK computation, we can read off the precise proper time at which the signal sent from the left boundary is registered by our observer's apparatus in the wormhole interior!

In Section \ref{sec:4}, we present the same replica computation from the perspective of the Euclidean path integral of JT gravity. In Section \ref{sec:4.1}, we argue that the probe in the Euclidean path integral can be effectively understood as a localized injection of a fixed $SL(2,R)$ charge. The precise value of this charge constitutes UV data which we obtain from a microscopic SYK computation in Appendix \ref{app:d}. We explicitly construct the Euclidean wormhole solution dominating in the intermediate $S_{probe}$ regime in Section~\ref{sec:eucworm} using the method developed in \cite{Stanford:2020wkf}. 
The wormhole is supported by the localized couplings between the left and right boundaries generated by the entangling unitary $U_{sys+ref}$ after we trace out the reference. We show that the replica correlator computed in the bulk geodesic approximation exactly matches the microscopic SYK result in Section \ref{sec:4.3}. As anticipated, the length of this wormhole is controlled by the entropy of the probe and it pinches off in the limits $S_{probe}\sim N$ and $S_{probe}\to 0$ in two different ways, as shown in Fig.~\ref{fig:disconnected}, \ref{fig:degenerate}. It is precisely in the regime where the wormhole saddle dominates that the modular flow reliably takes us behind the horizon.

The Euclidean cylinder saddle found in Section~\ref{sec:4} is reminiscent of the one discussed in \cite{Stanford:2020wkf} and it hints, once again, at the important role played by the quenched ensemble average of the SYK couplings. Leveraging this intuition, we speculate in Section~\ref{sec:discussion} on how the analogous computation may work in more general setups and higher dimensions and conclude with some thoughts on interesting future applications of this method.

\section{A bulk infalling observer in SYK} \label{sec:2}

\subsection{Preparing the initial state} \label{sec:prepare}
In this paper, we wish to explicitly use the tool of \cite{Jafferis:2020ora} to access the behind the horizon region of two AdS$_2$ black holes connected by an Einstein-Rosen bridge, directly from the boundary quantum description. The first step in this process is to prepare the appropriate initial state, describing a wormhole geometry connecting two black hole exteriors, together with an ``observer'' inserted in the right asymptotic region whose microstates are entangled with an external reference.

An $AdS$ wormhole configuration is dual to a pair of identical holographic systems $l$ and $r$, dynamically decoupled ($H=H_l+H_r$) and in a special entangled state, the \emph{thermofield double state} \cite{maldacena2013cool}:
\be
    |\beta\rangle_{lr} \equiv \mathcal{Z}^{-\frac{1}{2}} \sum_a e^{-\frac{\beta E_a}{2}} |E_a\rangle_r |E_a\rangle_l =\mathcal{Z}^{-\frac{1}{2}}  e^{-\frac{\beta}{4}H} \bra 0_{lr}
\ee
where $|E_a\rangle_{l,r}$ are energy eigenstates of each system and $\bra 0_{lr}$ is the maximally entangled state of the two systems obeying $(H_l-H_r)\bra 0_{lr}=0$. For simpler notation, we will omit the subscript $lr$ in $\bra 0_{lr}$ from now on. For $AdS_2$, the dual boundary systems are two SYK models \cite{Maldacena:2017axo}. Each SYK model is a quantum mechanical system of $N$ Majorana fermions $\psi_{l,r}^{j}$
obeying Clifford algebra
\begin{equation}
\{\psi_{a}^{j},\psi_{b}^{k}\}=\d_{ab}\d^{jk},\quad a,b=l,r.
\end{equation}
The SYK Hamiltonian couples $1\ll q\ll N$ of them with coupling constants $J^{l,r}_{j_1 \cdots j_q}$ which are random variables drawn from a Gaussian ensemble:
\begin{align}
H_{l,r} & =i^{q/2}\sum_{1\leq j_{1}<\cdots<j_{q}\leq N}J_{j_{1}\cdots j_{q}}^{l,r}\psi_{l,r}^{j_{1}}\cdots\psi_{l,r}^{j_{q}}\\
\mathbb{E}_{J}\left[ J_{j_{1}\cdots j_{q}}^{l,r}\right]&=0\\
\mathbb{E}_{J}\left[\left(J_{j_{1}\cdots j_{q}}^{l,r}\right)^{2}\right]&=\f{2^{q-1}\mJ^{2}(q-1)!}{qN^{q-1}}=\f{J^{2}(q-1)!}{N^{q-1}} \label{eq:2.5}
\end{align}
The maximal entangled state is defined as
\begin{equation}
(\psi_{l}^{j}+i\psi_{r}^{j})\bra 0=0,\quad \forall j=1,\cdots,N
\end{equation}
which leads to $J_{j_{1}\cdots j_{q}}^{l}=i^{q}J_{j_{1}\cdots j_{q}}^{r}$. The state $|\beta\rangle_{lr}$ can be prepared via the standard SYK Euclidean path integral of Fig.~\ref{fig:EucTFD}. Its holographic representation is given by the path integral of JT gravity+matter over half of the hyperbolic disk $\mathbb{H}_2$.

\subsubsection*{Inserting the probe}
Suppose now we want to introduce a particle at the $t=0$ slice in the bulk, at some (regulated) geodesic distance $\rho$ from the right asymptotic boundary and initially at rest. We can do this simply by inserting a local operator in the path integral at a Euclidean time $\tau$ from the right endpoint (Fig.~\ref{fig:Euclidean-path-integral})
\begin{equation}
|\beta,\tau \rangle_{lr} = \mathcal{Z}^{-\frac{1}{2}}\,e^{-\frac{(\beta-\tau) H_l}{2}}e^{ -\frac{\tau H_r}{2}}O\,|0\rangle \label{probestate1}
\end{equation}
Assuming that $O$ is dual to a bulk field with large enough mass $(1\ll m_O \ll N)$, the operator in (\ref{probestate1}) inserts a classical particle in the bulk path integral that will propagate along the corresponding $\mathbb{H}_2$ geodesic (a semi-circle), until it hits the $t=0$ slice at distance $\rho$ from right asymptotic boundary and at a normal angle. This is precisely the initial state of interest and Lorentzian evolution will propagate the particle along an infalling geodesic, like in Fig. \ref{fig:Euclidean-path-integral}.

The formalism of \cite{Jafferis:2020ora}, however, requires our probe to have a large number of microstates which are entangled with an external reference system. Since the details of the reference do not matter, we can take it, for convenience, to be a system with $N$ free Dirac fermions $c_j$ and $c_j^\dagger$, which we initiate in the vacuum state $|v\rangle_{ref}$. We are then interested in a state of the type:
\begin{equation}
  |\beta,\tau \rangle_{l,r,ref} = \mathcal{Z}^{-\frac{1}{2}}\sum_{a} d_a \,e^{-\frac{(\beta-\tau) H_l}{2}}e^{ -\frac{\tau H_r}{2}}O_a\,|0\rangle\, O^{ref}_a|v\rangle_{ref} \label{probestate2}  
\end{equation}
where $d_a$ are complex coefficients. An explicit and computationally tractable example of such a state that we will use for our analysis, is one where the desired entanglement between the system and the reference is created by a unitary $U$, generated by a bi-local fermion operator:
\begin{align}
    |\beta_l,\beta_r;\delta\rangle &= \mathcal{Z}^{-\frac{1}{2}} \,e^{-\frac{\beta_l H_l}{2}}e^{ -\frac{\beta_r H_r}{2}} \,U(\d) |0\rangle |v\rangle_{ref} \label{probestate}\\
   U(\d)&= \exp\left[\sqrt{2}\d \sum_{j=1}^N\psi^j_r (c^\dagger_j+c_j)\right] \label{entanglingU}
\end{align}
and we set $\beta_l = \beta -\tau$, $\beta_r = \tau$. This state can be expressed in the form (\ref{probestate2}) by Taylor expanding the unitary, to get:
\begin{align}
     |\beta_l,\beta_r;\d\rangle& = \mathcal{Z}^{-\frac{1}{2}} \,e^{-\frac{\beta_l H_l}{2}}e^{ -\frac{\beta_r H_r}{2}}  \sum_{k=0}^{N} e^{-\frac{1}{2}\mu(\d) k} \sum_{I_k} \Gamma^r_{I_k}\, |0\rangle \, c^\dagger_{I_k}\, |v\rangle_{ref} \label{probestatesize} \\
     \mu(\d)& =\log\cot^2\d, ~~~~~I_k\equiv\{(i_1,\cdots,i_k)|1\leq i_1<\dots < i_k \leq N\}
\end{align}
where $c^\dagger_{I_k}\equiv c^\dagger_{i_1}\cdots c^\dagger_{i_k}$ generates fermion number basis of reference, and the Hermitian operators $\Gamma_{I_k} \equiv \Gamma^a_{i_1 i_2 \dots i_k} = 2^{k/2}i^{k(k-1)} \psi_a^{i_1} \dots \psi_a^{i_k} $  for $a=l,r$ are the ``size'' eigenoperators of \cite{Roberts:2018mnp, Qi:2018bje}. We will regard the state as perturbation on thermofield double and thus restrict to nonnegative $\mu(\s)$, which is equivalent to the coupling range $\d\in[0,\pi/4]$.

Tracing out the reference yields a reduced density matrix for the SYK$_l\times$SYK$_r$ system which reads:
\begin{align}
    \rho_{\beta_l,\beta_r,\mu} &=\mathcal{Z}^{-1}\, e^{-\frac{\beta_l H_l}{2}-\frac{\beta_r H_r}{2}}\,\,\sum_{k=0}^N e^{-\mu(\d)k}\sum_{I_k}\Gamma^r_{I_k}|0\rangle\langle 0|\Gamma^r_{I_k}  \,\, e^{-\frac{\beta_l H_l}{2}-\frac{\beta_r H_r}{2}} \nn\\
    &=\mathcal{Z}^{-1}\, e^{-\frac{\beta_l H_l}{2}-\frac{\beta_r H_r}{2}}\,\, e^{-\mu(\d) S} \,\, e^{-\frac{\beta_l H_l}{2}-\frac{\beta_r H_r}{2}} \label{proberho}
\end{align}
where 
\be
 S= \f 1 2\sum_{j=1}^N\left(1 + 2 i \psi^j_l \psi_r^j\right) \label{size}
\ee
is the ``size'' operator, defined and explored in a series of recent works \cite{Qi:2018bje, Nezami:2021yaq, Haehl:2021emt, Jian:2020qpp, Lensky:2020ubw, Gao:2019nyj, Lucas:2018wsc, Schuster:2021uvg}. It is clear from \eqref{proberho} that the entropy of probe $S_{probe}$ (which is the same as the entropy of $\r_{\beta_l,\beta_r,\mu}$) is $O(N)$ for $\mu(\d)\sim O(1)$. We are interested in probes that can be regarded as relatively small excitations of the thermofield double state, to avoid significant backreaction on the AdS$_2$ wormhole geometry we are trying to explore. We will, therefore, consider sufficiently small values $\d$, however, not small enough for the excitation to be approximated by a single fermion insertion. In this case, $S_{probe}$ is intermediate as illustrated in Fig. \ref{fig:connected}. More precisely, we will work in the limit $e^{-\mu(\d)}\ll 1$ and  $q, N, \beta \mJ \to \infty$ with $q/N\to 0$.\footnote{Technically, because of the large $\mu(\d)$ regime that we are interested in, it is illegal to approximate \eqref{proberho} as $\mZ^{-1}\exp\left(-\b_l H_l-\b_r H_r-\mu(\d)S\right)$ by combining three exponents, which differs our modular flow from the evolution in eternal traversable wormholes \cite{Maldacena:2018lmt}.} The parametric regime in which our calculation is valid is discussed in detail in Appendix \ref{app:b}.

\subsection{Setting up the SYK computation} \label{sec:2.2}
According to the prescription of \cite{Jafferis:2020ora}, modular flow of a right exterior bulk operator $O_r(s)=\rho^{-is}\phi_r \rho^{is}$, where $\rho$ is the left-right density operator (\ref{proberho}), translates $\phi_r$ along the geodesic of our infalling probe while keeping its geodesic distance from it fixed, with the modular time $s$ being proportional to the proper time along the worldline (Fig.~\ref{fig:HKLL}). We must emphasize that this prescription has certain important caveats discussed and resolved in \cite{Jafferis:2020ora} which, however, will not be relevant in this work. A central objective of this paper is to explicitly apply this proposal to holographically reconstruct operators in the black hole interior, in SYK.

An infalling observer's geodesic crosses the horizon of the 2-sided wormhole after a finite amount of proper time. Beyond this point, it is in causal contact with part of the left asymptotic boundary, which allows signals from the left boundary to reach our observer and influence their measurements. Such causally propagating signals are reflected in the appearance of a non-vanishing (anti-)commutator between left boundary operators $O_l(t)$ and right operators $O_r(s)$ that have been translated along the infalling geodesic.

We can, therefore, test the validity of this reconstruction in the black hole interior by computing quantum mechanically the correlator \eqref{Wintro} with average over all Majorana fermions
\begin{equation}
    W(s,t)= \f 1 N \sum_{j=1}^{N}\text{Tr}\left(\rho\, \{ \rho^{-is} \,\psi^j_r\,\rho^{is} ,\, \psi^j_l(t) \}\right) \label{commutator}
\end{equation}
which should be exponentially small for some finite range of $s$ and sharply reach a peak at some finite $s$. This peak signals that the flowed operator $\rho^{-is} \,\psi_r\,\rho^{is}$ has entered the bulk lightcone of the left boundary operator $\psi_l(t)$ (see Fig. \ref{fig:7a} and \ref{fig:7b}). More general modular flowed correlators of bulk exterior operators can be obtained from (\ref{commutator}) by smearing the fermions in boundary time with the known HKLL kernel (see Fig. \ref{fig:HKLL} and Section \ref{bulkbehind}). As we will show, the SYK solution to (\ref{commutator}) exactly matches semi-classical bulk computation reviewed in Section \ref{sec: bulkexpect}, in a parametric regime of $\beta_l,\beta_r,\mu$ we specify.

\subsection{Bulk semiclassical expectation} \label{sec: bulkexpect}

We start with a discussion of what the correlation function (\ref{commutator}) is expected to be, if the bulk interpretation of modular flow as proper time translations along the probe's worldline in the bulk dual is correct.
The two sided black holes spacetime is just a portion of
global AdS$_{2}$. We can describe AdS$_{2}$ as the hypersurface \cite{Maldacena:2016upp}
\begin{align}
-Y_{-1}^{2}-Y_{0}^{2}+Y_{1}^{2}&=-1
\end{align}
in a 3-dimensional embedding space with metric
\begin{align}
    ds^{2}=-dY_{-1}^{2}-dY_{0}^{2}+&dY_{1}^{2} \label{eq:2.17}
\end{align}

\begin{figure}
\begin{centering}
\subfloat[ $\xi=1$\label{fig:7a}]{\begin{centering}
\includegraphics[height=5cm]{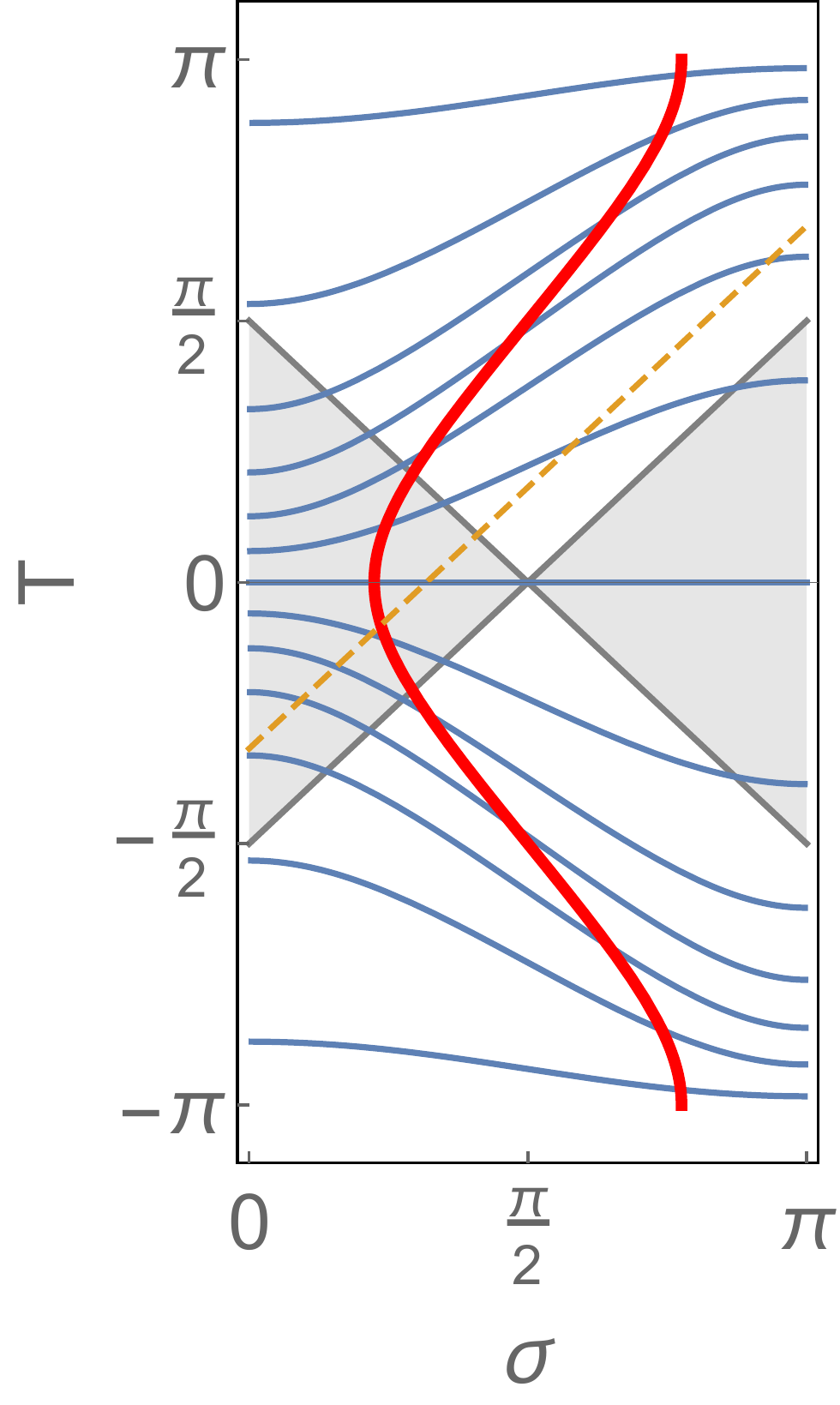}
\par\end{centering}
}\subfloat[ $\xi=-1$\label{fig:7b}]{\begin{centering}
\includegraphics[height=5cm]{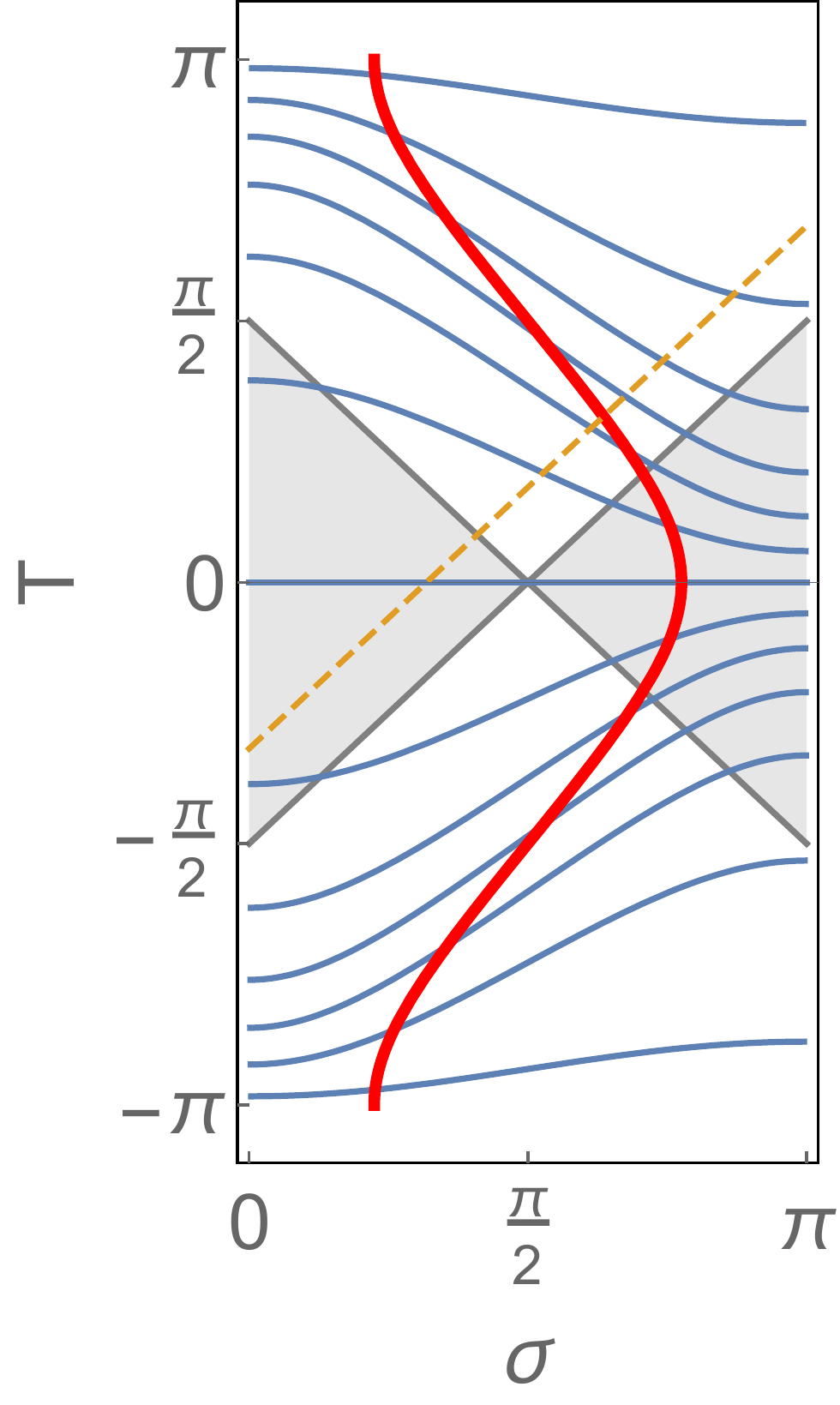}
\par\end{centering}
}\subfloat[\label{fig:7c}]{\begin{centering}
\includegraphics[height=5cm]{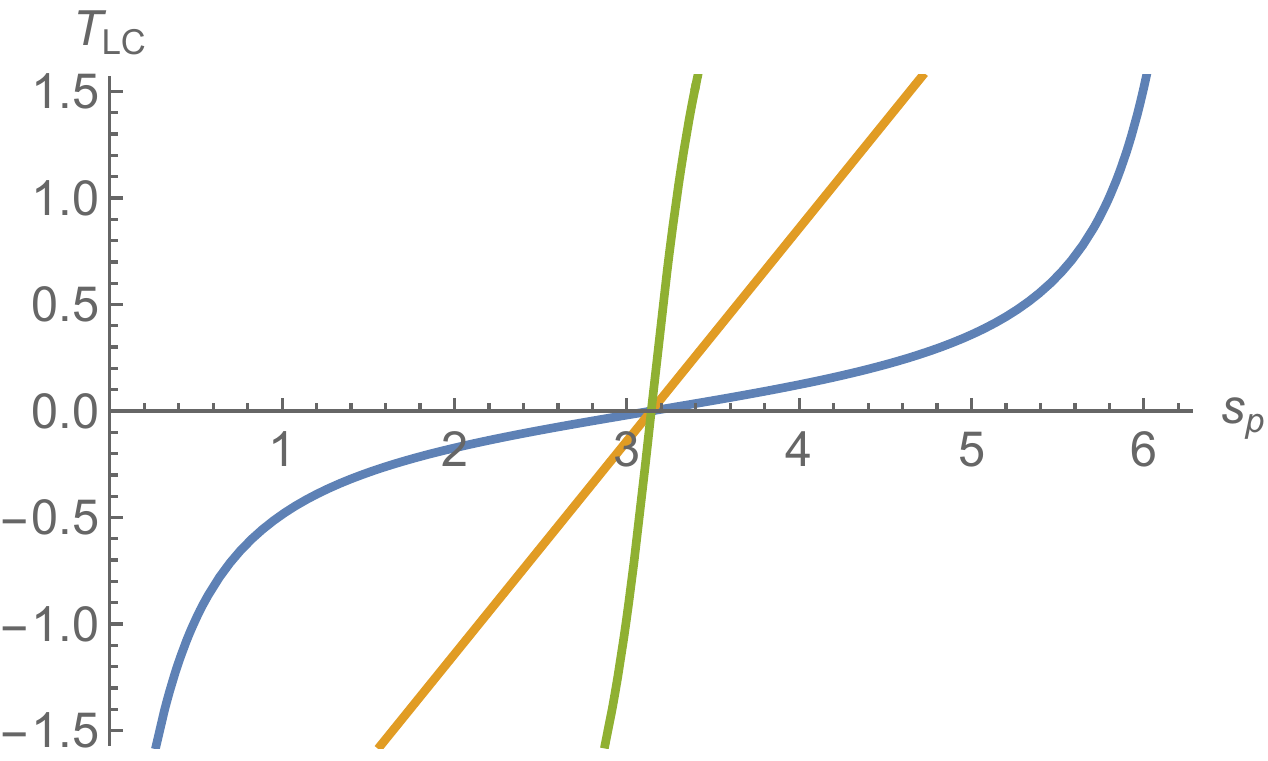}
\par\end{centering}
}
\par\end{centering}
\caption{(a)(b) The worldline of probe (red curve) and spatial geodesics with
equal $s_{p}$ separation and orthogonal to it (blue curves) in the
two sided big black holes spacetime. The two shaded regions are left
and right wedge respectively. The yellow dashed line is null and shot
from left boundary from $T=-1$. We see clearly that the probe takes
more proper time in (b) than (a) to reach the lightcone of the yellow
line. (c) The location of past lightcone location $T_{LC}$ on left
boundary of an atmosphere operator on right boundary after proper
time $s_{p}$ evolution. Blue, yellow and green curves are for $\xi=2,0,-2$.
\label{fig:(a)-The-worldline}}
\end{figure}
Parametrizing this surface as
\begin{equation}
Y_{-1}=\sin T\csc\s,\quad Y_{0}=\cos T\csc\s,\quad Y_{1}=-\cot\s
\end{equation}
yields the global AdS$_{2}$ metric 
\begin{equation}
ds^{2}=\f{-dT^{2}+d\s^{2}}{\sin^{2}\s},\quad\s\in[0,\pi],\,\,\,T\in\R
\end{equation}
The causal wedges of the left and right boundary in thermofield double
state (shaded regions in Fig. \ref{fig:7a} and Fig. \ref{fig:7b})
only extend for $T\in[-\pi/2,\pi/2]$ and the local boundary time $t_{l,r}$ is
defined as \cite{Maldacena:2016upp}
\begin{equation}
\tan\f T2=\tanh\f{\pi}{\b}t_{l,r}\label{eq:114}
\end{equation}
where $\b$ is the temperature of the thermofield double state.

AdS$_2$ has an $SO(2,1)$ symmetry whose embedding space representation reads:
\begin{align}
    M_1(x)=\begin{pmatrix}1 &0 & 0\\ 0
 & \cosh x & \sinh x\\
0  & \sinh x & \cosh x
\end{pmatrix},~ M_2(y) =\begin{pmatrix} \cosh y & 0 & \sinh y\\
 0 & 1 & 0\\
  \sinh y & 0 & \cosh y
\end{pmatrix},~
 M_3(\theta)&=\begin{pmatrix}
\cos\theta & -\sin\theta & 0 \\
\sin\theta  & \cos\theta &  0 \\
 0 &0 &1 
\end{pmatrix} \label{sl2rembedding}
\end{align}
The simplest timelike geodesic in AdS$_2$ is the worldline $\sigma =\frac{\pi}{2}$. In embedding coordinates this reads $U^{\mu}Y_{\mu}=0$ for $U^{\mu}=(0,0,1)$. Any other timelike geodesic can be obtained from this one by an $SO(2,1)$ transformation
\begin{equation}
U^{\mu}[M_1]^{\kappa}_{\mu}(\xi)[M_3]^\nu_\kappa(c)Y_{\nu}=0,\quad 
\end{equation}
The general timelike geodesic can be expressed as
\begin{equation}
\cos\s=r\cos (T-c),\qquad r\in(-1,1),\,\,\,c\in [-\pi,\pi] \label{eq:geod}
\end{equation}
where we set $\tanh\xi=r$. The parameter $c$ sets the timeslice at which the geodesic is instantenously at rest, in the global AdS frame. For a state prepared by a Euclidean path integral over the half disk, we should take $c=0$. The limits $r\ra\pm1$ correspond to null geodesics. On the $T=c=0$
slice, positive/negative $r$ corresponds to probe starting from left/right wedge, respectively (Fig. \ref{fig:7a} and Fig. \ref{fig:7b}).


\subsubsection*{Proper time flow}
The next step is to define a local bulk atmosphere operator 
by shooting a spacelike geodesic orthogonally from our probe's worldline at the initial time, and following it for proper length $\l$. We then want to propagate this operator along the timelike geodesic's proper time while keeping its relative location and angle to the probe's geodesic fixed. This is a natural choice of foliation related to the probe and is identical to the one used in \cite{Gao:2021uro} for the discussion of phase space variables of JT
gravity with dynamical EOW branes.

The spacelike geodesics orthogonal to $\sigma =\frac{\pi}{2}$
are $T=T_{0}$ for any $T_{0}$. In embedding space this reads $V^{\mu}Y_{\mu}=0$ with $V^{\mu}=(\cos T_{0},-\sin T_{0},0)$.
An initial bulk operator located at $(T,\s)=(0,\s_{0})$ is at a geodesic distance from the probe equal to
\begin{align}
\l=\int_{\pi/2}^{\s_{0}}\f{d\s}{\sin\s}=\f 12\log\f{1-\cos\s_{0}}{1+\cos\s_{0}}
\implies \cos\s_{0}=-\tanh\l\label{eq:109}
\end{align}
Propagation along the $\sigma= \frac{\pi}{2}$ geodesic for proper time $s_p=T_0$ then simply shifts the bulk operator to the global AdS point $(T_0,\s_{0})$.

Propagation along a general probe's geodesic (\ref{eq:geod}) can be obtained by a simple $SO(2,1)$ transformation of the above, since AdS isometries preserve both geodesic lengths and relative angles. Restricting our attention to probes that are at rest at global time $T=0$, the AdS location of a bulk operator at distance $\ell$ from the probe, translated along the geodesic (\ref{eq:geod}) for proper time $s_p=T_0$ is given by $(T_{b},\s_{b})$ determined by the equation
\begin{align}
Y_{\mu}&=[M_1]_{\mu}^{\nu}(\xi)Y_{\nu}^{(b)}\nn\\
 \Rightarrow\, &(\sin T_{b}\csc\s_{b},\cos T_{b}\csc\s_{b}\cosh\xi-\cot\s_{b}\sinh\xi,\cos T_{b}\csc\s_{b}\sinh\xi-\cot\s_{b}\cosh\xi)\nonumber \\
 & =(\sin T_{0}\csc\s_{0},\cos T_{0}\csc\s_{0},-\cot\s_{0}).
\end{align}
Using (\ref{eq:109}), we can solve that
\begin{align}
\cot\s_{b} & =\cos s_{p}\sinh\xi\cosh\l-\sinh\l\cosh\xi\label{eq:120}\\
\tan T_{b} & =\f{\sin s_{p}}{\cos s_{p}\cosh\xi-\tanh\l\sinh\xi}\label{eq:121}
\end{align}
The past lightcone of this atmosphere operator meets the left boundary
at $T_{LC}=T_{b}-\s_{b}$ which reads
\begin{align}
T_{LC} & =\arctan\f{\sin s_{p}}{\cos s_{p}\cosh\xi-\tanh\l\sinh\xi}-\arctan\f 1{\cos s_{p}\sinh\xi\cosh\l-\sinh\l\cosh\xi} \nn\\
& =2 \arctan \left(\f{\tan\f{s_{p}}2-e^{\l}}{e^{\xi}(1+e^{\l}\tan\f{s_{p}}2)} \right) \label{eq:140}
\end{align}
where the second arctan in the first line takes values in $[0,\pi]$. We plot $T_{LC}$ as a function of proper time $s_p$ in Fig. \ref{fig:7c} for an atmosphere operator near the right boundary, $\l\ra\infty$. From the plot, we see that only a finite range of $s_p$ leads to $T_{LC}\in[-\pi/2,\pi/2]$ as expected. Using \eqref{eq:114}, this lightcone crossing time corresponds to the left boundary time
\be
t_{LC}=\f{\b}{\pi}\arctanh\f{\tan\f{s_{p}}2-e^{\l}}{e^{\xi}(1+e^{\l}\tan\f{s_{p}}2)} \label{eq:1.30}
\ee

\subsubsection*{Matching to the path integral parameters} 
It is useful to express the lightcone crossing time (\ref{eq:1.30}) in terms of the parameters appearing in the path integral preparation of the state (Fig. \ref{fig:Euclidean-path-integral}). For this, we need to work out the relation between $\xi$ and $\b_{l,r}$ by considering the the purple curve in Fig. \ref{fig:Euclidean-path-integral} as a Euclidean geodesic. To compute this, it is convenient to switch to a different global coordinate system of EAdS$_2$
\begin{equation}
ds^{2}=d\r^{2}+\sinh^{2}\r d\tau^{2} \label{globaleads}
\end{equation}
where 
\begin{equation}
\cosh\r=\cosh T\csc\s,\quad\tan\tau=-\sinh T\sec\s
\end{equation}
The parameter $\tau$ in (\ref{globaleads}) is an angular coordinate on $\mathbb{H}_2$ and it is related to the Euclidean time of the boundary path integral $\tau_{\partial}$ via:
\begin{equation}
    \tau = \frac{2\pi \tau_{\partial}}{\beta}
\end{equation}
The geodesic that is orthogonal to the $T=0$
slice at $T=0$, $\s=\arccos r$ is (assuming $r\in[0,1]$)
\begin{equation}
\f{1+\sinh^{2}\r\sin^{2}\tau}{(\cosh\r+\sinh\r\cos\tau)^{2}}=\f{1+r}{1-r}
\end{equation}
and it meets the EAdS boundary ($\r\ra+\infty$) at Euclidean time
\begin{equation}
\tau=\pm\pi \d_{r}=\pm\arccos(-r)
\end{equation}
where we defined $\d_{l,r}\equiv\b_{l,r}/\b$. This leads to
\begin{equation}
\tan\pi\d_{l}/2=\cot\f{\arccos(-r)}2=\sqrt{\f{1-r}{1+r}}=e^{-\xi} \label{eq:1.35}
\end{equation}
where in the last step we used identity $\tanh\xi=r$.

\subsection{A replica-trick for modular flowed correlators} \label{sec:replica}
The rest of this paper is devoted to performing the computation of (\ref{commutator}) in two different ways and demonstrating its exact match to the bulk expectation \eqref{eq:1.30} with the parameter relation \eqref{eq:1.35}. Both of them rely on employing a replica trick: We first consider the correlator
\begin{align}
    W^{k,s}_{ab}(\tau_1, \tau_2)= \f 1 N \sum_{j=1}^N\text{Tr}\left[ \rho^{k-s} \psi^j_a(\tau_1) \rho^s \psi^j_b(\tau_2)\right]/\Tr\r^k \label{replicacorrelator}
\end{align}
where $\psi^j_a(\tau) \equiv e^{H_a \tau}\,\psi^j_a\, e^{-H_a \tau}$ and $a,b\in\{l,r\}$; we then obtain (\ref{commutator}) from (\ref{replicacorrelator}) via an appropriate analytic continuation in $k,s$ and $\tau_{1,2}$ with $(a,b)=(r,l)$. Here we average over all Majorana fermions in the SYK model. The replica correlator (\ref{replicacorrelator}) corresponds the SYK propagator on the ``necklace'' diagram (Fig.~\ref{fig:Necklace-diagram.-Splitting}). It is also important to remember that since SYK is a theory with random couplings, the correlator $W_{ab}^{k,s}$ refers to the statistical average over $J_{i_1,i_2,\dots, i_q}$ of the RHS, which we left implicit in  (\ref{replicacorrelator}). This fact will play a key role in our analysis and we will explicitly restore this ensemble average in formulas where it is important.

Before diving into the technical analysis of (\ref{replicacorrelator}), it is illuminating to first consider two extreme limits of the computation: $\mu \to 0$ and $\mu\to \infty$. 

\subsubsection*{(a) $\mu \to 0$ limit} 
Recalling the expression (\ref{proberho}) for the system's density matrix, we see that $\mu \to 0$ results in $e^{-\mu S}\to \mathbb{I}_{lr}$ and the state factorizes to a product of two thermal states for the left and right systems separately, with inverse temperatures $\beta_l$ and $\beta_r$ respectively

\begin{equation}
    \rho_{\beta_l,\beta_r, \mu\to 0} \to \mathcal{Z}^{-1} e^{-\beta_l H_l} e^{-\beta_r H_r} \label{mutozero}
\end{equation}
This limit corresponds to $\delta \to \frac{\pi}{4}$ in (\ref{probestate}) which yields a maximally entangled state between the probe and the reference, with $S_{probe}\to O(N)$. The factorization of $\rho$ in this limit implies that introducing a probe with a very large entropy destroys the correlations between SYK$_l$ and SYK$_r$ and by extension the common geometric interior of the AdS$_2$ wormhole we wish to probe.

The replica correlation function of interest, i.e. (\ref{replicacorrelator}) for $a=l$ and $b=r$, then becomes:
\begin{equation}
    W^{k,s}_{rl}(\tau_1, \tau_2) \overset{\mu\to 0}{\rightarrow} \frac{1}{N \mathcal{Z}} \sum_{j=1}^N \mathbb{E}_J\left[\text{Tr}_r[e^{-\beta_r k H_r} \psi_r^j(\tau_1)]\text{Tr}_l[e^{-\beta_l k H_l} \psi_l^j(\tau_2)] \right] \label{wmuto0}
\end{equation}
where we explicitly restored the (quenched) average over the random couplings $J_{j_1,j_2,\dots, j_q}$ implicit in all SYK computations. In the bulk, the computation of (\ref{wmuto0}) is dominated by the Euclidean gravitational path integral on two disconnected disks with circumferences $\beta_l k$ and $\beta_r k$ respectively (Fig.~\ref{fig:disconnected}), with the appropriate boundary fermion insertions on each side. This factorized contribution leads to an identically vanishing commutator (\ref{commutator}) for all $s,t$, consistently with the expectation that inserting a large entropy probe results in a long and potentially non-geometric wormhole and, as a consequence, the probe never enters a region that can be causally influenced by the left boundary. 

\subsubsection*{(b) $\mu \to \infty$ limit} The opposite limit, $\delta \to 0 \Rightarrow \mu(\d)\ra\infty$, in turn, yields $e^{-\mu(\d)S}\ra \bra{0} \ket{0}$ up to normalization and $\rho_{\beta_l,\beta_r,\infty}$ approaches the projector onto the thermofield double state $|\beta\rangle$ with inverse temperature $\beta = \beta_l +\beta_r$:
\begin{equation}
    \rho_{\beta_l,\beta_r,\mu \to \infty} \to |\b\rangle\langle \b| 
\end{equation}
The replica correlation function (\ref{replicacorrelator}) then reduces to:
\begin{equation}
    W^{k,s}_{rl}(\tau_1, \tau_2) \overset{\mu\to \infty}{\rightarrow} \begin{cases} \frac{1}{N} \sum_{j=1}^N \mathbb{E}_J\left[\langle \b| \psi_r^j(\tau_1) \psi_l^j(\tau_2) |\b\rangle \,\langle\b|\b\rangle^{k-1}\right] \, , \quad &\text{if: }s=0\\
    \frac{1}{N} \sum_{j=1}^N \mathbb{E}_J\left[\langle \b| \psi_r^j(\tau_1)|\b\rangle \langle \b| \psi_l^j(\tau_2) |\b\rangle \,\langle\b|\b\rangle^{k-2}\right] \, , \quad &\text{if: }s\neq 0
    \end{cases} \label{wmutoinfty}
\end{equation}
The bulk replica computation in this regime is dominated by a product of $k$ disconnected hyperbolic disks, each having a circumference $\b$ (Fig.~\ref{fig:degenerate}). Once again, this results in a vanishing commutator (\ref{commutator}) since this is physically the case of a probe with infinitesimally small entropy $S_{probe}\to 0$ and, thus, trivial modular flow.

\subsubsection*{(c) intermediate $\mu$} The two limits above make it clear that modular flow can only be interesting in the intermediate $\mu$ regime, when the probe has an entropy that is finite but small compared to that of the ambient black hole. We can gain some intuition for the behavior of the replica correlator for finite $\mu$, by approaching it from the $\mu\to 0$ side. First notice that $W^{k,s}_{rl}$ can be expressed as
\begin{align}
&W_{rl}^{k,s}(\tau_{1},\tau_{2})
= \f 1{N\mZ}\sum_{j=1}^{N}\E_{J}\left[\Tr[\r^{k-s}\psi_{r}^{j}(\tau_{r})\r^{s}\psi_{l}^{j}(\tau_{2})\right]\nonumber \\
 = & \f 1{N\mZ}\sum_{j=1}^{N} \E_J\left[\Tr\left[{\cal T} \left\{ e^{-k\b_{l}H_{l}-k\b_{r}H_{r}}\left(\prod_{\nu=0}^{k-1} e^{-\mu S(\nu+1/2)}\right) \psi_{r}^{j}(\tau_{1}+s\b_{r})\, \psi_{l}^{j}(\tau_{2}) \right\}\right]\right] \label{wfinitemu}  
\end{align}
where we defined $S(x)= e^{(\beta_l H_{l}+\beta_r H_r) x}\, S\,e^{-(\beta_l H_{l}+\beta_r H_r) x} $ and $\psi_{l,r}^j(x)=e^{H_{l,r} x}\,\psi_{l,r}^j\, e^{-H_{l,r} x}$, the operator $S$ is the size operator defined in (\ref{size}), ${\cal T}$ denotes Euclidean time ordering and the variables $\tau_{l,r}$ are restricted to the interval $\tau_{l,r}\in [0,\beta_{l,r}]$. As we take $\mu \to 0 $ in (\ref{wfinitemu}) we explicitly recover (\ref{wmuto0}).

The bulk AdS computation of (\ref{wfinitemu}) gets contributions from all geometries consistent with the boundary conditions set by ``necklace'' diagram (Fig.~\ref{fig:necklaceBC}). The two JT saddles of interest are: (a) the product of two disconnected hyperbolic geometries with disk topology and total perimeter lengths $k\beta_l$ and $k\beta_r$ respectively (Fig.~\ref{fig:disconnected}) and (b) the Euclidean wormhole geometry with cylindrical topology connecting the left and right boundaries (Fig.~\ref{fig:connected}). The latter is supported by the backreaction of the localized $\r_0=e^{-\mu S}$ insertions, since minimizing the corresponding potential energy $V(\mu) = \mu  \sum_{\nu=0}^{k-1}\langle S(\nu+1/2)\rangle$ favors large correlations between SYK$_l$ and SYK$_r$. The disconnected contribution cannot give rise to a non-trivial left-right commutator after analytic continuation. It is, therefore, the Euclidean wormhole saddle that describes the physics of our probe crossing the lightcone of the left boundary fermion ---when it dominates.

At small $\mu$, the insertion of $\r_0$ in (\ref{wfinitemu}) can be expanded perturbatively about $\mu=0$, and described as the insertion of $l$-$r$ bi-local operators, of low dimension. The backreaction of these bi-locals is small and thus a Euclidean wormhole supported by them would be very long, with a large JT gravity action, hence the disconnected geometry dominates (\ref{wfinitemu}). The computation in large $q$ SYK model in Appendix \ref{app:e} shows that, as $\mu$ increases, the backreaction of $\r_0$ on the Euclidean geometry leads, on the one hand, to a slow and bounded decrease of the action of the disconnected contribution, and, on the other hand, to a linear decrease of the action of the wormhole contribution (Fig. \ref{fig:phasetras}), whose length decreases as well. At a critical value $\mu_c$ the two saddles exchange dominance and the dominant contribution to (\ref{wfinitemu}) is given by the boundary-to-boundary propagator about the Euclidean wormhole geometry of Fig.~\ref{fig:connected}. The critical value $\mu_{cr} \sim 2\b \mJ/q^2$ is derived in Appendix \ref{app:e} for the large $q$ SYK model at low temperature. In the rest of this paper, we will only focus on $\mu >\mu_{cr}$ and this connected wormhole phase.

In Section \ref{sec:4}, we explicitly construct this bulk solution and the relevant propagator and show that its analytic continuation leads, indeed, to a modular flow consistent with the proper time translation interpretation discussed in Section~\ref{sec: bulkexpect}. The computation breaks down for very large values of $\mu$, when the wormhole pinches off to $k$ disconnected disks (Fig. \ref{fig:degenerate}).

\section{Replica computation in SYK} \label{sec:3}
In this Section, we perform the computation of (\ref{replicacorrelator}) and its analytic continuation by working directly on the boundary quantum theory and finding an approximate solution to the large $q$ SYK dynamics on the ``necklace" diagram (Fig.~\ref{fig:Necklace-diagram.-Splitting}). We make all our approximations explicit and bound the errors in our analysis and its parametric regime of validity in Appendix \ref{app:b}.  

\subsection{Large $q$ SYK on ``necklace" diagram} \label{sec:3.1}
\begin{figure}
\begin{centering}
\includegraphics[width=6cm]{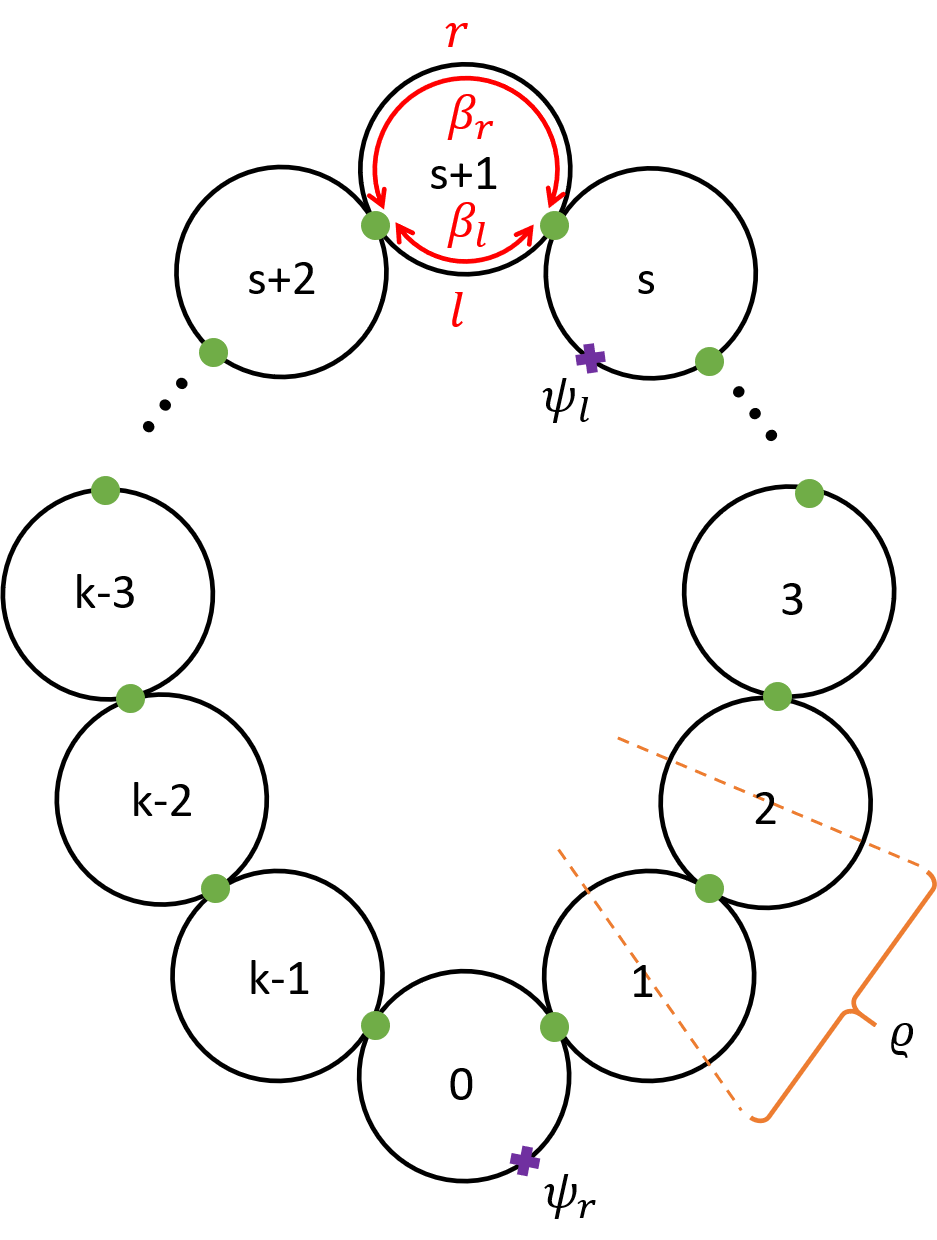}
\par\end{centering}
\caption{Necklace diagram. Splitting every circle into $l$ and $r$ on which
the system evolves with SYK Hamiltonian $H_{l,r}$ respectively. Each
green dot means insertion of $\protect\r_{0}$. \label{fig:Necklace-diagram.-Splitting}}
\end{figure}

As discussed in Section \ref{sec:prepare}, the density matrix of interest is, up to normalization:
\begin{equation}
\r=e^{-(\b_{l}H_{l}+\b_{r}H_{r})/2}\r_{0}e^{-(\b_{l}H_{l}+\b_{r}H_{r})/2}
\end{equation}
where
\begin{equation}
\r_{0}\equiv \exp\left(-i\mu \sum_{j=1}^{N}\psi_{l}^{j}\psi_{r}^{j}\right)
\end{equation}
We need to compute the correlation functions $W^{k,s}_{ab}(\tau_1, \tau_2)$ of $\psi_{a}^{j}$ \eqref{replicacorrelator} with
$k$ copies of $\r$ for positive integrer $k$ and nonnegative integer $s$ with $0\leq s\leq k$.
This amounts to computing correlation functions on the ``necklace" diagram
of Fig. \ref{fig:Necklace-diagram.-Splitting}.

Let us first compute
the correlation functions in the infinite $q$ limit, when both $l$ and $r$ SYK model Hamiltonians are zero.
In this case, the correlation is only affected by the insertion of
$\r_{0}$. Therefore, the correlation function is piecewise constant
and depends only on which circles the two fermions are located.
The first step, it to note the following identity
\begin{equation}
\r_0^{-1}\begin{pmatrix}\psi_{l}^{j}\\
i\psi_{r}^{j}
\end{pmatrix}\r_0=\begin{pmatrix}\cosh\mu & -\sinh\mu\\
-\sinh\mu & \cosh\mu
\end{pmatrix}\begin{pmatrix}\psi_{l}^{j}\\
i\psi_{r}^{j}
\end{pmatrix}\equiv M\begin{pmatrix}\psi_{l}^{j}\\
i\psi_{r}^{j}
\end{pmatrix}\label{eq:9}
\end{equation}
which means that whenever fermion crosses $\r_{0}$, the correlation
function is rotated by the matrix $M$. Let us define the 2 by 2 correlation
matrix as
\begin{equation}
g(s)=\f 1N\sum_{j=1}^{N}\Tr\left[\r_{0}^{k-s}\begin{pmatrix}\psi_{l}^{j}\\
i\psi_{r}^{j}
\end{pmatrix}\r_{0}^{s}\begin{pmatrix}\psi_{l}^{j}\\
i\psi_{r}^{j}
\end{pmatrix}^{\T}\right] \bigg/ \Tr\r_0^k
\end{equation}
in which we multiplied $\psi_{r}$ by $i$ for later convenience. For $s=0$, it is clear that
\begin{equation}
g(0)=\begin{pmatrix}\f 12 & -x\\
x & -\f 12
\end{pmatrix}\implies g(s)=M^{s}g(0)\label{eq:11}
\end{equation}
for some $x$ to be determined. The periodicity of the trace implies
that
\begin{equation}
g(k)=M^{k}\begin{pmatrix}\f 12 & -x\\
x & -\f 12
\end{pmatrix}=g(0)^{\T}=\begin{pmatrix}\f 12 & x\\
-x & -\f 12
\end{pmatrix}
\end{equation}
This can be easily solved by
\begin{equation}
x=\f 12\tanh\f{k\mu}2
\end{equation}
Plugging this solution back in (\ref{eq:11}), we have
\begin{equation}
g(s)=\f 1{2\cosh\f{k\mu}2}\begin{pmatrix}\cosh\f{(k-2s)\mu}2 & -\sinh\f{(k-2s)\mu}2\\
\sinh\f{(k-2s)\mu}2 & -\cosh\f{(k-2s)\mu}2
\end{pmatrix}
\end{equation}

Now let us move on to the SYK Hamiltonian. The necklace diagram describes the Euclidean
path integral of two SYK models on two different circles: the
$l$ circle has circumstance of $k\b_{l}$ and the $r$ circle has circumstance
of $k\b_{r}$. However, these two circles are not decoupled from each
other. The coupling comes from two sources: one is the identical
random coupling $J^{l,r}$, and the other is the localized insertion
of $\r_{0}$ after Euclidean evolution for $\b_{l,r}$. We will adopt
a hybrid treatment for these two types of couplings. For the former,
we integrate over the random coupling $J^{l,r}$ and manifest the
interaction between two circles; for the latter, we use (\ref{eq:9})
to transform the coupling into a specific gluing boundary condition
for correlations. It is crucial that the random couplings $J^{l,r}$ are identical for all replicas  and this leads to the quenched ensemble average when we integrate over $J^{l,r}$, otherwise the correlation between different replicas will be trivial. This quenched ensemble average is also important in the bulk and has been shown to be related to wormholes in recent studies \cite{Saad:2019lba, Engelhardt:2020qpv}. We will discuss more on this in Section \ref{sec:5.1}.


After integrating over random couplings, we have the following bilocal
effective action
\begin{equation}
S=-\f N2\log\det(\del_{\tau}\d_{ab}-\S_{ab})+\f N2\int_{0}^{k\b_{a}}d\tau\int_{0}^{k\b_{b}}d\tau'\left[\S_{ab}(\tau,\tau')G_{ab}(\tau,\tau')-\f{J^{2}}qs_{ab}G_{ab}(\tau,\tau')^{q}\right] \label{eq:action}
\end{equation}
where 
\begin{equation}
s_{ab}=\begin{pmatrix}1 & i^{q}\\
i^{q} & 1
\end{pmatrix}
\end{equation}
and $G_{ab}$ is the time ordered correlation function
\begin{equation}
G_{ab}(\tau_{1},\tau_{2})=\f 1N\sum_{j=1}^{N}\avg{\mT\psi^j_{a}(\tau_{1})\psi^j_{b}(\tau_{2})}_{necklace}
\end{equation}
which has the symmetry
\be
G_{ab}(\tau_1,\tau_2)=-G_{ba}(\tau_2,\tau_1) \label{eq:3.12}
\ee
It is important here to define a time ordering $\mT$ on the ``necklace" diagram, as follows. The ordering of fermions
with the same subscript ($a=b$) is as usual; for those with different subscripts $a\neq b$,
we first order them according to which necklace circle they are on,
and in case they are on the same circle we take the ordering as it is.

Taking variations of $\S$ and $G$ in \eqref{eq:action},
we have the equations of motion
\begin{equation}
G=(\del_{\tau}\d_{ab}-\S_{ab})^{-1},\quad\S_{ab}(\tau,\tau')=J^{2}s_{ab}G_{ab}(\tau,\tau')^{q-1}
\end{equation}
From the definition, we see that $G_{ab}$ is related to $g_{ab}$ by appropriate factor of $i$. To have a simpler notation later, we will define a parallel version of $G_{ab}$ with $\psi_r\ra i\psi_r$ and denoted by $\hat{g}_{ab}$. In the large $q$ limit, we make the standard assumption that the solution has the form
\begin{equation}
\hat{g}_{ab}(\tau_1,\tau_2)=g_{ab}(s)e^{\s_{ab}(\tau_1,\tau_2)/(q-1)},\quad s\equiv\left\lfloor \tau_{1}/\b_{a}\right\rfloor -\left\lfloor \tau_{2}/\b_{b}\right\rfloor \geq 0
\end{equation}
whose definition for $s<0$ is given by symmetry \eqref{eq:3.12}. At leading order in $1/q$, the equations of motion read
\begin{equation}
\del_{1}\del_{2}\s_{ab}(\tau_{1},\tau_{2})\pm2\mJ^{2}(2g_{ab}(s))^{q-2}e^{\s_{ab}(\tau_{1},\tau_{2})}=0
\end{equation}
with $+$ sign for $ab=ll,rr$ and $-$ sign for $ab=lr,rl$. This
is a piecewise Liouville equation whose general solution is
\begin{equation}
e^{\s_{ab}(\tau_{1},\tau_{2})}=\f{f'(\tau_{1})g'(\tau_{2})}{\mJ^{2}(2g_{ab}(s))^{q-2}(1\pm f(\tau_{1})g(\tau_{2}))^{2}}\label{eq:35}
\end{equation}
where $f$ and $g$ could be chosen differently
on different circles. Any solution of the above type has an $SL(2)$
symmetry
\begin{equation}
f\ra sl(f)\equiv\f{a+bf}{c+df},\quad g\ra sl_{\ad}(g)\equiv\f{d\mp cg}{\pm(-b\pm ag)},\quad bc-ad=1\label{eq:sym}
\end{equation}
We will use $\simeq$ to denote two pairs of function $(f,g)$ related
by this $SL(2)$ symmetry.

Since we are looking for a piecewise solution for $\s_{ab}$ and translation of both fermions for integer number of circles along the ``necklace" diagram does not change the solution, we will use a simpler notation by denoting $\s_{ab}^{s}(\tau_{1},\tau_{2})$ for $\s_{ab}(\tau_{1}+s\b_{a},\tau_{2})$
where $\tau_{1,2}\in[0,\b_{a,b}]$ from now on.

At every junction, the gluing boundary condition requires that 
\begin{align}
\hat{g}_{ab}(s\b_{a+},\tau_{2}) & =M_{ac}\hat{g}_{cb}(s\b_{a-},\tau_{2}),\quad s=1,\cdots,k\\
\hat{g}_{ab}(\tau_{1},s\b_{b+}) & =M_{bc}\hat{g}_{ac}(\tau_{1},s\b_{b-}),\quad s=1,\cdots,k
\end{align}
In terms of $\s^s_{ab}$, these conditions become
\begin{align}
e^{\s^{s+1}_{aa}(0,\tau)/q_-}-e^{\s^{s}_{aa}(\b_a,\tau)/q_-} & =\f{\sinh\mu\sinh\f{(k-2s)\mu}2}{\cosh\f{(k-2(s+1))\mu}2}\left(e^{\s^{s}_{aa}(\b_a,\tau)/q_-}-e^{\s^{s}_{\bar a a}(\b_{\bar a},\tau)/q_-}\right)\label{eq:21}\\
e^{\s^{s+1}_{a\bar a}(0,\tau)/q_-}-e^{\s^{s}_{a\bar a}(\b_a,\tau)/q_-} & =\f{\sinh\mu\cosh\f{(k-2s)\mu}2}{\sinh\f{(k-2(s+1))\mu}2}\left(e^{\s^{s}_{a\bar a}(\b_a,\tau)/q_-}-e^{\s^{s}_{\bar a \bar a}(\b_{\bar a},\tau)/q_-}\right)\label{eq:22}\\
e^{\s^{s-1}_{aa}(\tau,0)/q_-}-e^{\s^{s}_{aa}(\tau,\b_a)/q_-} & =\f{\sinh\mu\sinh\f{(k-2s)\mu}2}{\cosh\f{(k-2(s-1))\mu}2}\left(e^{\s^{s}_{a\bar a}(\tau,\b_{\bar a})/q_-}-e^{\s^{s}_{a a}(\tau,\b_{a})/q_-}\right)\label{eq:23}\\
e^{\s^{s-1}_{a\bar a}(\tau,0)/q_-}-e^{\s^{s}_{a\bar a}(\tau,\b_{\bar a})/q_-} & =\f{\sinh\mu\cosh\f{(k-2s)\mu}2}{\sinh\f{(k-2(s-1))\mu}2}\left(e^{\s^{s}_{a a}(\tau,\b_a)/q_-}-e^{\s^{s}_{a \bar a}(\tau,\b_{\bar a})/q_-}\right)\label{eq:24}
\end{align}
where $\bar a$ means ``$\neq a$" and $q_-\equiv q-1$. 
A special solution to the twist boundary condition
is to assume that both the left and the right hand sides of these conditions are separately zero.
This would mean that at each junction, all $\s_{ab}$ coincide. As we explain in Appendix \ref{app:a}, this is impossible to achieve using the configurations \eqref{eq:35}. Nevertheless, a somewhat relaxed gluing condition of this form will be used as an approximation in Section \ref{sec:approxsoln}, leading to a replica propagator that solves the SYK equations, up to a small error
in the large $\b,\mu, q$ limit. 

\subsection{Symmetries of $\s^s_{ab}$} \label{sec:symm}
In order to construct our SYK solution, it is helpful to understand the symmetries the propagator on the ``necklace'' diagram needs to satisfy. 

First, note that $\hat{g}_{ab}$ is real, which can be easily shown using the definition of SYK Hamiltonian and  $\r_0$ and using the Grassmann algebra. The complex conjugate of the replica correlator then satisfies
\begin{equation}
\Tr(\r^{k-s}\psi_{a}(\tau_{1})\rho^{s}\psi_{b}(\tau_{2}))^{*}=\Tr(\r^{k-s}\psi_{b}(-\tau_{2})\rho^{s}\psi_{a}(-\tau_{1}))
\end{equation}
which implies that
\begin{equation}
\s_{ab}^{s}(\tau_{1},\tau_{2})=\s_{ba}^{s}(\b_{b}-\tau_{2},\b_{a}-\tau_{1})\label{eq:62}
\end{equation}
Physically, we can understand this condition as a KMS condition along the each circle in the ``necklace" diagram. We will refer to this as the ``small KMS symmetry".

There is another symmetry for $s\neq0,k$ which becomes evident by noting that
\begin{equation}
\Tr(\r^{k-s}\psi_{a}(\tau_{1})\rho^{s}\psi_{b}(\tau_{2}))=\Tr(\rho^{s}\psi_{b}(\tau_{2})\r^{k-s}\psi_{a}(\tau_{1}))
\end{equation}
which implies 
\begin{equation}
\s_{ab}^{s}(\tau_{1},\tau_{2})=\s_{ba}^{k-s}(\tau_{2},\tau_{1})\label{eq:64}
\end{equation}
Together with (\ref{eq:62}) we have
\begin{equation}
\s_{ab}^{s}(\tau_{1},\tau_{2})=\s_{ab}^{k-s}(\b_{a}-\tau_{1},\b_{b}-\tau_{2})\label{eq:44-2}
\end{equation}
Physically, we can understand this condition as a KMS condition for the whole ``necklace" loop, which we dub the ``big KMS symmetry". For $s=0$ and $a\neq b$, we have
\begin{align}
\Tr(\r^{k}\psi_{a}(\tau_{1})\psi_{b}(\tau_{2}))&=-\Tr(\rho^{k}\psi_{b}(\tau_{2})\psi_{a}(\tau_{1}))\nn\\
\implies\s_{lr}^{0}(\tau_{1},\tau_{2})&=\s_{rl}^{0}(\tau_{2},\tau_{1})=\s_{lr}^{0}(\b_{l}-\tau_{1},\b_{r}-\tau_{2})\label{eq:65}
\end{align}
which extends (\ref{eq:44-2}) to the $s=0$ case. For the case $a=b$ becomes
\begin{align}
\s_{aa}^{0}(\tau_{1},\tau_{2}) & =\s_{aa}^{0}(\tau_{2},\tau_{1})\label{eq:43-2}\\
\s_{aa}^{0}(\tau,\tau) & =0\label{eq:44-3}
\end{align}
$\sigma_aa$, however, is not smooth along $\tau_{1}=\tau_{2}=\tau$
due to the coincident fermions. Instead, we may use (\ref{eq:43-2})
and (\ref{eq:44-3}) of $\tau_{1}\geq\tau_{2}$ to define the case
of $\tau_{1}\leq\tau_{2}$.

\subsection{Approximate solution} \label{sec:approxsoln}

The analysis of Appendix \ref{app:a} highlights the difficulty of finding an exact large $q$ solution that satisfies all twist boundary conditions \eqref{eq:21}-\eqref{eq:24} and also respects all symmetries discussed in Section \ref{sec:symm}. We will, therefore, make a strategic retreat and look for an approximate solution, whose error will later bound.

We are interested in the regime of large $\mu$ where the correlation functions in the same circle of the ``necklace'', say $s=0$, should be quite close to those in thermofield double state.
We will thus build our approximate solution for finite $\mu$ by starting with the thermofield double solution ($\mu\to \infty$). A special case of our
twisted boundary condition is to assume that $\s^s_{ab}$ is continuous
at all junctions. This means that all LHS of \eqref{eq:21}-\eqref{eq:24} are zero. Of course, this condition alone does not guarantee the RHS of \eqref{eq:21}-\eqref{eq:24} are also zero, but we can work with this assumption regardless and confirm at the end of the computation that the violation of the twisted boundary conditions is much smaller than $1/q$ in the low temperature
limit. Moreover, as analyzed in Appendix \ref{app:a} the ``big KMS symmetry" seems to be the main obstacle for obtaining an exact solution. As a fix, we construct an approximate solution by first finding a solution that violates the ``big KMS symmetry'' and then adding its KMS image
\begin{align}
\hat{g}_{ab}(s\b_{a}+\tau_{1},\tau_{2}) & \app g_{ab}(s)e^{\s_{ab}^{s}(\tau_{1},\tau_{2})/q}+g_{ba}(k-s)e^{\s_{ba}^{k-s}(\tau_{2},\tau_{1})/q}\label{eq:50}
\end{align}
for $0\leq s\leq k$ and then copy this solution antiperiodically
for other $s$. Of course, this approximation does not solve the
Liouville equation but we expect it to be very close to the real
solution in the low temperature limit. A similar argument was used in 
\cite{Saad:2018bqo}. Taking this approximation automatically satisfies the ``big KMS symmetry" (\ref{eq:64}).
We also show that our solution of $\s_{ab}^{s}$ guarantees the ``small KMS symmetry" (\ref{eq:62}). 

Let us first write down the solution for infinite $\mu$. In this case,
$\r_{0}$ reduces back to the projector onto the EPR state and any $s\neq0$ correlation
function is zero. For $s=0$, the correlation function is same as
that in a thermofield double state with temperature $\b=\b_{l}+\b_{r}$.
The solution is well known
\begin{align}
e^{\s_{ll}(\tau_{1},\tau_{2})} & =e^{\s_{rr}(\tau_{1},\tau_{2})}=\f{\w^{2}}{\mJ^{2}\cos^{2}\w(\tau_{12}-\b/2)}\label{eq:51}\\
e^{\s_{rl}(\tau_{1},\tau_{2})} & =e^{\s_{lr}(\tau_{1},\tau_{2})}=\f{\w^{2}}{\mJ^{2}\cos^{2}\w(\tau_{1}+\tau_{2}-\b/2)}\label{eq:52}
\end{align}
with 
\begin{equation}
\w=\mJ\cos\w\b/2
\end{equation}
One can easily check that this solution satisfies the symmetries (\ref{eq:62})
and (\ref{eq:65}).

For the case of large but finite $\mu$ we may still use the
aforementioned solution for $\s_{ab}^{0}$. To obtain the solution for $\s_{ab}^{s}$ in the other circles of the ``necklace'' 
we will assume continuity across the junctions 
\begin{equation}
\s_{ab}^{s}(\b_{a},\tau)=\s_{ab}^{s+1}(0,\tau),\quad\s_{ab}^{s}(\tau,0)=\s_{ab}^{s+1}(\tau,\b_{b})\label{eq:55-1}
\end{equation}
This condition is sufficient for obtaining all correlation functions, as we will show shortly. As usual, each solution $\sigma^s_{ab}$ of the Liouville equation is characterized by a pair of functions. By the argument of Appendix \ref{app:a}, the continuity condition leads to the following function choices
\be
\s_{ll}^s:=(f_s,f),~\s_{rr}^s:=(\bar{h}_s,h),~\s_{rl}^s:=(h_s,f)
\ee
where all functions $f_{s},h_{s},\bar{h}_{s},f,h$ are related by SL(2) transformations. In particular, the solution \eqref{eq:51} and \eqref{eq:52} correspond to
\begin{equation}
f=h=\tan\w\tau,\quad f_{0}=h_{0}=\bar{h}_{0}=\tan\w(\tau-\b/2)
\end{equation} 
The goal now is to use the continuity requirement to obtain this family of $SL(2,R)$ transformed functions in terms of the known $f,h,f_0,h_0,\bar{h}_0$.

\subsubsection*{$\protect\s_{ll}^{s}$ and $\protect\s_{rr}^{s}$ }

Let us first focus on $\s_{ll}^{s}$. We define
\begin{equation}
f_{s}=u_{s}+v_{s}\tan(\w\tau+\g_{s}),\quad\mJ_{s}\equiv\mJ(2g_{ll}(s))^{q-2}=\mJ\left[\f{\cosh\f{(k-2s)\mu}2}{\cosh\f{k\mu}2}\right]^{q-2}\label{eq:38-1}
\end{equation}
where $\{u_{k},v_{k},\g_{k}\}$ are three parameters characterizing
the $SL(2)$ transformation.

With this definition, we have 
\begin{equation}
e^{\s_{ll}^{s}(\tau_{1},\tau_{2})}=\f{\w^{2}v_{s}}{\mJ\mJ_{s}(\cos(\w\tau_{1}+\g_{s})(\cos\w\tau_{2}+u_{s}\sin\w\tau_{2})+v_{s}\sin\w\tau_{2}\sin(\w\tau_{1}+\g_{s}))^{2}}
\end{equation}
The boundary condition (\ref{eq:55-1}) can be solved by
\begin{align}
u_{s+1} & =\tan(\w\b_{l}+\g_{s})-\f 12\a_{s}v_{s}\sin2\g_{s+1}\sec^{2}(\w\b_{l}+\g_{s})\label{eq:41-1}\\
v_{s+1} & =\a_{s}v_{s}\cos^{2}\g_{s+1}\sec^{2}(\w\b_{l}+\g_{s})\label{eq:vrecur}\\
\tan\g_{s+1} & =\tan\g_{s}+\a_{s}v_{s}\sin\w\b_{l}\sec\g_{s}\sec(\w\b_{l}+\g_{s})\label{eq:grecur}\\
u_{s} & =(1-v_{s})\tan(\w\b_{l}+\g_{s})\label{eq:44-4}
\end{align}
where 
\begin{equation}
\a_{s}\equiv\mJ_{s+1}/\mJ_{s}=\left[\f{\cosh\f{(k-2(s+1))\mu}2}{\cosh\f{(k-2s)\mu}2}\right]^{q-2}
\end{equation}
which has symmetry $\a_{s}\a_{k-s-1}=1$. Note that in this solution,
(\ref{eq:41-1})-(\ref{eq:grecur}) are recurrence relation and (\ref{eq:44-4})
is a self-consistency condition for each $s$. In particular, one can
check that (\ref{eq:44-4}) holds at every level of the recurrence if it is satisfied initially.
Using (\ref{eq:44-4}) we can write $\s_{ll}^{s}$ as
\begin{align}
e^{\s_{ll}^{s}(\tau_{1},\tau_{2})}= & \f{\w^{2}v_{s}\cos^{2}(\w\b_{l}+\g_{s})}{\mJ\mJ_{s}\left[\cos(\w\tau_{1}+\g_{s})\cos(\w(\tau_{2}-\b_{l})-\g_{s})+v_{s}\sin\w\tau_{2}\sin\w(\tau_{1}-\b_{l})\right]^{2}}\label{eq:46-1}
\end{align}
In particular, $s=0$ corresponds to $v_{0}=1$ and $\g_{0}=-\w\b/2$.
One can easily check that this solution obeys symmetry (\ref{eq:62}). 

As $l$ and $r$ are identical systems, we can repeat above analysis
to $\s_{rr}^{s}$. The solution will be the same as $\s_{ll}^{s}$
but with replacement $\b_{l}\ra\b_{r}$ and parameters $\{u_{s},v_{s},\g_{s}\}\ra\{\bar{u}_{s},\bar{v}_{s},\bar{\g}_{s}\}$
related to $\bar{h}_{s}$.

\subsubsection*{$\protect\s_{rl}^{s}$ and $\protect\s_{lr}^{s}$ }

Solving $\s_{rl}^{s}$ is quite similar. We define
\begin{equation}
h_{s}=\tilde{u}_{s}+\tilde{v}_{s}\tan(\w\tau+\tilde{\g}_{s}),\quad\tilde{\mJ}_{s}\equiv\mJ(2g_{rl}(s))^{q-2}=\mJ\left[\f{\sinh\f{(k-2s)\mu}2}{\cosh\f{k\mu}2}\right]^{q-2}\label{eq:38}
\end{equation}
Taking ansatz (\ref{eq:38}) into (\ref{eq:35}), we have
\begin{equation}
e^{\s_{rl}^{s}(\tau_{1},\tau_{2})}=\f{\w^{2}\tilde{v}_{s}}{\mJ\tilde{\mJ}_{s}(\cos(\w\tau_{1}+\tilde{\g}_{s})(\cos\w\tau_{2}-\tilde{u}_{s}\sin\w\tau_{2})-\tilde{v}_{s}\sin\w\tau_{2}\sin(\w\tau_{1}+\tilde{\g}_{k}))^{2}}
\end{equation}
The boundary condition (\ref{eq:55-1}) can be solved by
\begin{align}
\tilde{u}_{s+1} & =\cot\w\b_{l}-\cos\g_{s}\csc\w\b_{l}\sec(\w\b_{r}+\tilde{\g}_{s})-\f 12\tilde{\a}_{s}\tilde{v}_{s}\sec^{2}(\w\b_{r}+\tilde{\g}_{s})\sin2\tilde{\g}_{s}\label{eq:41}\\
\tilde{v}_{s+1} & =\tilde{\a}_{s}\tilde{v}_{s}\cos^{2}\tilde{\g}_{s+1}\sec^{2}(\w\b_{r}+\tilde{\g}_{s})\label{eq:vtrecur}\\
\tan\tilde{\g}_{s+1} & =\tan\tilde{\g}_{s}-\tilde{\a}_{s}\tilde{v}_{s}\sin\w\b_{l}\sec\tilde{\g}_{s}\sec(\w\b_{r}+\tilde{\g}_{s})\label{eq:gtrecur}\\
\tilde{u}_{s} & =\cot\w\b_{l}-\cos\tilde{\g}_{s}\csc\w\b_{l}\sec(\w\b_{r}+\tilde{\g}_{s})-\tilde{v}_{s}\tan(\w\b_{r}+\tilde{\g}_{s})\label{eq:44}
\end{align}
where 
\begin{equation}
\tilde{\a}_{s}\equiv\tilde{\mJ}_{s+1}/\tilde{\mJ}_{s}=\left[\f{\sinh\f{(k-2(s+1))\mu}2}{\sinh\f{(k-2s)\mu}2}\right]^{q-2}
\end{equation}
which has symmetry $\tilde{\a}_{s}\tilde{\a}_{k-s-1}=1$. Again, in
this solution, (\ref{eq:41})-(\ref{eq:gtrecur}) are recurrence relation
and (\ref{eq:44}) is a self-consistency condition for each $s$. Using (\ref{eq:44}) we can write $\s_{rl}^{s}$
as
\begin{align}
e^{\s_{rl}^{s}(\tau_{1},\tau_{2})}= & \f{\w^{2}\tilde{v}_{s}\cos^{2}(\w\b_{r}+\tilde{\g}_{s})\sin^{2}\w\b_{l}}{\mJ\tilde{\mJ}_{s}}\left[\cos(\w\tau_{1}+\tilde{\g}_{s})(\cos\tilde{\g}_{s}\sin\w\tau_{2}\right.\nn\\
&\left. -\cos(\w\b_{r}+\tilde{\g}_{s})\sin\w(\tau_{2}-\b_{l}))-\tilde{v}_{s}\sin\w\b_{l}\sin\w\tau_{2}\sin\w(\tau_{1}-\b_{r})\right]^{-2}\label{eq:46}
\end{align}
In particular, $s=0$ corresponds to $\tilde{v}_{0}=1$ and $\tilde{\g}_{0}=-\w\b/2$. 

To get $\s_{lr}^{s}$, we can simply use symmetry (\ref{eq:62}).
However, we also need to check this procedure is consistent with our
boundary condition (\ref{eq:55-1}) that defines above recurrence
sequence. This turns out to be the case simply because (\ref{eq:55-1})
also respects the symmetry (\ref{eq:62}). In other words, taking $ab=rl$
in (\ref{eq:55-1}) together with the symmetry (\ref{eq:62}) exactly leads to
$ab=lr$ in (\ref{eq:55-1}). 

\subsubsection*{Approximate solution of recurrence}
Solving these recurrence relations exactly and in closed form is a difficult task. Instead, we will leverage the observation that these recurrence sequences converge very fast and can be well approximated by their continuous version which are second order differential equations. Solving the differential equations leads to an approximate solution of the recurrence sequence and also offers a closed form  which is required for the subsequent analytic continuation we want to perform. We perform this computation in Appendix \ref{app:slvrec} and present the result here.

Let us define the recurrence variables
\begin{align}
y_{s} & =\f{\cos(\w\b_{l}+\g_{s})}{\cos\g_{s}},\quad x_{s}=v_{s}\sec^{2}\g_{s},\quad\lambda=\sin^{2}\w\b_{l}\\
\tilde{y}_{s} & =\f{\cos(\w\b_{r}+\tilde{\g}_{s})}{\cos\tilde{\g}_{s}},\quad\tilde{x}_{s}=\tilde{v}_{s}\sec^{2}\tilde{\g}_{s},\quad\tilde{\lambda}=\sin\w\b_{l}\sin\w\b_{r}
\end{align}
Their continuous versions obeying the aforementioned differential equations are denoted by exchanging the subscript $s$ for a variable $s$, e.g. $y_s\ra y(s)$ etc. In the large $\mu$ limit, the solution is
\begin{align}
y(s)&=\begin{cases}
\a^{1/2}\exp [c_{1}\coth(c_{1}s+b_{1})] & s\leq\left\lfloor k/2\right\rfloor \\
\a^{-1/2}\exp[ c_{2}\coth(c_{2}s+b_{2})] & s>\left\lfloor k/2\right\rfloor 
\end{cases}\label{eq:80-x}\\
x(s)&=\begin{cases}
\f{x_{0}\sinh^{2}b_{1}}{\sinh^{2}(c_{1}s+b_{1})} & s\leq\left\lfloor k/2\right\rfloor \\
\f{x_{0}\sinh^{2}b_{1}\sinh^{2}(c_{2}\left\lfloor k/2\right\rfloor +b_{2})}{\sinh^{2}(c_{1}\left\lfloor k/2\right\rfloor +b_{1})\sinh^{2}(c_{2}s+b_{2})} & s>\left\lfloor k/2\right\rfloor 
\end{cases}\label{eq:84-x}\\
\tilde{y}(s)&=\begin{cases}
\a^{1/2}\exp[\tilde{c}_{1}\tanh(\tilde{c}_{1}s+\tilde{b}_{1})] & s\leq\left\lfloor k/2\right\rfloor \\
\a^{-1/2}\exp[\tilde{c}_{2}\tanh(\tilde{c}_{2}s+\tilde{b}_{2})] & s>\left\lfloor k/2\right\rfloor 
\end{cases}\label{eq:86-x}\\
\tilde{x}(s)&=\begin{cases}
\f{\tilde{x}_{0}\cosh^{2}\tilde{b}_{1}}{\cosh^{2}(\tilde{c}_{1}s+\tilde{b}_{1})} & s\leq\left\lfloor k/2\right\rfloor \\
\f{\tilde{x}_{0}\cosh^{2}\tilde{b}_{1}\cosh^{2}(\tilde{c}_{2}\left\lfloor k/2\right\rfloor +\tilde{b}_{2})}{\cosh^{2}(\tilde{c}_{1}\left\lfloor k/2\right\rfloor +\tilde{b}_{1})\cosh^{2}(\tilde{c}_{2}s+\tilde{b}_{2})} & s>\left\lfloor k/2\right\rfloor 
\end{cases}\label{eq:84-1-x}
\end{align}
in which $\a\equiv e^{-\mu(q-2)}\app e^{-\mu q}$ and other parameters are defined as
\begin{align}
c_{1} & =\log(y_{\infty}/\a^{1/2}),\quad b_{1}=\text{arccoth}(\log(y_{0}/\a^{1/2})/\log(y_{\infty}/\a^{1/2}))\\
c_{2} & =\log(y_{\infty}\a^{1/2}),\quad b_{2}=\text{arccoth}\left(\f{\log\a+c_{1}\coth(c_{1}\left\lfloor k/2\right\rfloor +b_{1})}{c_{2}}\right)-c_{2}\left\lfloor k/2\right\rfloor \\
\tilde{c}_{1} & =\log(\tilde{y}_{\infty}/\a^{1/2}),\quad\tilde{b}_{1}=\text{arctanh}(\log(\tilde{y}_{0}/\a^{1/2})/\log(\tilde{y}_{\infty}/\a^{1/2}))\\
\tilde{c}_{2} & =\log(\tilde{y}_{\infty}\a^{1/2}),\quad\tilde{b}_{2}=\text{arctanh}\left(\f{\log\a+\tilde{c}_{1}\tanh(\tilde{c}_{1}\left\lfloor k/2\right\rfloor +\tilde{b}_{1})}{\tilde{c}_{2}}\right)-\tilde{c}_{2}\left\lfloor k/2\right\rfloor 
\end{align}
where $y_\infty$ and $\tilde{y}_\infty$ are limit values of recurrence sequences for which a closed form is presented in \eqref{eq:yinfty2} and \eqref{eq:yinfty}. However, their exact formula is not needed since they can reliably be approximated as $y_1$ and $\tilde{y}_1$ in large $\b$ and small $\a$ limit.

In terms of $x(s)$, $y(s)$, $\tilde{x}(s)$ and $\tilde{y}(s)$, the large $q$ solution becomes 
\begin{align}
 g_{ll}(s)& e^{\s_{ll}^{s}(\tau_{1},\tau_{2})/q}
=\f 12\left(\w\lambda \mJ^{-1}x(s)^{1/2}y(s)[(\sin\w(\b_{l}-\tau_{1})+y(s)\sin\w\tau_{1})\right.\nn\\
&\left. \times (\sin\w\tau_{2}+y(s)\sin\w(\b_{l}-\tau_{2}))-\lambda x(s)\sin\w\tau_{2}\sin\w(\b_{l}-\tau_{1})]^{-1}\right)^{2/q} \\
g_{rl}(s)&e^{\s_{rl}^{s}(\tau_{1},\tau_{2})/q}
=  \f{\sgn(g_{rl}(s))}2\left(\w\tilde{\lambda}\mJ^{-1}\tilde{x}(s)^{1/2}\tilde{y}(s)[(\sin\w(\b_{r}-\tau_{1})+\tilde{y}(s)\sin\w\tau_{1})\right. \nn\\
&\left.\times(\sin\w\tau_{2}+\tilde{y}(s)\sin\w(\b_{l}-\tau_{2}))+\tilde{\lambda}\tilde{x}(s)\sin\w\tau_{2}\sin\w(\b_{r}-\tau_{1})]^{-1}\right)^{2/q}
\end{align}
For $\s_{rr}^{s}$ and $\s_{lr}^{s}$, we can simply switch $\b_{l}\leftrightarrow\b_{r}$.
Note that to get $\s_{lr}^{s}$, we can also use symmetry (\ref{eq:62}),
which turns out to be the same as the swap $\b_{l}\leftrightarrow\b_{r}$.

It is worth recalling at this point, that the solution we obtained above is an approximate one, in a number of different ways. First and foremost, this solution does not exactly satisfy the twisted gluing conditions at the junctions of the ``necklace'' diagram. In Appendix \ref{app:slvrec}, we confirm that the errors of this approximation, namely the deviation of the RHS of \eqref{eq:21}-\eqref{eq:24} from zero, are much smaller than $1/q$ in the large $\mu,\b$ limit (see Fig.~\ref{fig:The-errors-of}). In Appendix \ref{app:b}, we present a further systematic analysis of the errors introduced by all the approximations we make, in order to justify the validity of our solution in large $\mu,\b$ limit.

\subsection{Analytic continuation} \label{sec:3.4}

\begin{figure}
\begin{centering}
\includegraphics[width=6cm]{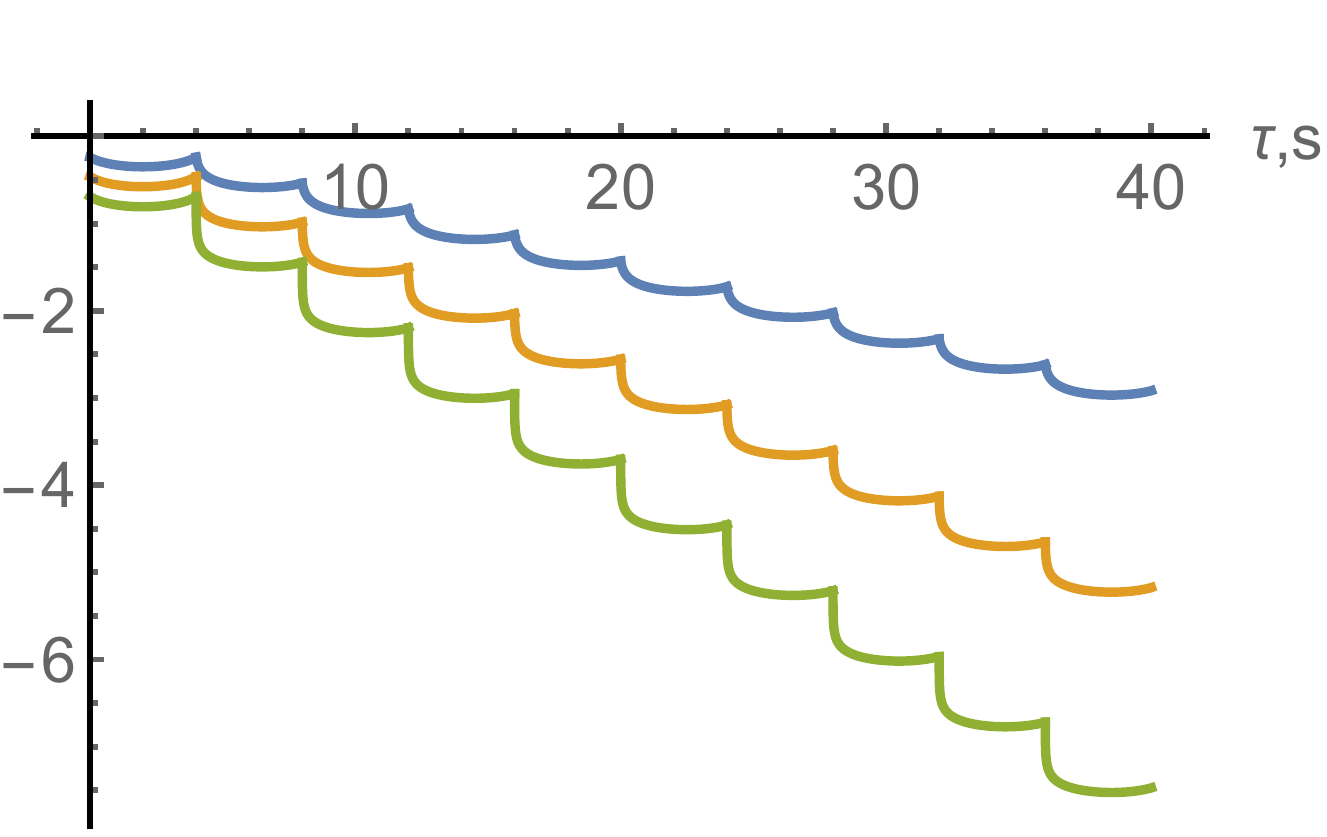}
\par\end{centering}
\caption{The plot of $\protect\s_{rl}^{s}(\tau,\protect\b_{l}/2)/q$, where
different $s$ are joined together in order. Blue, yellow and green
curves are for $\protect\mJ=20,200,2000$ respectively. We see the
correlation decays exponentially as $s$ increases and the decay is
stronger when we increase $\protect\mJ$. Here other parameters are
$\protect\b_{l}=1$, $\protect\b_{r}=4$, $\protect\a=1/500$, $q=20$
and $k=9$. \label{fig:The-plot-of}}
\end{figure}

Let us now return to the physical question of interest. The quantity we want to compute is the causal correlator \eqref{commutator}, which we restate for convenience
\begin{equation}
W(s,t)=\f 1 N\sum_{j=1}^N \Tr\left(\r\{\r^{-is}\psi^j_{r}\r^{is},\psi^j_{l}(t)\}\right)
\end{equation}
The right SYK operator is evolved with the modular Hamiltonian $\rho^{is}$ --- which is expected to be the SYK dual of the proper time evolution along the infalling probe's worldline. The anti-commutator with the left boundary insertion is intended to detect the moment $\rho^{-is} \psi_r \rho^{is}$ crosses the bulk lightcone of $\psi_l(t)$ in the wormhole interior. 

The causal propagator can be obtained from the imaginary part of Euclidean ``necklace'' diagram correlation function
$\hat{g}_{rl}$ we computed in the previous Section
\begin{align}
 & W(s,t)
= 2\Im\hat{g}_{rl}(is\b_{r}+\b_{r}/2,it+\b_{l}/2)
\end{align}
To obtain this imaginary part, we need to analytically continue two parameters, $k$ and $s$. We do this using the following prescription. We first
analytically continue $s$ to pure imaginary $is$ while keeping $k$ a positive odd integer greater than 1. Then we continue $k$ to
$1$. Taking $s\ra is$ first means that we
should take the $s<\left\lfloor k/2\right\rfloor $ case of our $x,y,\tilde{x},\tilde{y}$
for $\s_{ab}^{s}$ and the other case for $\s_{ab}^{k-s}$ in \eqref{eq:50}. Then taking $k\ra1$ sets $\left\lfloor k/2\right\rfloor =0$
which leads to 
\begin{align}
b_{2} & =\text{arccoth}(\log(y_{0}\a^{1/2})/\log(y_{\infty}\a^{1/2}))\\
\tilde{b}_{2} & =\arctanh(\log(\tilde{y}_{0}\a^{1/2})/\log(\tilde{y}_{\infty}\a^{1/2}))
\end{align}

The causal correlator $W(s,t)$ then reads:
\begin{align}
 W(s,t)&= 2\Im g_{rl}(is)\left(e^{\s_{rl}^{is}(\b_{r}/2,\b_{l}/2+it)/q}+e^{\s_{rl}^{1-is}(\b_{r}/2,\b_{l}/2-it)/q}\right)\nonumber \\
= & \Im\left(\f{\w\tilde{\lambda}\tilde{x}(is)^{1/2}\tilde{y}(is)/(\mJ\sin\w\b_{r}/2)}{(1+\tilde{y}(is))(\sin\w(\b_{l}/2+it)+\tilde{y}(is)\sin\w(\b_{l}/2-it))+\tilde{\lambda}\tilde{x}(is)\sin\w(\b_{l}/2+it)}\right)^{2/q}\nonumber \\
 & +(t\leftrightarrow -t, \tilde{x}(is)\leftrightarrow \tilde{x}(1-is),\tilde{y}(is)\leftrightarrow \tilde{y}(1-is))\label{eq:102}
\end{align}
where for $\tilde{x}(is)$ and $\tilde{y}(is)$ we use $\tilde{c}_{1},\tilde{b}_{1}$
and for $\tilde{x}(1-is)$ and $\tilde{y}(1-is)$ we use $\tilde{c}_{2},\tilde{b}_{2}$.
Only the first term in (\ref{eq:102}) is important because the second term becomes small
in low temperature limit where $\b_{l,r}$ are both large (or equivalently
$\mJ$ is large). This can be seen already in the plot of the Euclidean
correlator before analytic continuation in Fig.~\ref{fig:The-plot-of}.
In this plot, the amplitude of correlation function decreases when
we increase (the real part of) $s$. 

\begin{figure}
\begin{centering}
\subfloat[\label{fig:6a}]{\begin{centering}
\includegraphics[width=6cm]{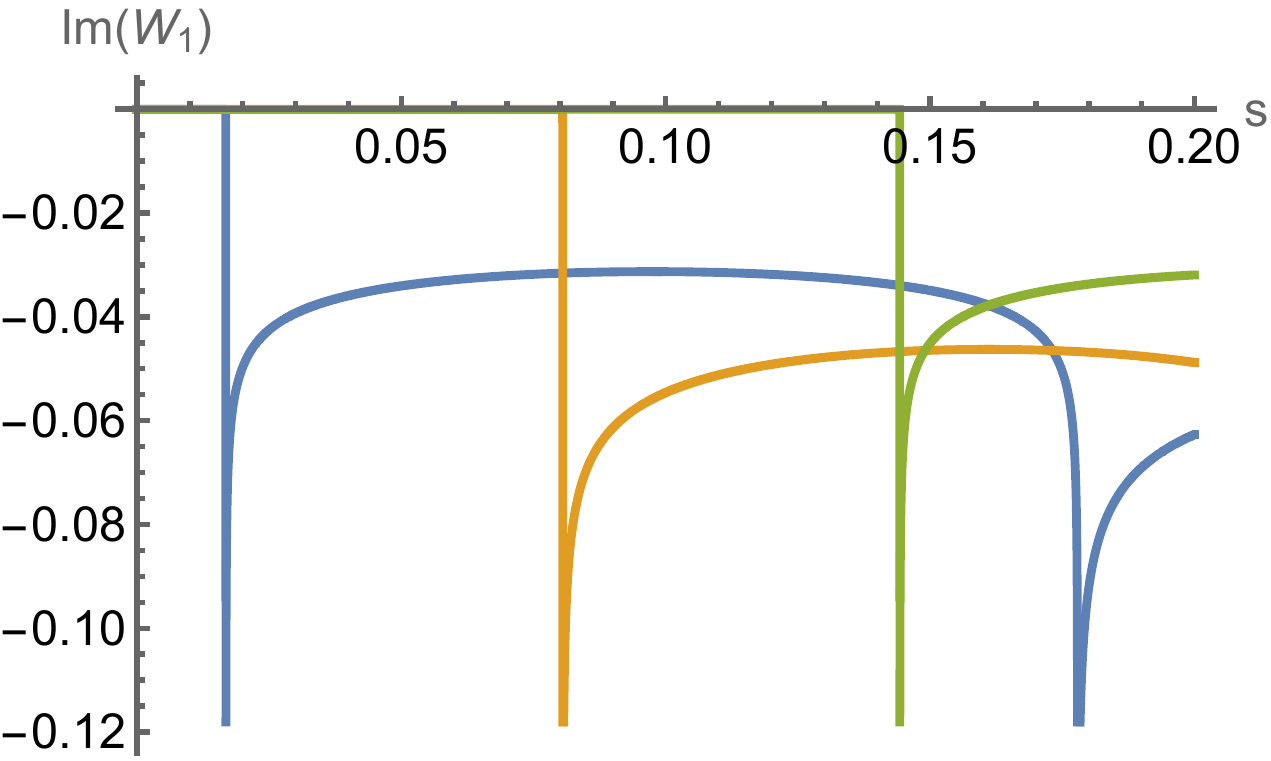}
\par\end{centering}
}\subfloat[]{\begin{centering}
\includegraphics[width=6cm]{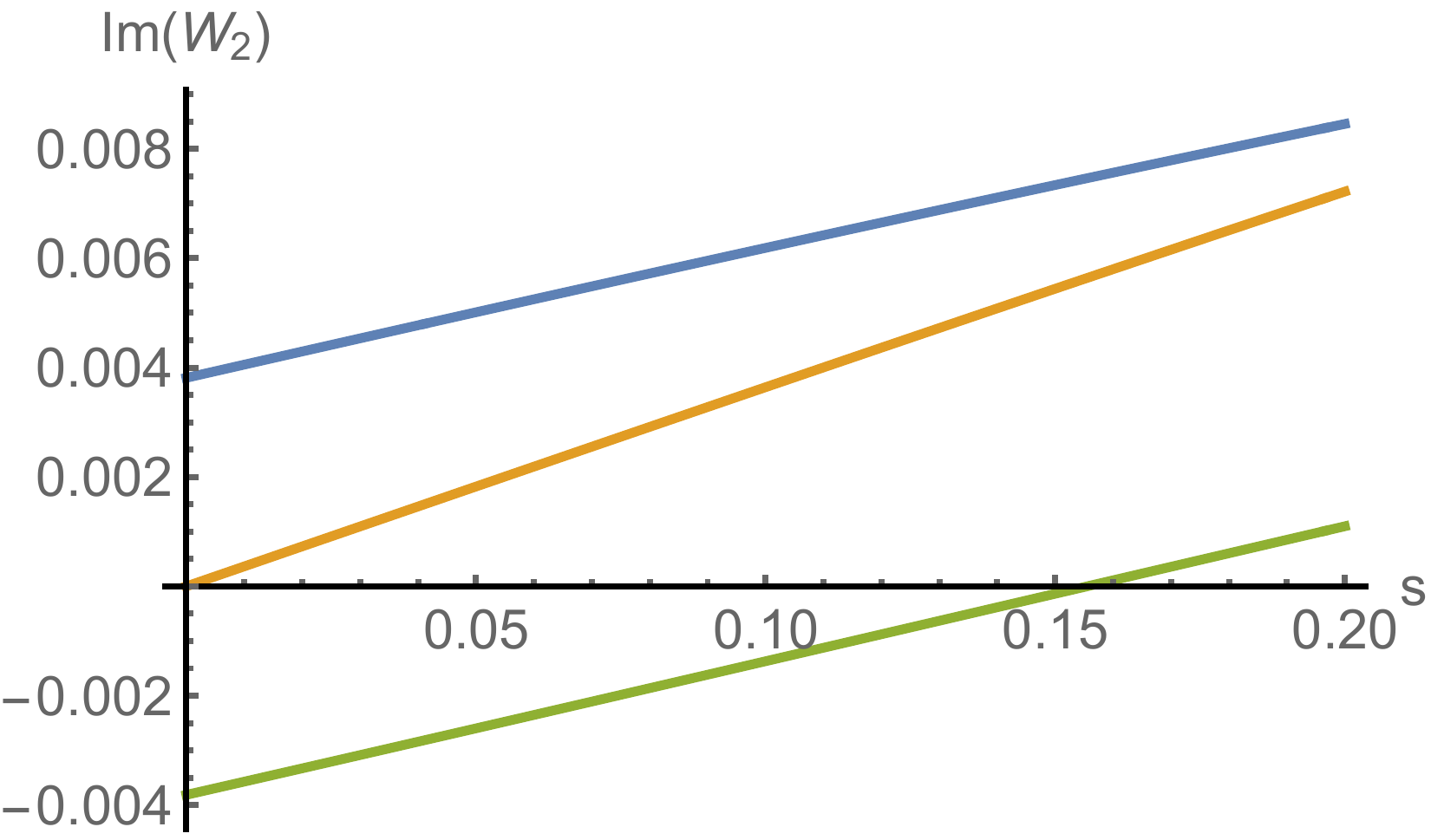}
\par\end{centering}
}\\
\subfloat[\label{fig:6c}]{\begin{centering}
\includegraphics[width=6cm]{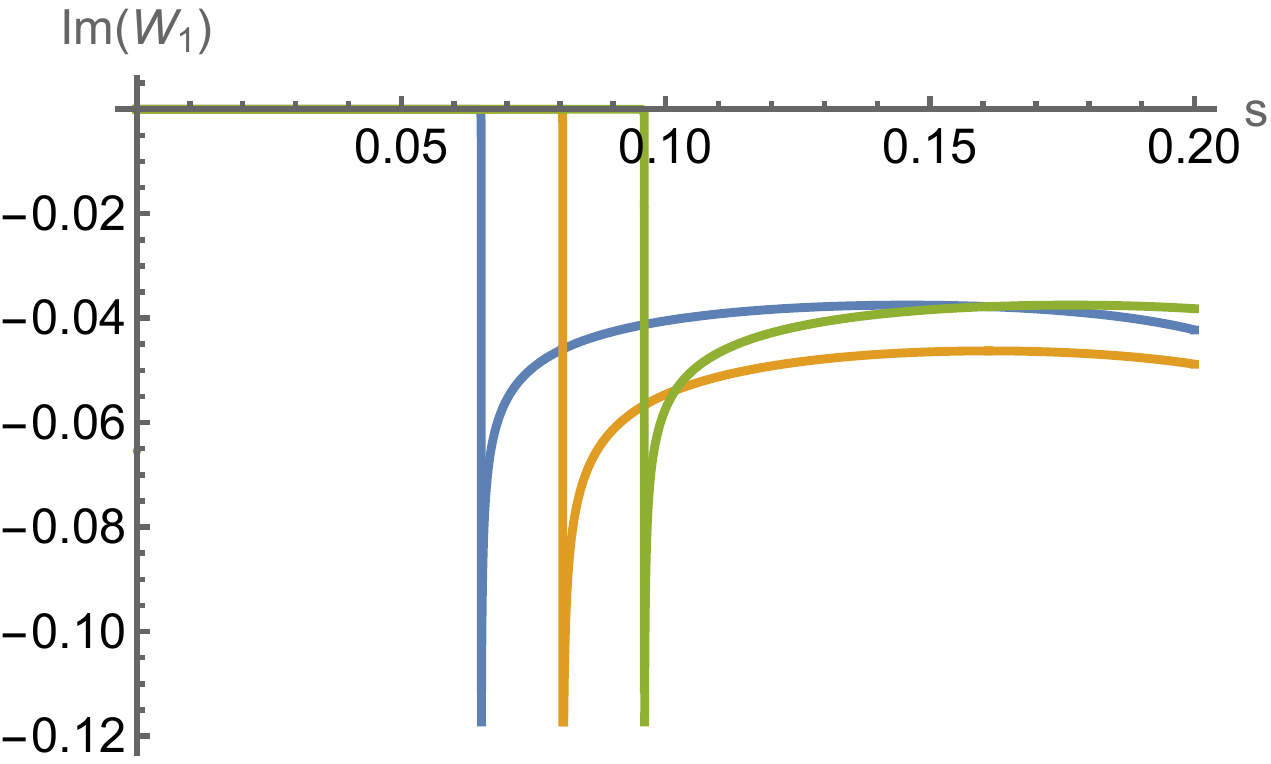}
\par\end{centering}
}\subfloat[]{\begin{centering}
\includegraphics[width=6cm]{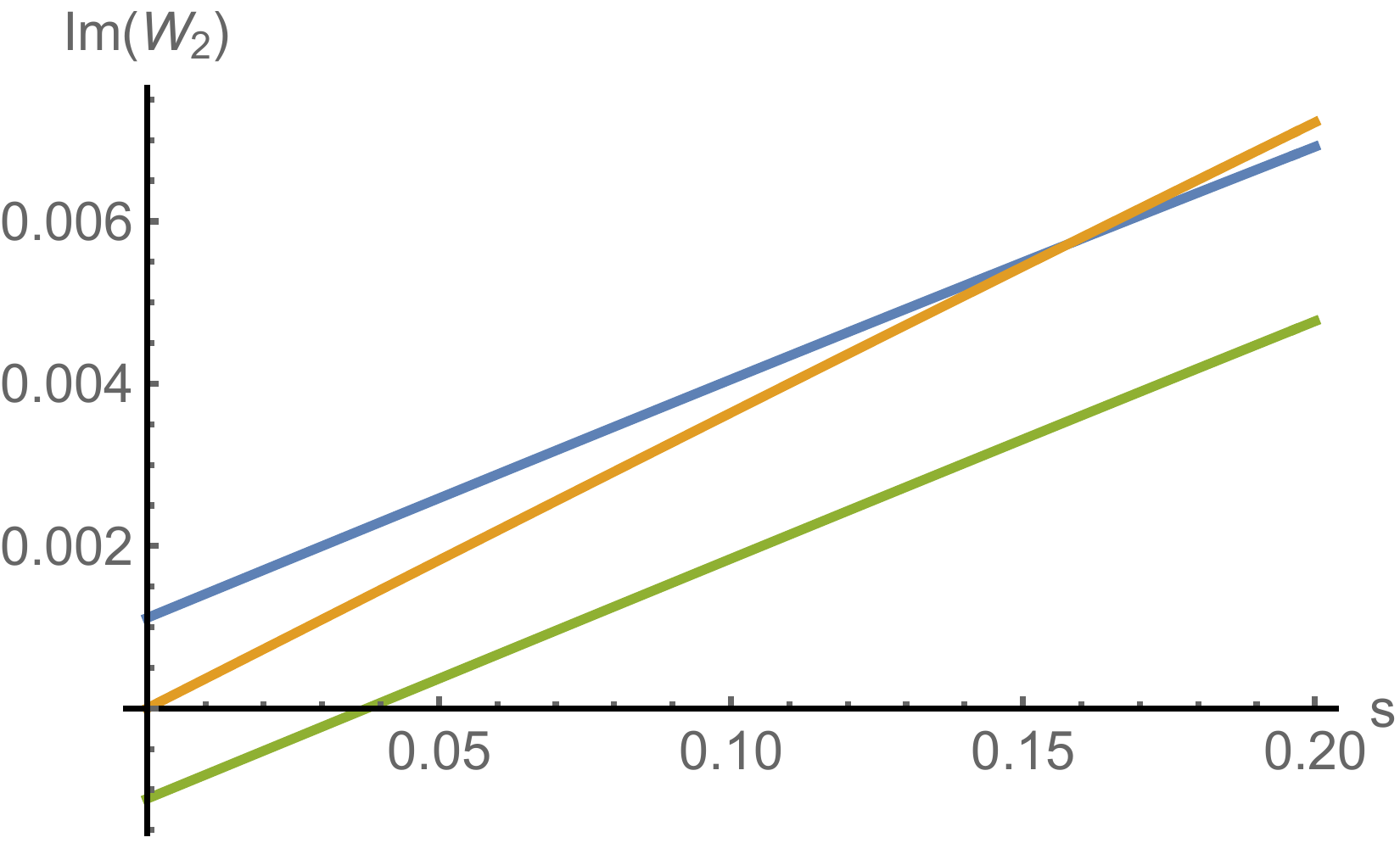}
\par\end{centering}
}
\par\end{centering}
\caption{The plot of $\Im W_{1,2}(s,t)$ with different $t$. Blue, yellow
and green curves are with $t=-3,0,3$ respectively. The parameters
for (a) and (b) are $\protect\b_{l}=1,\protect\b_{r}=4$ (injection
is in left side) and for (c) and (d) are $\protect\b_{l}=4,\protect\b_{r}=1$
(injection is in right side). Other parameters are $\protect\a=10^{-5}$,
$\protect\mJ=10^{6}$ and $q=12$. \label{fig:The-plot-of-1}}
\end{figure}
We can separate the two lines in (\ref{eq:102}) before taking imaginary part and denote them as $W_{1}$ and $W_{2}$
respectively. We plot their imaginary part in Fig.~\ref{fig:The-plot-of-1}. $\Im W_{2}$
is generally smaller than $\Im W_{1}$ as expected, therefore, we
can ignore it in the large $\mJ,\mu$ limit. The analysis of Appendix \ref{app:b} offers the following more accurate statement: $|W_2/W_1|\ll1$ if $\f{\beta \mJ \a^{1/2}}{\pi}\sin \pi \d_l=\eta$ for $\eta\ll 1$, $\eta\gg 1$ and $|\eta-1|\ll 1$, where we assume $\d_l\sim O(1)$. In other words, if $\beta \mJ$ and $\a^{-1/2}=e^{\mu q/2}$ are separate large scales or, alternatively, they are both large and fine tuned, $W_2$ becomes negligible. 

There is another
reason we should ignore $W_{2}$ that at $s=0$ the imaginary part
of $W$ should be zero for any $t$. Clearly, $W_{1}$ obeys this
rule as one can see it by plugging in the value of $\tilde{x}(0)$
and $\tilde{y}(0)$ from (\ref{eq:90}) and (\ref{eq:91}) but $W_{2}$
does not (unless $t=0$). This is an artifact of introducing the image
for correlator. But in some large $\mJ,\mu$ limit, this violation is small so
we may expect our approximation close to true solution in this regime.
This is similar to \cite{Saad:2018bqo} where the image term is ignored
in computation of the ramp of form factor in SYK model because it involves a long time.

The key observation is the existence of very sharp peaks of $\Im W_{1}$ at specific finite modular times $s$. In the bulk dual these should be interpreted as the infalling proper times at which $\psi_{r}$ enters
the light-cone of the left boundary insertion $\psi_{l}(t)$. In particular, as we increase $t$, the location
of peak moves towards large $s$, which is an important feature consistent with this interpretation. Furthermore, the blue curve
in Fig. \ref{fig:6a} has two peaks. If we plot $\Im W_{1}$ for a larger range of
$s$, we will see periodic peaks for all different $t$. We should
interpret these periodic peaks as $\psi_{r}$ entering the light-cone
of $\psi_{l}(t)$ many times because the AdS boundary condition reflects
null rays from $\psi_{l}(t)$ between two boundaries, causing the modular flowed operator to cross its lightcone an infinite number of times.

\subsection*{The location of the peak and the bulk lightcone}
We can compute the location of peaks in the expectation value of the modular flowed commutator as follows. In the low temperature/strong
coupling limit, we see that the sequence $\tilde{y}_{s}$ converges to its limit value extremely fast, Fig. \ref{fig:xtyt}.
We can, therefore, replace $\tilde{y}_{\infty}$ with $\tilde{y}_{1}$
without affecting the result. Focusing on large SYK coupling $\mJ$, we can obtain the solution
\begin{align}
\w & =\f{\pi}{\b}\left(1-\f 2{\b\mJ}+O(1/\b^{2}\mJ^{2})\right)\\
\tilde{y}_{0} & \app\f{\b\mJ}{\pi}\sin\pi\d_{l},\quad\tilde{y}_{1}\app\tilde{y}_{0}(1+\a),\quad\tilde{y}_{s}\app\tilde{y}_{s-1}(1+O(\a(\a\tilde{y}_{0}^{-2})^{s-1}))\label{eq:104}\\
\tilde{x}_{0} & \app\b^{2}\mJ^{2}/\pi^{2},\quad\tilde{x}_{s}\app O(\a(\a\tilde{y}_{0}^{-2})^{s-1})\label{eq:105}
\end{align}
Note that the last equation estimates how close $\tilde{y}_{1}$ to
$\tilde{y}_{\infty}$ in the small $1/\mJ$ and $\a$ limit. Using this
formula, we have
\begin{equation}
\tilde{c}_{1}\app\log\left(\f{\b\mJ}{\pi\a^{1/2}}\sin\pi\d_{l}\right),\quad\tilde{b}_{1}\app\f 12\log(\tilde{c}_{1}/\a)\label{eq:106}
\end{equation}
which are both large numbers. This means that the analytically
continued function $\tilde{y}(is)$ is oscillating very quickly and with a small amplitude around a large mean value $\tilde{y}(0)$.
Therefore, we can simply replace all $\tilde{y}(is)$ as $\tilde{y}_{0}$
in $W_{1}$ and get
\begin{equation}
W_{1}(s,t)\app\left(\f{(2\pi\sin\pi\d_{l}/2)/(\b\mJ)}{X(s)\sin\w(\b_{l}/2-it)+X(s)^{-1}\sin\w(\b_{l}/2+it)}\right)^{2/q}\label{eq:107}
\end{equation}
where
\begin{equation}
X(s)=\cos\tilde{c}_{1}s+i\tanh\tilde{b}_{1}\sin\tilde{c}_{1}s\ra e^{i\tilde{c}_{1}s}
\end{equation}
where, in last step, we also took the large $\tilde{b}_{1}$ limit as suggested
by (\ref{eq:106}). With this approximation, we see clearly that $W_{1}$
is real for small $s$ until the denominator vanishes at modular time
\begin{equation}
s=\f 1{2\tilde{c}_{1}}\left(\pi+2\arctan\f{\tanh\pi t/\b}{\tan\pi\d_{l}/2}+2\pi\N\right)\label{eq:109-1}
\end{equation}
which determines the location of the peaks in Fig. \ref{fig:6a} and
Fig. \ref{fig:6c}. Here $2\pi\N$ counts for all periodic peaks.

In the following, we only focus on the first peak that corresponds
to choosing $0\in\N$. Clearly, (\ref{eq:109-1}) is a monotonically
increasing function of $t$ as expected. For $t=0$, the peak location
is fixed at $s=\pi/(2\tilde{c}_{1})$ and is independent on the value
of $\d_{l}$. This feature is also illustrated by the yellow curves
in Fig. \ref{fig:6a} and Fig. \ref{fig:6c}, where the slight distinction
is due to subleading corrections. On the other hand, taking a reflection
$t\ra-t$ flips the value of $s$ symmetrically around $\pi/(2\tilde{c}_{1})$.

This result matches exactly with the bulk expectation Fig.~\ref{fig:7c} in Section~\ref{sec: bulkexpect}. Indeed, the $\l\ra\infty$ limit of \eqref{eq:1.30} reduces to \eqref{eq:109-1}, if we identify
\be
s=\f 1{2\tilde{c}_{1}}s_{p}
\ee
According to \cite{Jafferis:2020ora}, the modular time parameter $s$ should be interpreted as the bulk proper time in  units of the inverse temperature of
probe black hole $\b_{probe}/(2\pi)$. The matching condition above defines the effective temperature of our probe, produced by the entangling unitary \eqref{entanglingU} in Section~\ref{sec:prepare}, which reads:
\be
\b_{probe}=4\pi\tilde{c}_{1}=4\pi \log\left(\f{\b\mJ}{\pi\a^{1/2}}\sin\pi\d_{l}\right) \label{betaprobeSYK}
\ee
This offers an explicit confirmation of the validity of the proposal of \cite{Jafferis:2020ora} in the setup explored in this work.

A feature of our SYK result that is at odds with the proposal of \cite{Jafferis:2020ora}, when taken at face value, is the fact that the modular flow associated with the probe we initiated in the right exterior gives consistent results even when it is used to evolve SYK$_l$ operators (see Fig. \ref{fig:6a}) This is far outside the expected regime of validity of the modular time/proper time connection: The arguments presented in \cite{Jafferis:2020ora} only guarantee a coincidence of the two operators when acting on operators in the vicinity of the probe. The reason for the extended regime of validity of the prescription in our setup is the emergent SL$(2,R)$ symmetry of SYK which underlies the solution for the modular flowed correlator we studied.

\subsection{Bulk fields behind horizon} \label{bulkbehind}

In the previous Subsection we studied the modular flow of a right boundary Majorana fermion; this is an operator at an infinite geodesic distance $\ell \to \infty$ from the infalling probe's worldline. We can generalize the discussion to the modular flow of a bulk field at a finite distance $\ell$ from  the probe.

We can achieve this by expressing a bulk
fermion, localized in the right exterior region on the initial $T=0$ slice, in terms of boundary fermions using the usual HKLL reconstruction \cite{Hamilton:2005ju, Hamilton:2006az, Hamilton:2006fh}. The metric in the right Rindler
wedge of eternal AdS$_2$ black hole reads
\begin{equation}
ds^{2}=d\r^{2}-\left(\f{2\pi}{\b}\right)^{2}\sinh^{2}\r dt^{2}
\end{equation}
and the bulk spinor field is expressed as an integral over the Majorana operators of SYK$_r$ as
\begin{equation}
\chi(\r,t)=\int_{D(t_*)}dt'K(\r,t;t')\psi_{r}(t') \label{eq:hkll}
\end{equation}
where the integral range $D(t_*)=[-t_{*},t_{*}]$ only includes the boundary
time-strip that is spacelike separated from $(\r,t)$.\footnote{In 2D, a bulk spinor has two components but a boundary spinor only has one. Therefore, the bulk spinor reconstructed via \eqref{eq:hkll} has a specific polarization \cite{Lensky:2020ubw}.} In for $t'\in D(t_*)$,
$K(\r,t;t')$ is a real function. See Fig. \ref{fig:HKLL} as an illustration.  The relevant AdS$_2$ kernel, at leading order in $1/N$,
was derived in \cite{Lensky:2020ubw}.

The modular flow of the bulk spinor $\chi$ is
\begin{equation}
\chi_{s}(\r,t)\equiv\r^{-is}\chi(\r,t)\r^{is}=\int_{D(t_*)}dt'K(\r,t;t')\r^{-is}\psi_{r}(t')\r^{is}
\end{equation}
Let us take $t=0$ and some arbitrary finite $\r$. After an amount $s$ of evolution with the infalling modular Hamiltonian, the
causal correlation between $\chi_{s}(\r,0)$ and $\psi_{l}(t)$ reads
\begin{equation}
\avg{\{\chi_{s}(\r,0),\psi_{l}(t)\}}=\int_{D(t_*)}dt'K(\r,0;t')\cdot2\Im\left(g_{rl}(is)e^{\s_{rl}^{is}(\b_{r}/2+it',\b_{l}/2+it)/q}\right)\label{eq:136}
\end{equation}
where we have, once again, omitted the sum over ``big KMS'' images in the SYK result for the commutator. Even without using the
specific form of $K$, we can already read off the modular time at which the commutator (\ref{eq:136}) becomes
nonzero: It is the value of $s$ for which the largest $t'$
hits the lightcone of left insertion $\psi_{l}(t)$. Using the same approximation
as (\ref{eq:107}), we have 
\begin{align}
\hat{W}(s,t;t') & \equiv2\Im\left(g_{rl}(is)e^{\s_{rl}^{is}(\b_{r}/2+it',\b_{l}/2+it)/q}\right)\nonumber \\
 & =\Im \left(\f{(\pi\sin\pi\d_{l})/(\b\mJ)}{e^{is_{p}/2}\cos\w(\b_{l}/2-it')\sin\w(\b_{l}/2-it)+c.c}\right)^{2/q} \label{eq:3.85}
\end{align}
A bulk spinor located at distance $\l$ away from the probe on the
$T=0$ slice, is located at the AdS$_2$ point (\ref{eq:120}) and (\ref{eq:121})
with $s_{p}=0$. This operator is supported on the asymptotic boundary over the time strip $D(t_*)$ with
\begin{equation}
t_{*}=\f{\b}{\pi}\arctanh(e^{\xi-\l})\label{eq:138}
\end{equation}
On the other hand, the pole of $\tilde{W}(s,t;t_{*})$ is at 
\be
\Re\left(e^{is_{p}/2}\sin\w(\b_{r}/2+it_{*})\sin\w(\b_{l}/2-it)\right)=0
\ee
which can be solved to find:
\begin{equation}
t=\f{\b}{\pi}\arctanh\f{\tan\f{\pi\d_{l}}2\tanh\f{\pi t_{*}}{\b}\tan\f{s_{p}}2-1}{\cot\f{\pi\d_{l}}2\tan\f{s_{p}}2+\tanh\f{\pi t_{*}}{\b}}=\f{\b}{\pi}\arctanh\f{\tan\f{s_{p}}2-e^{\l}}{e^{\xi}(1+e^{\l}\tan\f{s_{p}}2)}\label{eq:139}
\end{equation}
where we used (\ref{eq:1.35}) and (\ref{eq:138}) in the second step. This result exactly matches with bulk expectation \eqref{eq:1.30}.

\paragraph{Locality of bulk modular flow} Using \eqref{eq:3.85}, we can show that modular flow preserves the locality of the field $\chi(\r,t)$ in the bulk. The key fact is that, in the regime where (\ref{eq:3.85}) is valid, our modular flow reduces to an $SL(2)$ isometry $U(s)$ of AdS$_2$. Specifically, it is the symmetry that fixes a particular timelike bulk geodesic (what we referred to previously as our probe's trajectory) and moves $\chi$ from $(\r,t)$ to $U(s)(\r,t)$ with reference to that geodesic, just as described in section \ref{sec: bulkexpect}
\be
\chi_s(\r,t)=\chi(U(s)(\r,t)) \label{eq:3.89}
\ee
In embedding coordinates, this $U(s)$ transformation can be expressed in a simple way
\be
U(s)\cdot Y=M_1(\xi)^{-1} \cdot M_3(-s_p) \cdot M_1(\xi)\cdot Y
\ee
where $M_i$ are given by \eqref{sl2rembedding}. 

To understand why this is so, note that bulk correlation functions between two points, $Y_a$ and $Y_b$, in AdS$_2$ are functions of geodesic length $\l$ between them which is, in turn, given by $\cosh\l=-Y_a\cdot Y_b$. One can easily show that \eqref{eq:3.85} is proportional to $(-Y_a\cdot U(s)\cdot Y_b)^{-2/q}$, with $Y_a$ at the left AdS boundary and $Y_b$ at the right, using the Rindler coordinate representation for $Y_\mu$
\be
Y_{-1}=\sinh \r \sinh \f{2\pi}{\b}t,~~ Y_0=\cosh\r,~~ Y_1=\pm\sinh\r\cosh  \f{2\pi}{\b} t
\ee
where plus (minus) sign is for left (right) Rindler wedge. Now recall that the HKLL reconstruction of a bulk field is uniquely determined by the mode expansion of the dual boundary operator and the bulk equation of motion. Since the latter is invariant under $SL(2)$ isometry, the modular evolution of a boundary operator $\psi$ is uniquely extended to that of a bulk field and, therefore, acts on it exactly as \eqref{eq:3.89}.

\section{Replica computation in EAdS$_2$} \label{sec:4}
In this section, we compute the modular flowed commutator (\ref{commutator}) using the replica trick (\ref{replicacorrelator}) for the bulk JT gravity path integral. As discussed in Section \ref{sec:replica}, there are two classical geometries that contribute to the replica correlator $W_{ab}^{k,s}(\tau_1,\tau_2)$, shown in Fig.~\ref{fig:disconnected} and \ref{fig:connected}. However, only the Euclidean wormhole contribution can lead to a non-trivial anticommutator between $\psi_l(t)$ and $\rho^{-is}\psi_r \rho^{is}$. We, therefore, start by constructing the relevant bulk wormhole saddle. We then compute the boundary-to-boundary propagator in this geometry and analytically continue it to obtain the desired anticommutator $W(s,t)$, finding exact agreement with (\ref{eq:109-1}). In Appendix \ref{app:e} we specify the parameter regime in which the wormhole saddle is indeed the dominant contribution, deriving the regime of validity of our path integral analysis.

\subsection{The replica path integral in JT gravity} \label{sec:4.1}
Our starting point will be expression (\ref{wfinitemu}) for the finite $\mu$ replica correlator of interest which we repeat here for convenience 
\begin{align}
&W_{rl}^{k,s}(\tau_{1},\tau_{2})
= \f 1{N\mZ}\sum_{j=1}^{N}\E_{J}\left[\Tr[\r^{k-s}\psi_{r}^{j}(\tau_{r})\r^{s}\psi_{l}^{j}(\tau_{2})\right]\nonumber \\
 = & \f 1{N\mZ}\sum_{j=1}^{N} \E_J\left[\Tr\left[{\cal T} \left\{ e^{-k\b_{l}H_{l}-k\b_{r}H_{r}}\left(\prod_{\nu=0}^{k-1} e^{-\mu S(\nu+1/2)}\right) \psi_{r}^{j}(\tau_{1}+s\b_{r})\, \psi_{l}^{j}(\tau_{2}) \right\}\right]\right] \label{wfinitemu2}  
\end{align}

The head-on holographic computation of (\ref{wfinitemu2}) in the large $\mu$ regime we are interested in is tricky. The difficulty lies in pinning down the precise deformation to the JT gravity action introduced by the potential term $\mu \sum_{\nu=0}^{k-1} S(\nu+1/2)$ when $\mu\gg 1$.\footnote{For small $\mu$, each $e^{-\mu S}$ insertion can be effectively replaced by $e^{-\mu \langle S\rangle}$ with the expectation value computed about the given bulk geometry. This approximation is not valid at large $\mu$} We will make progress by exploiting the fact that the insertions of $\r_0=e^{-\mu S}$ are localized on the boundary. This means that the bulk gravity action is the standard JT action, describing the familiar Schwarzian dynamics of a pair of boundary particles,\footnote{one for SYK$_l$ and one for SYK$_r$} almost everywhere in the bulk, except in the region near the $\r_0$ insertions which have the physical effect of pulling the two boundary particles closer together, as discussed in Section \ref{sec:replica}. Such localized kicks of the boundary particle's trajectory can be effectively parametrized by the change they induce in its $SL(2,R)$ charge. Focusing on the effect of $\r_0$ on this charge amounts to looking only at its gravitational backreaction. The precise value of the $SL(2,R)$ charge $M$ associated to each insertion $\r_0$, however, is UV information that needs to be computed microscopically using an SYK analysis.

Our strategy for this computation will, therefore, be the following: We look for a solution of pure JT gravity with a cylindrical topology, connecting two boundaries of total renormalized proper lengths $k\beta_l$ and $k\beta_r$, respectively, with $k$ localized insertions of $\r_0$ at the appropriate points which we effectively treat as kicks with $SL(2,R)$ Casimir $M$. This yields a geometry that depends on the parameters $\beta_l,\beta_r,k$ and $M$. We compute the boundary-to-boundary propagator for fermions at arbitrarty replica separations $s$ in this geometry using the geodesic approximation and analytically continue it to $k\to 1$ and $s\to is$ to obtain the modular flowed correlator of interest. Finally, we use the microscopic solution of the previous Section to evaluate our effective  charge $M$ in terms of the SYK parameters $\mu, {\cal J}, q$ and import it in the solution. The final result for the modular flowed correlator precisely matches the SYK computation of the previous Section.

\subsection{The Euclidean wormhole solution} \label{sec:eucworm}
\begin{figure}
    \centering
    \includegraphics[width=16cm]{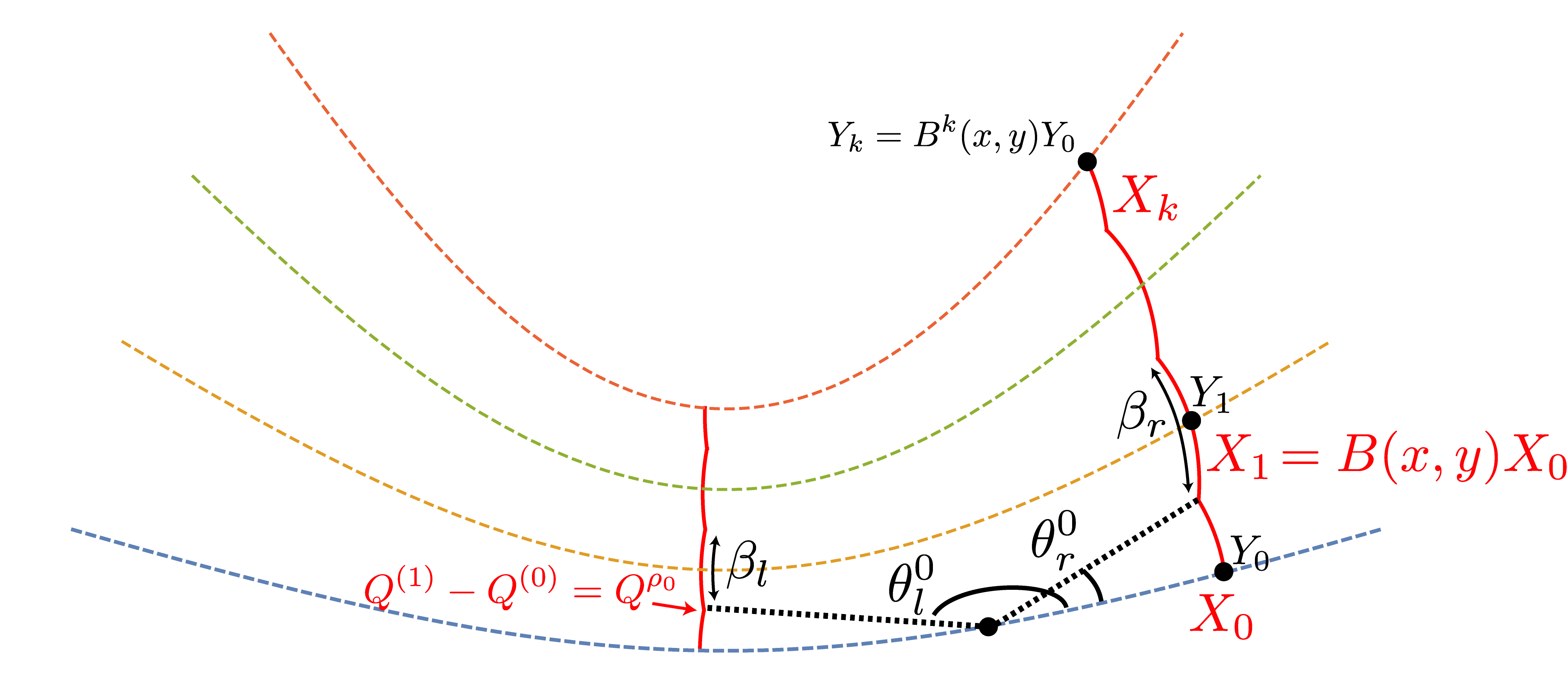}
    \caption{The Euclidean wormhole geometry dominating the bulk JT path integral with ``necklace'' diagram boundary conditions at intermediate values of $S_{probe}$, constructed from the patch of the hyperbolic plane $\mathbb{H}_2$ between the two solid red curves. Each red colored segment is an arc of a circle $X_n$, $n=0,1, \dots k$ (\ref{sl2rimages}), which are related to each other via iterative applications of the $SL(2,R)$ transformation $B(x,y)$ (\ref{replicaboost}). Blue, yellow, green and red dashed curves are hyperbolic geodesics that define the diameters of each circles $X_n$, whose intersection with $X_n$ is chosen as angular starting point $Y_n$ of each circle respectively. The boost parameters $x,y$ and the radius $\rho$ of the circles are fixed by demanding that the arc lengths between circle intersections are $\beta_l$ and $\beta_r$ for the left and right boundaries respectively, and that the local kicks at the intersections, caused by the attractive force exerted by the $\rho_0$ insertions, correspond to changes of the boundary's $SL(2,R)$ charge by an amount $Q^{\r_0}$, with $(Q^{\r_0})^2 =-M^2$ being UV data obtained from a microscopic SYK calculation (Appendix~\ref{app:d}) and given by (\ref{eq:Mmatch}). The cylindrical topology is obtained by taking the quotient of $\mathbb{H}_2$ with respect to the action of $B^k(x,y)$, essentially identifying the geodesic diameters defining $Y_0$ and $Y_k$.}
    \label{fig:wormhole}
\end{figure}

Since any solution to the JT gravity equations is locally hyperbolic, the wormhole solution we are looking for can be understood as a patch of $\mathbb{H}_2$, endowed with the topology of a cylinder by a subsequent identification with respect to an isometry of $\mathbb{H}_2$. Our goal is, therefore, to identify the right patch of EAdS$_2$ and the relevant isometry used to compactify it. The appropriate patch and its identification is shown in Fig.~\ref{fig:wormhole}.
The rules of the construction are simple and were already discussed in \cite{Stanford:2020wkf}. The JT dynamics in our case describes a pair of boundary particles that propagate according to the Schwarzian dynamics for proper lengths $\beta_l$ and $\beta_r$, respectively, before getting interrupted by a local insertion of $\r_0$. Forgetting about the effect of the latter for the moment, the solution of the Euclidean Schwarzian equations of motion is well known and it describes a circular boundary particle trajectory in EAdS$_2$. Using the embedding space coordinate of EAdS$_2$
\be
X_0^\mu(\theta) = \{\sinh\rho \sin \theta,\sinh\rho \cos\theta, \cosh\rho \}
\ee
with the same $(-,-,+)$ signature metric \eqref{eq:2.17}, this trajectory can be written as
\be
X_0^\mu(\t) Q^{(0)}_\mu  = \frac{1}{2\epsilon},~\epsilon\to 0  \label{eq:5.5}
\ee
with $SL(2,R)$ charge
\be
Q^{(0)}_\mu= \{ 0,0,\frac{1}{2\epsilon\cosh\rho}\}\label{sl2rcharge}
\ee
where the radius of the circle $\rho$ is meant to be taken to infinity simultaneously with $\epsilon\to 0$ so that:
\begin{equation}
    2\pi\epsilon \sinh \rho \to \beta_E \label{infiniterho}
\end{equation}
The remaining parameter $\beta_E$ characterizes the solution and is related to the energy of the state via the thermodynamic relation $\beta_E =\frac{\pi^2}{E^2}$.

In the case at hand, this circular trajectory is interrupted by the $\r_0$ insertions. To understand their effect, let us first select a diameter of $X_0$, intersecting with $X_0$ at two points with one point labelled as $Y_0$, with respect to which we will measure angular locations. It turns out that for coordinate given in \eqref{eq:5.5}, we can choose $\t=0$ for $Y_0$. Starting from $Y_0$, the left and right boundary particles are initiated at $\theta=\pi$ and $\theta=0$, respectively, and then propagate along the two converging circular arcs of $X_0$ for proper lengths $\beta_l/2$ and $\beta_r/2$. At that point, their evolution is modified by the presence of $\r_0$ which, as explained in \cite{Stanford:2020wkf}, acts as a ``kick'' on both left and right boundary trajectories with $SL(2,R)$ Casimir $M$. The kick makes them start moving along arcs of a new EAdS circle $X_1$, which intersects $X_0$ at the location of the insertion but whose $SL(2,R)$ charge is shifted by the charge of the operator,  $Q^{(1)}_\mu = Q^{(0)}_\mu+ Q^{\r_0}_\mu$, where $(Q^{\r_0})^2 =-M^2$. See Fig. \ref{fig:wormhole} as an illustration.

Since all circles on hyperbolic space are related by $SL(2,R)$ transformations, this new circular trajectory can be described as:
\begin{equation}
    X^\mu_1(\theta) = [B(x,y)]^\mu_\nu X^\nu_0(\theta)
    \end{equation}
where $B(x,y)$ is some 2-parameter element of $SL(2,R)$. As all circles are defined as the first equation of \eqref{eq:5.5}, it is equivalent to say that the new circular trajectory is defined with the new charge $Q^{(1)}=Q^{(0)}\cdot B(x,y)$. The reason for the 2-parameters $x,y$ is that together with $\beta_E$ they account for the 3 physical parameters of our problem, $\beta_l,\beta_r,M$. The goal is then to determine the precise transformation $B(x,y)$ and the value of $\beta_E$ given $\beta_l,\beta_r,M$. 

\subsubsection*{Gluing conditions} 
The conditions on $B(x,y)$ and $\beta_E$ are simple to describe: (a) The intersection points of $X_1$ with $X_0$ must be at angular locations $\theta^{l,r}_0 $ (with respect to starting point $Y_0$) such that the corresponding (renormalized) arc lengths of $X_0$ match the left and right inverse temperature parameters $\beta_{l,r}$ (Fig.~\ref{fig:wormhole}):
\begin{align}
   \frac{\beta_E}{2\pi} \theta^{r}_0  &= \frac{\beta_r}{2} \\
    \frac{\beta_E}{2\pi} (\pi-\theta^{l}_0 ) &= \frac{\beta_l}{2} \label{matching1}
\end{align}
and (b) the $SL(2,R)$ charge must be conserved at the intersection point, which can be ensured by:
\begin{align}
    \left(Q^{(0)}\cdot B(x,y) - Q^{(0)}\right)^2=-M^2 \label{matching2-1}
\end{align}
The boundary particles will then begin to follow $X_1$ starting from its intersection points with $X_0$, located at angular locations $\theta_1^r=-\theta_0^r$ and $\theta_1^l=-\theta_0^l$ (with respect to the starting point $Y_1= B(x,y) \cdot Y_0$ of $X_1$) for proper lengths $\beta_r$ and $\beta_l$ before encountering another operator insertion with a similar effect. The same story will then be repeated $k$ times.

The two conditions above can be satisfied by the $SL(2,R)$ transformation:
\begin{equation}
    B(x,y)=M_1(-x)\cdot M_2(y) \cdot M_1(x) \label{replicaboost}
\end{equation}
where the generators $M_i$, $i=1,2,3$ of $SL(2,R)$ in embedding space were defined in (\ref{sl2rembedding}). Taking \eqref{replicaboost} into \eqref{matching2-1}, we have 
\be
\cosh x \sinh \frac{y}{2} =\frac{\beta_E M}{2\pi}\label{matching2}
\ee
The intersections of the circles $X_0(\theta_0)$ and $X_1(\theta_1)$ are at the angular locations $\theta_{0}^{r,l}$ that solve the equation:
\be
\coth(y/2)= \cosh x \coth\r \csc\t_0^{r,l} + \sinh x \cot\t_0^{r,l}
\ee
Setting $\theta_{0}^{r,l}$ equal to (\ref{matching1}) amounts to 2 constraints on the 3 undetermined parameters of our solution $x,y$ and $\rho$ (equivalently $\beta_E$) in terms of $\beta_l,\beta_r$. The last constraint that allows us to solve the system comes from further imposing (\ref{matching2}). 

Iterating the procedure $k$ times is straightforward, by virtue of the homogeneity of EAdS$_2$: The sequence of $SL(2,R)$ transformed circles
\begin{equation}
    X_n= B^n(x,y)\cdot X_0 = M_1(-x)\cdot M_2(ny)\cdot M_1(x)\cdot X_0 \, ,\quad n=0,\dots k-1 \label{sl2rimages}
\end{equation}
are guaranteed to intersect each other at angular locations $\theta_n^{r,l}= \pm \theta_0^{r,l}$ (with respect to the $n$-th starting point $Y_n=B^n\cdot Y_0$) ensuring that the proper length of the arcs $I^r_n=[-\theta_n^r,\theta_n^r]$ and $I^l_n=[\theta_n^l,2\pi -\theta_n^l]$, $n=1,\dots, k-1$ between subsequent intersections is always $\beta_r$ and $\beta_l$ respectively. Fig.~\ref{fig:wormhole} shows the resulting patch of EAdS relevant for a wormhole with $k=3$.

\subsubsection*{Compactification} 
The final step, is to compactify this patch of hyperbolic space to obtain a solution with cylindrical topology. This is, also, straightforward since the entire configuration was constructed by subsequent applications of an $SL(2,R)$ transformation: We simply identify the diameter defining $Y_0$ of the initial circle $X_0$ with the diameter defining $Y_k=B^k(x,y)Y_0$ of the final one $X_k$ ---namely, we quotient $\mathbb{H}_2$ by the action of $B^k(x,y)=M_1(-x)\cdot M_2(ky)\cdot M_1(x)$. This completes the construction of the Euclidean wormhole saddle of the replica JT path integral.

\subsection{Modular flowed correlator} \label{sec:4.3}
Having constructed the Euclidean wormhole solution, we can return to the computation of 
\begin{equation}
W_{rl}^{k,s}(\tau) = \text{Tr} [\rho^{k-s} \psi_r \rho^s \psi_l(\tau) ]
\end{equation}
and we will take $s<k/2$ without loss of generality. The boundary correlator of conformal dimension $\D$ is given by $\propto\cosh^{-\D} \l$ where $\l$ is the geodesic distance of two boundary points \cite{Maldacena:2017axo}. We can account for the cylindrical topology of the bulk configuration by employing the method of images:
\begin{equation}
W_{rl}^{k,s}(\tau)  \sim \sum_{m=0}^\infty \frac{1}{\cosh^{\Delta_\psi} \ell(P^{(0)}_l(\tau), P_r^{(s +m k)})} + \sum_{m=0}^\infty \frac{1}{\cosh^{\Delta_\psi}\ell( P^{(0)}_l(\tau), P_r^{(k-s +m k)})} \label{wormhole2pt}
\end{equation}
where $\ell(\cdot, \cdot)$ is the length of the shortest geodesic connecting the 2 points in the Euclidean wormhole, ${\Delta_\psi}=1/q$ is the dimension of a Majorana fermion and $P_{l,r}^{(m)}$ are the embedding space coordinates of the left or right fermion insertions on the ``necklace" diagram:
 \begin{align}
P^{(0)}_l(\tau)&= X_0(\pi-\tau)\, ,\quad \tau \in [-\theta_0^r,\theta_0^r]\\
 P^{(m)}_r&= X_m(0) = B^m(x,y) X_0(0) \label{wormholepoints}
  \end{align}
The second term in \eqref{wormhole2pt} that involves $k-s$ separation of ``necklace circles" comes from the geodesics connecting two boundary point from the other circular direction on cylindrical topology. This is the same idea we used to sum over images in \eqref{eq:50} in order to ensure the ``big KMS symmetry''. As in the SYK computation of Section \ref{sec:4}, let us focus on the dominant contribution to (\ref{wormhole2pt}) which, after the final analytic continuation to $k\to 1$ $s\to i s$, comes from the shortest wormhole geodesic, $\ell(P_l^{(0)}(\tau), P_r^{(s)})$. As long as $M\ll N$, we can approximate the length of the latter by the embedding space formula $\cosh \ell(P^{(0)}_l, P_r^{(s)})= P^{(0)}_l\cdot P_r^{(s)}$ and the replica 2-point function becomes
\begin{equation} W_{rl}^{k,s}(\tau) \approx \frac{1}{(X_0(\pi-\tau)\cdot B^s(x,y)\cdot X_0(0))^{\Delta_\psi}}\, , \quad \tau\in [-\theta_0^r,\theta_0^r] \label{replica2ptsimple}
\end{equation}
Since the dependence of the function (\ref{replica2ptsimple}) on the replica separation $s$ is through $M_2(sy)$, which is analytic in $s$, we can directly continue $k\to 1$, $s\to i s$ and $\tau\ra  2\pi i t/\b_E$. After a straightforward computation, the modular flowed correlation function under the limit of \eqref{infiniterho} is
\begin{align}
    W(s,t)  &=2^{-{\Delta_\psi}}\left(\frac{\beta_E\mathcal{J}}{2\pi}\right)^{-2{\Delta_\psi}}\left(e^{-x}\cos\frac{ys}{2}\cosh \frac{\pi t}{\b_E}+\sin \frac{ys}{2} \sinh \frac{\pi t}{\b_E}  \right)^{-2{\Delta_\psi}} \label{wormholeWexp}
\end{align}
wehre we removed the overall factor proportional to $\e^{2\D_\psi}$ as a normalization choice. Here the replica-symmetric wormhole geometry parameters $x,y,\beta_E$ are fixed by the parameters $\beta_l,\beta_r,\mu$ of the SYK state via the conditions discussed in the previous Section. In the large $\b_E$ limit, the latter admit the simple solution:
\begin{align}
   \t_0^l&=\t_0^r \iff \b_E=\b_l+\b_r\\
    \tanh x& \approx \cos\left(\frac{\pi\beta_l}{\beta_E} \right)\Rightarrow e^{-x}= \tan \frac{\pi \delta_l}{2} \label{wormholex}\\
   \sinh\frac{y}{2}&\approx \frac{\beta_E M}{2\pi}\frac{1}{\cosh x} \Rightarrow y\sim 2 \log \frac{\beta_E M \sin \pi \delta_l}{\pi } \label{wormholey}
\end{align}
where we defined $\d_l=\b_l/\b_E$ similarly as before. Note that (\ref{wormholex}) exactly matches (\ref{eq:1.35}) of the semiclassical particle analysis so the wormhole parameter $x$ corresponds to the boost parameter $x=\xi$. 

The modular flowed 2-sided correlation function (\ref{wormholeWexp}) will develop a branch cut and thus give rise to a non-trivial anticommutator (\ref{commutator}) at the modular time:
\begin{equation}
    s= \frac{2}{y}\text{arccot}\left(-e^{x}\tanh\frac{\pi t}{\b_E}\right)=\f 2 y \left[ \pi +\arctanh \left(\f {\tanh\frac{\pi t}{\b_E}} {\tan \f {\pi \d_l} {2}}\right) \right] \label{lightconebulk}
\end{equation}
which exactly matches with \eqref{eq:109-1}, with the identification of $y$ with $\tilde{c}_1/2$ and $\b_E$ with $\b$. This determines the probe's effective temperature to be
\begin{equation}
    \beta_{probe}=2\pi y \approx 4\pi \log \frac{\beta_E M \sin \pi \delta_l}{\pi} \label{betaprobebulk}
\end{equation}
This value of $\beta_{probe}$ is consistent with the SYK expression for the normalization of the probe's clock (\ref{betaprobeSYK}), after matching the $SL(2,R)$ charge $M$ to the SYK parameter 
\be
M= \mJ e^{\mu q/2} \label{eq:Mmatch}
\ee
This precise value of the $SL(2,R)$ charge (\ref{eq:Mmatch}) introduced by $\r_0$, which was deduced here from consistency, can indeed be obtained directly from a microscopic SYK computation, as we show in Appendix \ref{app:d}. The $SL(2,R)$ charge of $\r_0$ increases as we dial up $\mu$, consistent with the expectation that as $\mu\to \infty$, $\r_0$ approaches a projector onto the maximally entangled state between $l$ and $r$ causing the wormhole to pinch off and split into $k$ disconnected disks (Fig.~\ref{fig:degenerate}).

\section{Discussion}
\label{sec:discussion}

\subsection{Lessons for a general prescription for interior reconstruction} \label{sec:5.1}

In this paper, we utilized the framework of \cite{Jafferis:2020ora} in order to holographically reconstruct the degrees of freedom hidden behind the horizon of an $AdS_2$ black hole in Jackiw-Teitelboim gravity. Our motivation for this investigation was twofold: (a) provide an explicit application of the proposed interior reconstruction method in a setup that is under technical control and (b) identify the key ingredients of the computation that can clarify the relation of our approach to other interior reconstruction techniques, and may additionally offer clues for how to successfully apply the prescription in more interesting setups involving higher dimensional and possibly single-sided black holes.

\paragraph{Entanglement with reference couples the two exteriors via modular flow} The first noteworthy aspect of our construction that distinguishes it from previous works is the fact that we do not deform the boundary dynamics of the system in order to access the interior. It is well understood that turning on an explicit coupling between the two boundaries can lead to traversable wormholes \cite{Gao:2016bin} that allow some left excitations to causally reach the right boundary after a finite time \cite{Gao:2018yzk, Maldacena:2017axo}. Explicit couplings between the two sides can also be utilized in the AdS$_2$/SYK correspondence to construct approximate SYK duals of the bulk $SL(2,R)$ symmetry generators which can transport operators behind the horizon \cite{Lin:2019qwu}. Our conceptual contribution lies in demonstrating that the interior can be explored without such boundary Hamiltonian deformations, or even reference to a second asymptotic region.

Our construction, instead, relies on introducing a bulk probe whose microstates we entangle with an external reference. The preparation of this initial state is all the information we need to define the operator $\rho^{is}$ which transports local operators in relation to the bulk worldline our probe follows. We are essentially using the relative phases between our holographic system and the reference as an internal ``clock'' which allows us to specify the location of operator insertions in the bulk. This clock is relational in nature and is distinct from the boundary clock generated by the SYK Hamiltonians. 

It is, of course, true that the modular flow couples SYK$_l$ and SYK$_r$ which is why we can get a non-trivial anti-commutator $\{\psi_l, \rho^{-is}\psi_r \rho^{is}\}$ after sufficient $s$. However, this coupling is not an input but instead a consequence of the entanglement between our holographic system and the reference. The initial state determines the coupling between the 2 sides ---we are not allowed to pick it by hand. This two-sided coupling appearing in modular flow after tracing out a subsystem is reminiscent of the discussion of \cite{Chandrasekaran:2021tkb}. 

The conceptual advantage of this perspective is highlighted by imagining an application of our reconstruction to single-sided black hole interiors. In this case, there exist no second microscopic system describing a second exterior wormhole region; we have a single holographic CFT in a high energy state. The Hamiltonian deformation that could move us into the interior ---the analog of the $SL(2,R)$ generators of \cite{Lin:2019qwu}--- becomes unclear in this case (though see \cite{Kourkoulou:2017zaj,DeBoer:2019yoe,Brustein:2018fkr} and the recent interesting work \cite{Leutheusser:2021qhd} for suggestions) but our approach carries over unchanged. The situation is similar for 2-sided holographic systems in states dual to very long wormholes, where the 2-sided coupling required for propagating to the interior is exponentially complex \cite{Bouland:2019pvu}, or for the case of AdS black holes evaporating into an external reservoir, where the interior becomes part of the ``entanglement island'' of the radiation system at sufficiently long times. Hence, the application of our method to the aforementioned setups appears to us to be a very promising avenue for future work.

At this point, it is important to point out that the interior reconstruction method we explored is highly non-linear: Every initial state we prepare our system in, provides us with a generally different operator $\rho^{is}$, after tracing out the reference. This extreme non-linearity leads to a number of problems when one attempts to apply our prescription starting from general initial states. These problems were discussed in \cite{Jafferis:2020ora} and can be successfully addressed, as will be explained in an upcoming work \cite{LamprouJafferisdeBoer}.

\paragraph{Chaos and universality of the effective coupling} Both microscopic and Euclidean JT path integral analysis highlight the role of the emergent $SL(2,R)$ symmetry of the IR sector of SYK: The generator of the probe's modular flow effectively reduced, in the appropriate parameter regime, to an element of this $SL(2,R)$ algebra. This symmetry is only approximate and provides an effective description of the maximally chaotic dynamics of the quantum theory. In particular, the $SL(2,R)$ algebra can be organized into a boost element $B$ and its two eigen-operators, $P_\pm$ with eigenvalues $\pm i$
\begin{align}
    [B,P_\pm]=\pm iP_\pm \, , \quad [P_+, P_-] =i B
\end{align}
which grow exponentially under the boost flow $e^{iBt}$. Holographically, $B$ is linked to the IR action of SYK Hamiltonian, while $P_\pm$ characterize the exponentially growing disruption of correlations caused by small perturbations as a function of boundary time, due to the so-called scrambling phenomenon in chaotic systems \cite{Lin:2019qwu}. In fact, this very symmetry was the key principle that guided the construction of the effective theory of maximal chaos of \cite{Blake:2021wqj, Blake:2017ris}.

The prominent role of the $SL(2,R)$ symmetry in determining our modular flow, therefore, hints at a possible universality of the SYK modular evolution that takes us into the black hole interior ---a universality established by maximal chaos. As explained above, entangling a probe introduced in the right asymptotic region to a reference system results in a modular flow that couples the two asymptotic regions of the wormhole, after tracing out the reference. Maximal chaos then appears to imply a particular universal form for this effective coupling which is largely independent of the precise details of the probe we introduced: its scrambling ``potential'', characterized by the amount of $SL(2,R)$ charge the coupling injects, determines all the useful information about the modular flow, at least in the setup analyzed in this work, where all details of the exact microscopic coupling just amounts to tuning the value of the $SL(2,R)$ charge. It would be interesting to understand if maximal scrambling leads to a similarly universal modular flow in higher dimensions and whether it provides an avenue for connecting our approach to that of \cite{Leutheusser:2021qhd} and \cite{DeBoer:2019yoe}.

\paragraph{Ensemble average and operator randomness} The third important element of our construction was the quenched ensemble average over SYK couplings. In the microscopic treatment this was important for obtaining the Liouville equations dictating the fermion propagation on the ``necklace'' diagram, while it entered our bulk discussion via the appearance of the Euclidean wormhole saddle between the two boundaries.\footnote{Of course, in our setup the two asymptotic boundaries in the ``necklace" diagram are also coupled, as discussed above. This coupling is responsible for supporting this wormhole, in the sense that it allows it to become a saddle, and also ensures that it dominates in the appropriate regime. Nevertheless, the effect of the coupling can be understood as amplifying the wormhole contribution which exists irrespective of the coupling but is a non-perturbatively small, off-shell contribution to the path integral in its absence. }  

In an attempt to understand the physical role of this averaging in more general situations, let us return to our original setup from Section \ref{sec:prepare}: A thermofield double state of a pair of $0-$dimensional holographic quantum systems dual to an $AdS_2$ wormhole, which we entangle with an external reference in the completely general state
\begin{equation}
  |\beta,\tau \rangle_{l,r,ref} = \mathcal{Z}^{-\frac{1}{2}}\sum_{i} d_i \,e^{-\frac{\beta_l H_l}{2}}e^{ -\frac{\beta_r H_r}{2}}O_i\,|0\rangle\, O^{ref}_i |v\rangle_{ref} \label{probestate22}  
\end{equation}
where again $|0\rangle$ is the maximally entangled state of the two systems and written in energy basis is
\be
\bra 0 \propto \sum_\a \bra{E_\a} _l \bra{E_\a} _r
\ee
This time, however, we will not make any specific choice of operator basis, $O_i$, as we did in the main text.  Instead, we will treat the operators $O_i$ as \emph{random} matrices within an energy window $E\in [0, E_{cut}]$ with $E_{cut}\lesssim O(N)$. This is motivated by the Eigenstate Thermalization Hypothesis (ETH) \cite{DAlessio:2015qtq}, according to which
the energy basis matrix elements $[O_i]_{\alpha\bar{\alpha}}$ of simple operators $O_i$ in a chaotic theory have the form:
\begin{align}
[O_i]_{\alpha\bar{\alpha}} &=  e^{-\frac{S(E_\alpha +E_{\bar{\alpha}})}{2}}f_i(E_\alpha, E_{\bar{\alpha}}) R^i_{\alpha\bar{\alpha}} \label{eth}
\end{align}
where $R_{\alpha\bar{\alpha}}$ is to a good approximation a Gaussian random matrix with statistics
\begin{align} 
\mathbb{E}[R^i_{\alpha\beta}]\approx 0,~~~\mathbb{E}[R^i_{\alpha\beta}R^{i*}_{\alpha\beta}]\approx 1 \label{stat}
\end{align}
Here we make an extra simplifying assumption and treat the envelope function $f_i$ as an energy filter, restricting the matrix elements to a sufficiently low energy sector:
\begin{equation}
    f_i(E_\alpha, E_{\beta}) \approx \begin{cases} 1 \quad &E_{\alpha},E_{\beta} \lesssim O(N) \\
    0 \quad &\text{otherwise}
    \end{cases}
\end{equation}
Choosing $O^{ref}_i\bra{v}_{ref}$ to be an orthogonal basis in the reference and tracing out the latter yields the density matrix
\begin{equation}
\rho=\sum_i \left|d_i\right|^2 \,\,\left(e^{-\frac{\beta_l}{2} H_l } \,e^{-\frac{\beta_r}{2} H_r} \,O_i |0\rangle \langle 0| \, O_i^\dagger \,e^{-\frac{\beta_l}{2} H_l } \,e^{-\frac{\beta_r}{2} H_r}\right)
\end{equation}
whose matrix elements in the energy basis of the boundary systems read:
\begin{align}
\rho_{\alpha\bar{\alpha}, \beta\bar{\beta}}&=\,_l \langle E_{\alpha}| \,_{r}\langle E_{\bar{\alpha}}| \,\rho\, |E_{\beta}\rangle_l  |E_{\bar{\beta}}\rangle_r  \nonumber\\
&= \sum_{i\alpha\beta\bar{\alpha}\bar{\beta}} \left|d_i\right|^2 q_l^{\frac{E_{\alpha}+E_{\beta} }{2}} q_r^{\frac{E_{\bar{\alpha}}+E_{\bar{\beta}}}{2}}\, \,[O_i]_{\alpha\bar{\alpha}}[O_i]^*_{\beta\bar{\beta}} \label{JTrhoLR}
\end{align}
where we introduced for convenience the notation $q_{l,r}=e^{-\beta_{l,r}}$.

We can consider now the same replica correlation function $W_{rl}(k,s)$ we studied in this paper:
\begin{equation}
W_{rl}(k,s)= \text{Tr}\left[\rho^{k-s}\phi_r\,\rho^s \,\phi_l\right] \label{Wdiscussion}
\end{equation}
whose analytic continuation in $k$ and $s$ produces the modular flowed correlation function that holographically describes the proper time evolved bulk propagator. Plugging in (\ref{Wdiscussion}) the general expression for $\rho$, we obtain:
\begin{align}
&W_{rl}(k,s)= \sum_{i_1,i_2, \dots i_k} \left| d_{i_1} \right|^2\left| d_{i_2} \right|^2 \dots \left| d_{i_k} \right|^2\sum_{\{\alpha_j,\bar{\alpha}_j\}_{j=1}^k}q_l^{\frac{1}{2}(E_\gamma -E_{\a_{s+1}}) +\sum_{j=1}^k E_{\alpha_j}} q_r^{\frac{1}{2}(E_{\bar{\g}} -E_{\bar{\alpha}_1})+  \sum_{j=1}^k E_{\bar{\alpha}_j}}\nonumber\\
&\times [\phi_r]_{ \bar{\g}\bar{\alpha}_1}[O_{i_1}]_{\alpha_1,\bar{\alpha}_1}\, [O_{i_1}]^*_{\alpha_2,\bar{\alpha}_2}[O_{i_2}]_{\alpha_2,\bar{\alpha}_2}\, [O_{i_2}]^*_{\alpha_3,\bar{\alpha}_3} \dots [O_{i_s}]_{\gamma \bar{\alpha}_{s+1}}^* [\phi_l]_{ \gamma \alpha_{s+1}}\dots [O_{i_k}]_{\alpha_k,\bar{\alpha}_k}\, [O_{i_k}]^*_{\alpha_1\bar{\g}} \label{horribleexpression}
\end{align}
The only aspect of (\ref{horribleexpression}) that interest us is the pattern of index contractions which, when combined with the randomness of the matrix elements (\ref{stat}), can help us understand the two distinct limiting phases of our computation, corresponding to the saddle of Fig.~\ref{fig:degenerate} or that of Fig.~\ref{fig:disconnected}, when the entropy of the probe becomes infinitesimally small ($S_{probe}\to 0$) or maximal ($S_{probe}\to O(N)$) respectively.

The first phase is recovered by choosing the weight $|d_i|^2$ to have support only on a single operator, say the identity for simplicity, reducing (\ref{horribleexpression}) to:
\begin{equation}
W_{rl}(k,s) \approx \begin{cases} \langle 0| e^{-\frac{\beta_l+\beta_r}{2}H_l} \phi_r \phi_l e^{-\frac{\beta_r+\beta_l}{2} H_l }  |0\rangle \quad &s=0 \\
\text{Tr}_l[ e^{-(\beta_l +\beta_r) H_l} \phi_l]\,\text{Tr}_r[ e^{-(\beta_l +\beta_r) H_r} \phi_r] \quad &s\neq 0
\end{cases}
\end{equation}
which obviously leads to trivial modular flow after analytic continuation.

The second phase is reached by taking $|d_i|^2$ to be an almost homogenous weight over a large subset of operators. It is reasonable to assume that homogeneously summing over all random operators (\ref{eth}) in the theory effectively acts as an \emph{ensemble average} in the following sense: 
\begin{align} 
\sum_i |d_i|^2  \,R^i_{\alpha\beta}\approx 0,~~~\sum_i |d_i|^2 \,R^i_{\alpha\bar{\alpha}}R^{i*}_{\beta\bar{\beta}}\approx \delta_{\alpha \beta}\delta_{\bar{\alpha} \bar{\beta}} \label{stat2}
\end{align}
Note that this assumption is different from ETH because we are summing over a subset of matrices labelled by $i$. It is, however, motivated by it, and supported by the statistics of OPE coefficients in holographic CFT$_2$ discussed in the interesting recent works \cite{Collier:2019weq,Belin:2020hea, Belin:2021ryy}. Using the assumption (\ref{stat2}) in (\ref{horribleexpression}) and being mindful of the various index contractions, we find
\begin{equation}
W_{rl}(k,s) \approx \text{Tr}_l \left[ e^{-k\beta_l H_l } \phi_l\right]\,\text{Tr}_r \left[ e^{-k\beta_r H_r } \phi_r\right]
\end{equation}
which precisely matches the SYK result in the $S_{probe}\to O( N)$ limit (\ref{wmuto0}) corresponding to the disconnected bulk phase of Fig.~\ref{fig:disconnected}. Due to factorization of $W_{rl}$ the modular flow in this case is again trivial but for a different reason: The probe is too large, backreacting on the bulk wormhole and disconnecting the left and right exteriors.

As in our main text analysis, it is the intermediate regime that is of interest for probing the black hole interior using modular flow. The important feature of this intermediate regime in our SYK example was the existence of a coupling between the left and right systems in the Euclidean path integral which could support the bulk Euclidean wormhole saddle. Such a coupling in the general formalism sketched in this Section can appear by including deviations from the Gaussian statistics for the operator matrix elements (\ref{stat2}). In fact, it is well known that the Gaussian approximation is inconsistent with maximal chaos, as manifested in the exponential decay of out-of-time-order 4-point functions \cite{Foini:2018sdb}. Given the importance of the maximally chaotic dynamics of SYK in our work, it would be interesting to investigate whether the corrected operator statistics required for maximal scrambling suffice to support the Euclidean wormhole of Fig.~\ref{fig:connected} that enables us to modular flow into the interior. We leave a careful investigation of this question for future work.

\subsection{Collisions behind the horizon}
Our setup of modular flowed operator allows us to reconstruct bulk operators behind horizon in the reference frame of the infalling semiclassical probe. As the backreaction of the probe to geometry is negligible and its trajectory is well described by a geodesic, we can regard it as a free-falling classical apparatus that measures the scattering amplitude of collisons behind the horizon. 

To be more precise, let us imagine we start with incoming particles generated by a series of boundary operators $\phi_l^1(t_{l,1})\cdots\phi_l^{n_l}(t_{l,n_l})\phi_r^1(t_{r,1})\cdots\phi_r^{n_r}(t_{r,n_r})$ acting on thermofield double state. Here we assume the $n_{l,r}\ll N$ such that perturbation theory of scattering holds. This incoming state consists of $n_l$ particles shooting from left boundary and $n_r$ particles shooting from right boundary. At some latter time, these particles will collide behind horizon to some outgoing particles. However, because of the horizon, these outgoing particles are not visible to boundary observer, which is the main obstacle to understand physics behind horizon.

\begin{figure}
\begin{centering}
\subfloat[\label{fig:8a}]{\begin{centering}
\includegraphics[height=5cm]{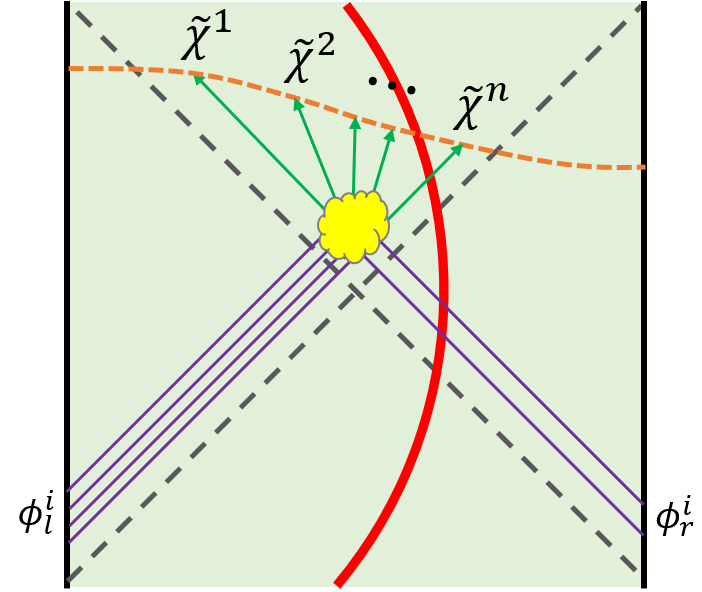}
\par\end{centering}
}\subfloat[\label{fig:8b}]{\begin{centering}
\includegraphics[height=5cm]{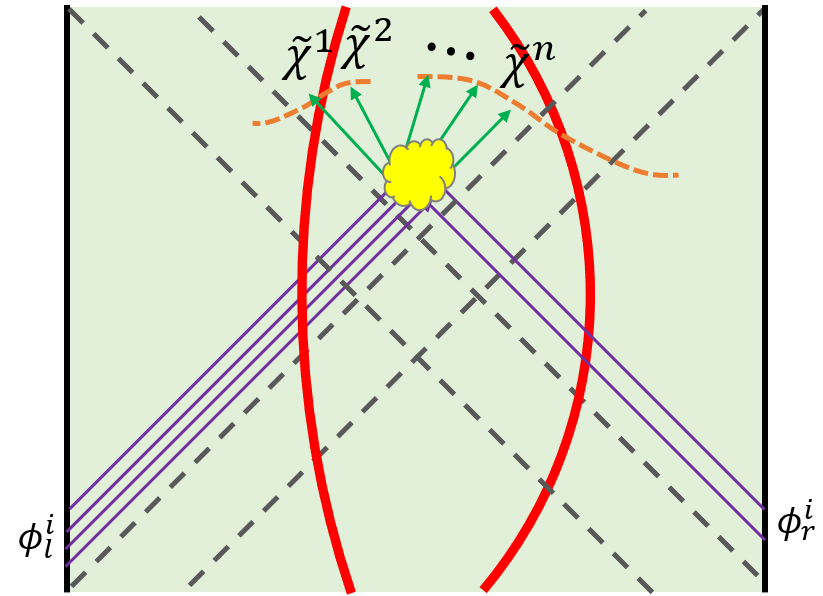}
\par\end{centering}
}
\par\end{centering}
\caption{Measure the scattering amplitude of boundary particles behind horizon. (a) is sending one probe to measure the amplitude to outgoing particles $\{\tilde{\chi}^1,\cdots,\tilde{\chi}^n\}$ on the whole Cauchy slice in thermofield double state, where $\tilde{\chi}^j(s)\equiv \r^{-is}\chi^j \r ^{is}$ are the modular flowed bulk operators. (b) is sending two probes in a more general spacetime (say long wormholes) to measure the amplitude because the ``atmosphere" of one probe can only extend to finite range. In both plots, red curves are worldlines of probe, orange dashed lines are the spatial slices (``atmosphere") related to the probe.}
\end{figure}

There is one way to study the outgoing particles by turning on some explicit coupling between two boundaries to form a traversable wormhole after all incoming particles are injected. The traversable wormhole opens a throat for outgoing particles and they could be seen by boundary observer. This proposal was studied in \cite{Haehl:2021tft} by computing six-point function in AdS$_2$. However, how many outgoing particles will be seen by boundary observer depends on the width of the throat opened by the traversable wormhole. Moreover, the negative energy from the explicit coupling to support the traversable wormhole will collide with the outgoing particles  and thus modulates the outgoing signal with details depending on the collision process.  

Alternatively, we can use our modular Hamiltonian to send the apparatus for outgoing particles into the horizon and measure the scattering amplitude without changing the geometry. We can study the following inner product
\be
\mA(\{\phi_l^i,\phi_r^j\}\ra\{\chi^k\})=\Tr\left(\r^{1-is}\chi^1\cdots\chi^n \r^{is} \phi_l^1(t_{l1})\cdots\phi_l^i(t_{li})\phi_r^1(t_{r1})\cdots\phi_r^j(t_{rj})\right) \label{eq:innerA}
\ee
where $\{\chi^k\}$ is a set of bulk operators initiated on global $t=0$ slice acting on thermofield double state with the probe $\r$. Note that the full set of $\chi^k$ could be reconstructed by HKLL method explained in Secrtion \ref{bulkbehind} by both left and right boundary data. Scanning all possible $\chi^k$ gives full information of the scattering amplitude of the collision among incoming particles behind horizon on a spatial slice related to the infalling probe after proper time $s \b _{probe}/(2\pi)$. See Fig.~\ref{fig:8a} for an illustration. Because we measure the scattering behind horizon directly, this approach also has advantage of not modulating the outgoing signal comparing to the method in \cite{Haehl:2021tft}. 

One might suspect that modulation still occurs because incoming and outgoing particles will collide with the probe when they intersect with the worldline of the latter. However, this is a subleading effect for the collision among particles because this scattering amplitude is proportional to the energy of the probe, which is low due to its worldline being far from boundary. One can already see this from the computation in Section \ref{bulkbehind} that the pole location of causal correlator for $\l>0$ does not contain Shapiro delay that one might have expected due to the collision between $\psi_l(t)$ with the probe before hitting $\chi_s$.

In more general spacetime, say long wormhole (e.g. \cite{Shenker:2013pqa, Goel:2018ubv} and also \cite{Brown:2019rox}), where we could only apply the modular flow to atmosphere operators that are close to the probe \cite{Jafferis:2020ora}, we can simply generalize above approach by including multiple probes with different worldlines to detect outgoing particles at different locations using the same inner product \eqref{eq:innerA} replacing $\r$ by the reduced density matrix for multiple probes  (Fig. \ref{fig:8b}). 


\section*{Acknowledgements} We would like to thank Jan de Boer, Daniel Jafferis, Arjun Kar, Ho Tat Lam, Adam Levine, Hong Liu, Mark Van Raamsdonk for 
stimulating and helpful discussions. PG is supported by the US Department of Energy grants DE-SC0018944 and DE-SC0019127. Both PG and LL are supported by the Simons foundation as members of the {\it It from Qubit} collaboration.

\appendix

\section{Analysis of twisted boundary conditions} \label{app:a}

Given the solution of Liouville equation, we will not be able to construct
a solution in which all $\s_{ab}$ meet at all $\Z\b_{a}$ points
and also respect all symmetries. First, requiring 
\begin{equation}
\s_{rl}^{s}(\b_{r},\tau)=\s_{rl}^{s+1}(0,\tau),\quad\s_{rl}^{s}(\tau,0)=\s_{rl}^{s+1}(\tau,\b_{l})\label{eq:43-1}
\end{equation}
for all $s$ is inconsistent with periodic condition $\s_{rl}^{k}=\s_{rl}^{0}$.
Above condition requires the function pair choice for $\s_{rl}^{s}$ be
$(h_{s},f)$ where the second function could be the same $f$. \footnote{It must be an $SL(2)$ of $f$, and by symmetry (\ref{eq:sym}) we
can choose it to be $f$.} Also, a careful check of this ansatz leads to 
\begin{equation}
h_{s}(\b_{r})=h_{s+1}(0)\label{eq:44-1}
\end{equation}
We must have $h_{s}$ and $f$ both to be monotonous function to guarantee
correlation function to be real (because of $1/q$ power of $e^{\s}$).
However, this obviously contradicts with (\ref{eq:44-1}) and $h_{k}=h_{0}$
because periodic function cannot be monotonous. Indeed, this argument
can be generalized to the case where difference of both sides of (\ref{eq:43-1})
is a constant, in which (\ref{eq:44-1}) still holds.

There are many other inconsistencies related to $\s_{ll}^{s}$ and
$\s_{rr}^{s}$. For $\s_{ll}^{s}$, the above periodic issue is avoid
by the reflection (\ref{eq:43-2}). By similar argument, boundary
condition
\begin{equation}
\s_{ll}^{s}(\b_{l},\tau)=\s_{ll}^{s+1}(0,\tau),~\s_{ll}^{s}(\tau,0)=\s_{ll}^{s+1}(\tau,\b_{l}),~\s_{ll}^{s}(\b_{l},\tau)=\s_{rl}^{s}(\b_{r},\tau),~\s_{ll}^{s}(0,\tau)=\s_{rl}^{s}(0,\tau)\label{eq:47}
\end{equation}
requires the function choice for $\s_{ll}^{s}$ to be $(f_{s},f)$
where all $f_{s}$ are related by $SL(2)$ transformations. The periodic
condition for $s=k$ leads to
\begin{equation}
(f_{k}(0),f(\tau))=(f_{0}(\tau),f(0))\implies f_{0}\simeq f
\end{equation}
Hence, each $f_{s}$ is some $SL(2)$ transformation of $f$. Taking
$f_{0}=\f{a+bf}{c+df}$ into UV condition (\ref{eq:43-2}) leads to
$f$ being in the form of $u+v\tan(\w\tau+\g)$. Indeed, any $SL(2)$
of $f$ is also in this form. 

Similarly, for $\s_{rr}^{s}$, we have 
\begin{equation}
\s_{rr}^{s}(\b_{r},\tau)=\s_{rr}^{s+1}(0,\tau),~\s_{rr}^{s}(\tau,0)=\s_{rr}^{s+1}(\tau,\b_{r}),~\s_{rr}^{s}(\tau,\b_{r})=\s_{rl}^{s}(\tau,\b_{l}),~\s_{rr}^{s}(\tau,0)=\s_{rl}^{s}(\tau,0)\label{eq:49}
\end{equation}
which leads to the function choice of $\s_{rr}^{s}$ to be $(\bar{h}_{s},h)$
where all $\bar{h}_{s}$ and $h_{s}$ are related by $SL(2)$. Moreover,
the periodic condition for $s=k$ and UV condition leads to $h_{s}\simeq h$
with $h$ in the same form as $f$ but with possibly different parameters.
Taking such tangent related functions, one can easily show that the
last two equations of (\ref{eq:47}) (or (\ref{eq:49})) that connect
$\s_{rl}^{s}$ with $\s_{ll}^{s}$ (or $\s_{rr}^{s}$) on two ends
cannot be satisfied. 

\section{Solving the recurrence} \label{app:slvrec}

There are two sequences to solve. To solve the recurrence, we first
define the following new variables
\begin{align}
y_{s} & =\f{\cos(\w\b_{l}+\g_{s})}{\cos\g_{s}},\quad x_{s}=v_{s}\sec^{2}\g_{s},\quad\lambda=\sin^{2}\w\b_{l}\\
\tilde{y}_{s} & =\f{\cos(\w\b_{r}+\tilde{\g}_{s})}{\cos\tilde{\g}_{s}},\quad\tilde{x}_{s}=\tilde{v}_{s}\sec^{2}\tilde{\g}_{s},\quad\tilde{\lambda}=\sin\w\b_{l}\sin\w\b_{r} \label{eq:B.2}
\end{align}
The recurrence (\ref{eq:vrecur}) and (\ref{eq:grecur}) can be rewritten
as 
\begin{equation}
x_{s+1}=\a_{s}x_{s}y_{s}^{-2},\quad y_{s+1}-y_{s}=-\a_{s}\lambda x_{s}y_{s}^{-1}\label{eq:74}
\end{equation}
and (\ref{eq:vtrecur}) and (\ref{eq:gtrecur}) can be rewritten as
\begin{equation}
\tilde{x}_{s+1}=\tilde{\a}_{s}\tilde{x}_{s}\tilde{y}_{s}^{-2},\quad\tilde{y}_{s+1}-\tilde{y}_{s}=\tilde{\a}_{s}\tilde{\lambda}\tilde{x}_{s}\tilde{y}_{s}^{-1}\label{eq:75}
\end{equation}
It follows that 
\begin{equation}
y_{s+1}/y_{s}-1=-\lambda x_{s+1},\quad\tilde{y}_{s+1}/\tilde{y}_{s}-1=\tilde{\lambda}\tilde{x}_{s+1}\label{eq:69}
\end{equation}
Taking them back to the second equations of (\ref{eq:74}) and (\ref{eq:75})
leads to a recurrence for $y_{s}$ and $\tilde{y}_{s}$ on themselves
\begin{equation}
\f{y_{s+1}/y_{s}-1}{y_{s}/y_{s-1}-1}=\a_{s}y_{s}^{-2},\quad\f{\tilde{y}_{s+1}/\tilde{y}_{s}-1}{\tilde{y}_{s}/\tilde{y}_{s-1}-1}=\tilde{\a}_{s}\tilde{y}_{s}^{-2}\label{eq:77}
\end{equation}

\begin{figure}
\begin{centering}
\subfloat[]{\begin{centering}
\includegraphics[width=4.8cm]{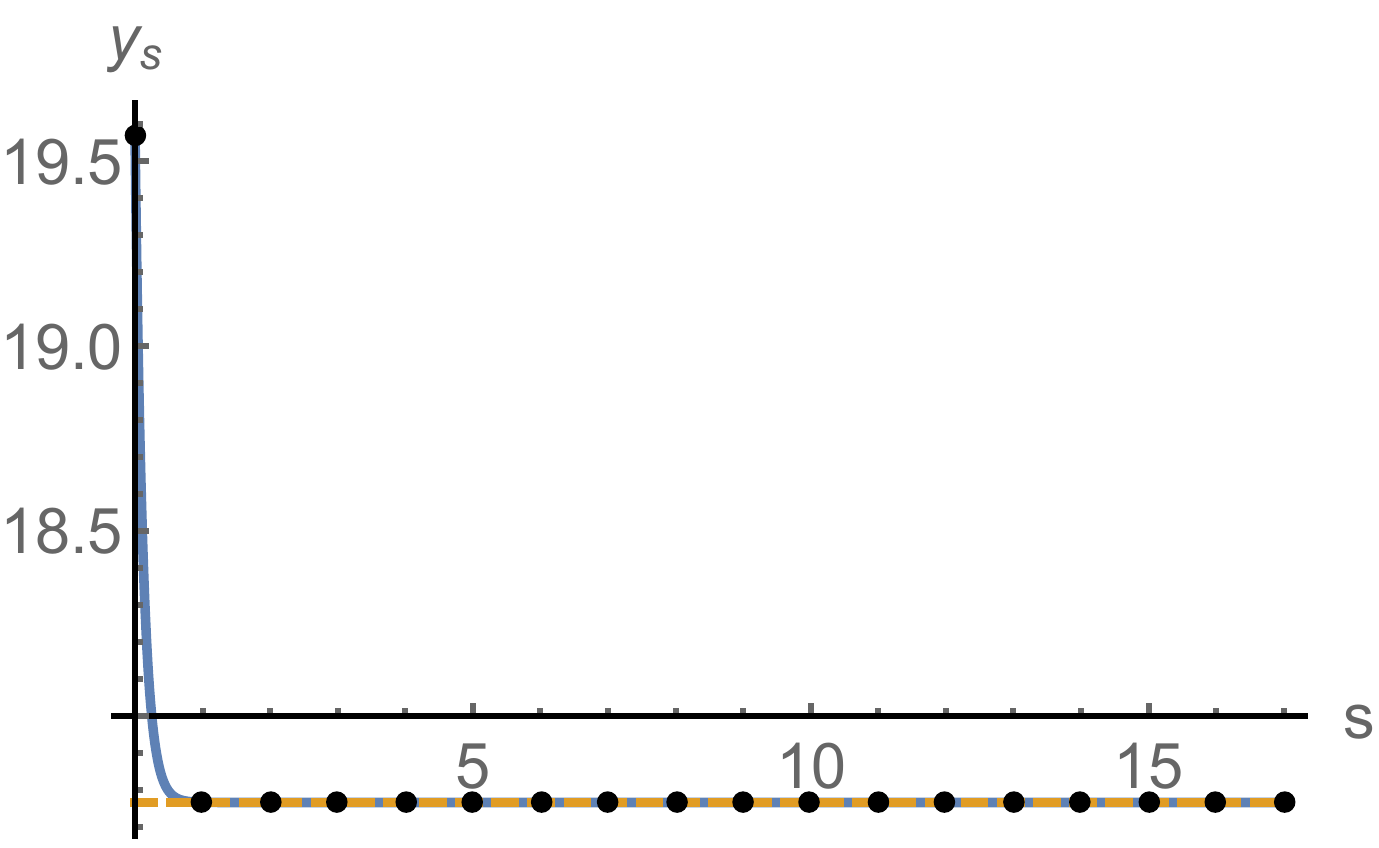}
\par\end{centering}
}\subfloat[]{\begin{centering}
\includegraphics[width=4.8cm]{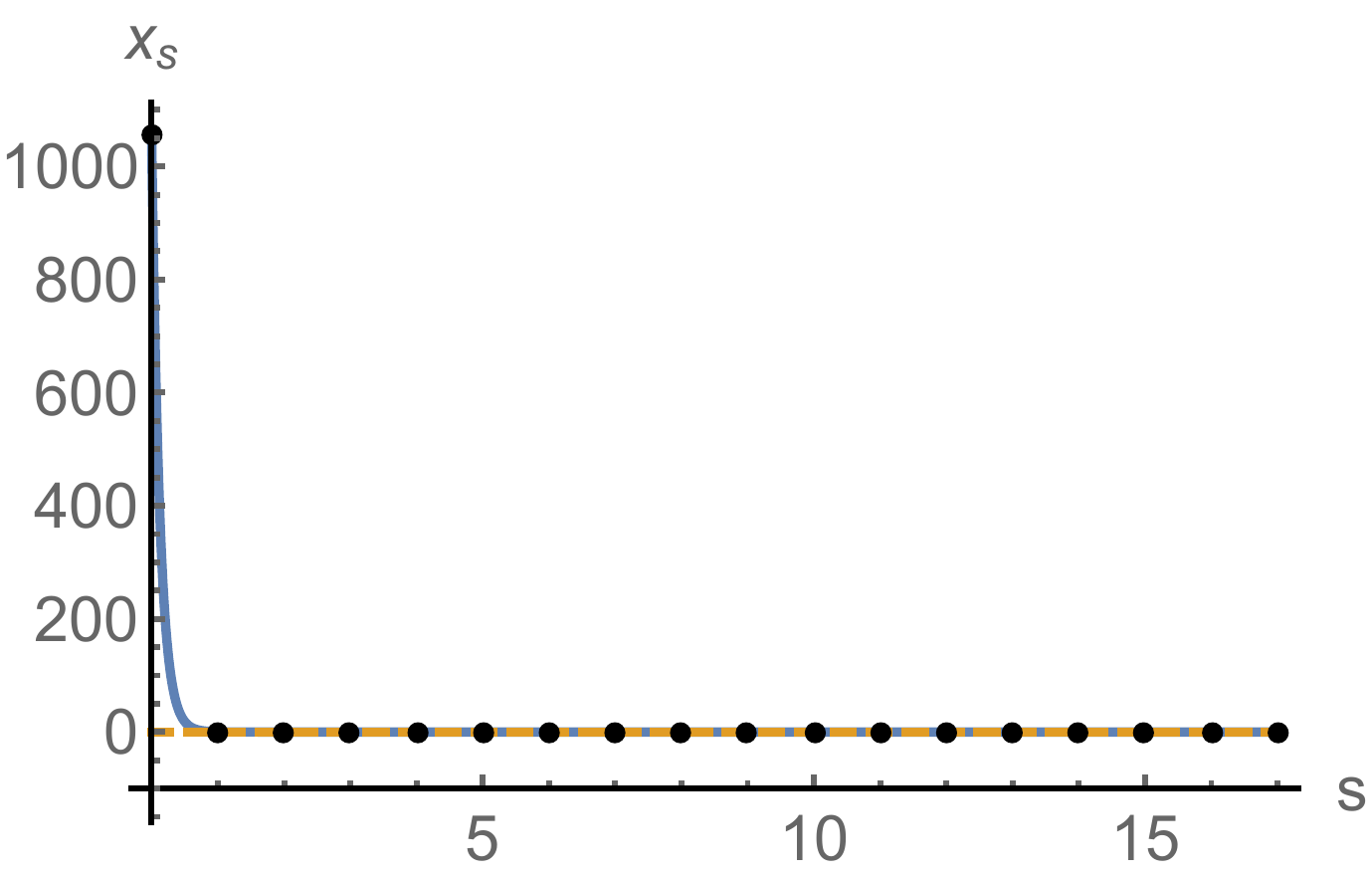}
\par\end{centering}
}\subfloat[]{\begin{centering}
\includegraphics[width=4.8cm]{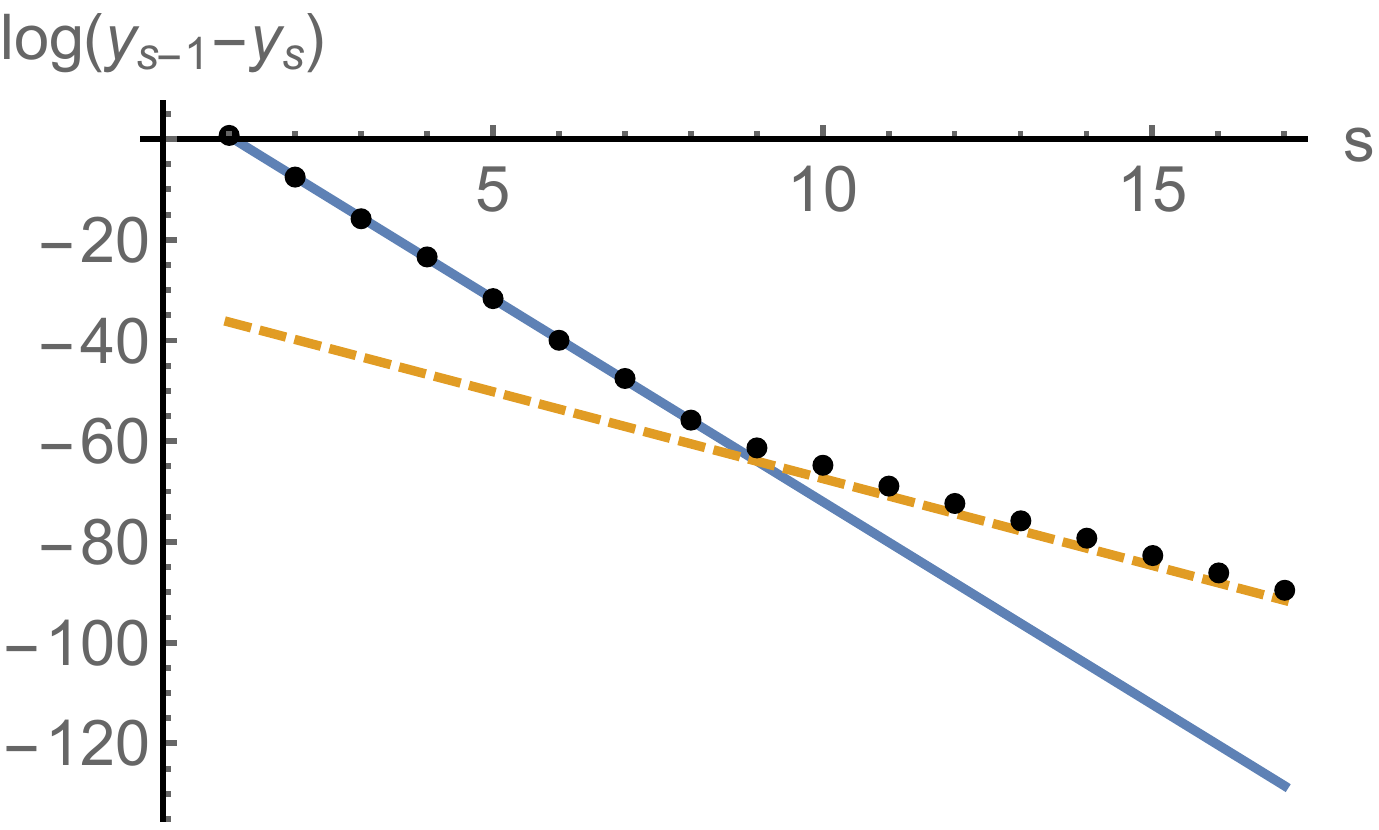}
\par\end{centering}
}
\par\end{centering}
\caption{(a) Exact solution of $y_{s}$ versus approximation $y(s)$. (b) Exact
solution of $x_{s}$ versus approximation $x(s)$. (c) Exact solution
of $\log(y_{s}-y_{s-1})$ versus approximation $\log(y(s)-y(s-1))$.
In all plots, the black dots are exact data, blue curve is the approximation
for $s<\left\lfloor k/2\right\rfloor $ and yellow dashed curve is
for $s>\left\lfloor k/2\right\rfloor $. We see that both $x_{s}$
and $y_{s}$ converge very fast and the approximations match very
well. The difference between the approximations of $s<\left\lfloor k/2\right\rfloor $
and $s>\left\lfloor k/2\right\rfloor $ are very small and only visible
when we check $y_{s}-y_{s-1}$ in log plot. Other parameters are $\protect\b_{l}=1$,
$\protect\b_{r}=4$, $\protect\mJ=20$, $\protect\a=1/10$ and $k=17$.
\label{fig:xy}}
\end{figure}
However, these recurrence cannot be solved explicitly. We assume $k$
to be an odd number. Let us take large $\mu$ case in which $\a_{s}$
and $\tilde{\a}_{s}$ become identical and piecewise constant
\begin{equation}
\a_{s}=\tilde{\a}_{s}\ra\begin{cases}
\a=e^{-\mu(q-2)} & s=0,\cdots,\left\lfloor k/2\right\rfloor -1\\
1 & s=\left\lfloor k/2\right\rfloor \\
1/\a & s=\left\lfloor k/2\right\rfloor +1,\cdots,k-1
\end{cases}\label{eq:78}
\end{equation}
Furthermore, we will solve (\ref{eq:77}) approximately by replacing
it with its differential version
\begin{equation}
(\log(\log y)')'=\log\a_{s}-2\log y\label{eq:79}
\end{equation}
where $y=y(s)$. This differential equation can be solved for each
piece where $\a_{s}$ is a constant as
\begin{equation}
y(s)=\begin{cases}
\a^{1/2}\exp \left[c_{1}\coth(c_{1}s+b_{1})\right] & s<\left\lfloor k/2\right\rfloor \\
\a^{-1/2}\exp\left[ c_{2}\coth(c_{2}s+b_{2})\right] & s>\left\lfloor k/2\right\rfloor 
\end{cases}\label{eq:80}
\end{equation}
Here we ignored the $s=\left\lfloor k/2\right\rfloor $ case because
it is just one point and not related to our later analytic continuation.
Here need to choose $c_{i}$ and $b_{i}$ to be real parameters because
(\ref{eq:74}) shows that $y_{s}$ is monotonically decreasing sequence.
To determine these four parameters, we will impose the following condtions.
For small $\a$, we find that $y$ decays to its limit value very
fast (see Fig. \ref{fig:xy}), we can use the limit value $y_{\infty}$
and initial value $y_{0}$ to fix $c_{1}$ and $b_{1}$. Here is a
caveat that the limit value $y_{\infty}$ should be defined as the
one using $\a_{s}=\a$ all along the sequence. But it turns out to
be the same as the limit value if we use (\ref{eq:78}) and take $k$
to infinity limit, which we denote as $y_{\infty}$. To fix $c_{2}$
and $b_{2}$, besides the limit value $y_{\infty}$, we also use the
continuity condition of $y(s)$ at $s=\left\lfloor k/2\right\rfloor $.
One can easily solve them as
\begin{align}
c_{1} & =\log(y_{\infty}/\a^{1/2}),\quad b_{1}=\text{arccoth}(\log(y_{0}/\a^{1/2})/\log(y_{\infty}/\a^{1/2}))\\
c_{2} & =\log(y_{\infty}\a^{1/2}),\quad b_{2}=\text{arccoth}\left(\f{\log\a+c_{1}\coth(c_{1}\left\lfloor k/2\right\rfloor +b_{1})}{c_{2}}\right)-c_{2}\left\lfloor k/2\right\rfloor 
\end{align}
The numerics in Fig. \ref{fig:xy} show that this approximation matches
with exact result pretty well. With solution (\ref{eq:80}), we can
take it into the first equation of (\ref{eq:74}) and find
\begin{equation}
x_{s}=\begin{cases}
x_{0}e^{-2c_{1}\sum_{i=0}^{s-1}\coth(c_{1}i+b_{1})} & s\leq\left\lfloor k/2\right\rfloor \\
x_{0}y(\lfloor k/2\rfloor ) e^{-2c_{1}\sum_{i=0}^{\left\lfloor k/2\right\rfloor -1}\coth(c_{1}i+b_{1})-2c_{2}\sum_{i=\left\lfloor k/2\right\rfloor +1}^{s-1}\coth(c_{2}i+b_{2})} & s>\left\lfloor k/2\right\rfloor 
\end{cases}
\end{equation}
Similarly, if we approximate the sum as integral (just like taking
recurrence sequence as differential equation), we get
\begin{equation}
x(s)=\begin{cases}
\f{x_{0}\sinh^{2}b_{1}}{\sinh^{2}(c_{1}s+b_{1})} & s<\left\lfloor k/2\right\rfloor \\
\f{x_{0}\sinh^{2}b_{1}\sinh^{2}(c_{2}\left\lfloor k/2\right\rfloor +b_{2})}{\sinh^{2}(c_{1}\left\lfloor k/2\right\rfloor +b_{1})\sinh^{2}(c_{2}s+b_{2})} & s>\left\lfloor k/2\right\rfloor 
\end{cases}\label{eq:84}
\end{equation}
For our approximation (\ref{eq:50}), we will use the two solutions
in (\ref{eq:80}) and (\ref{eq:84}) respectively in $\s_{ll}^{s}$
and $\s_{ll}^{k-s}$. In terms of $x(s)$ and $y(s)$, we have the large $q$ solution to be
\begin{align}
 g_{ll}(s)& e^{\s_{ll}^{s}(\tau_{1},\tau_{2})/q}
=\f 12\left(\w\lambda \mJ^{-1}x(s)^{1/2}y(s)[(\sin\w(\b_{l}-\tau_{1})+y(s)\sin\w\tau_{1})\right.\nn\\
&\left. \times (\sin\w\tau_{2}+y(s)\sin\w(\b_{l}-\tau_{2}))-\lambda x(s)\sin\w\tau_{2}\sin\w(\b_{l}-\tau_{1})]^{-1}\right)^{2/q} 
\end{align}

\begin{figure}
\begin{centering}
\subfloat[]{\begin{centering}
\includegraphics[width=4.8cm]{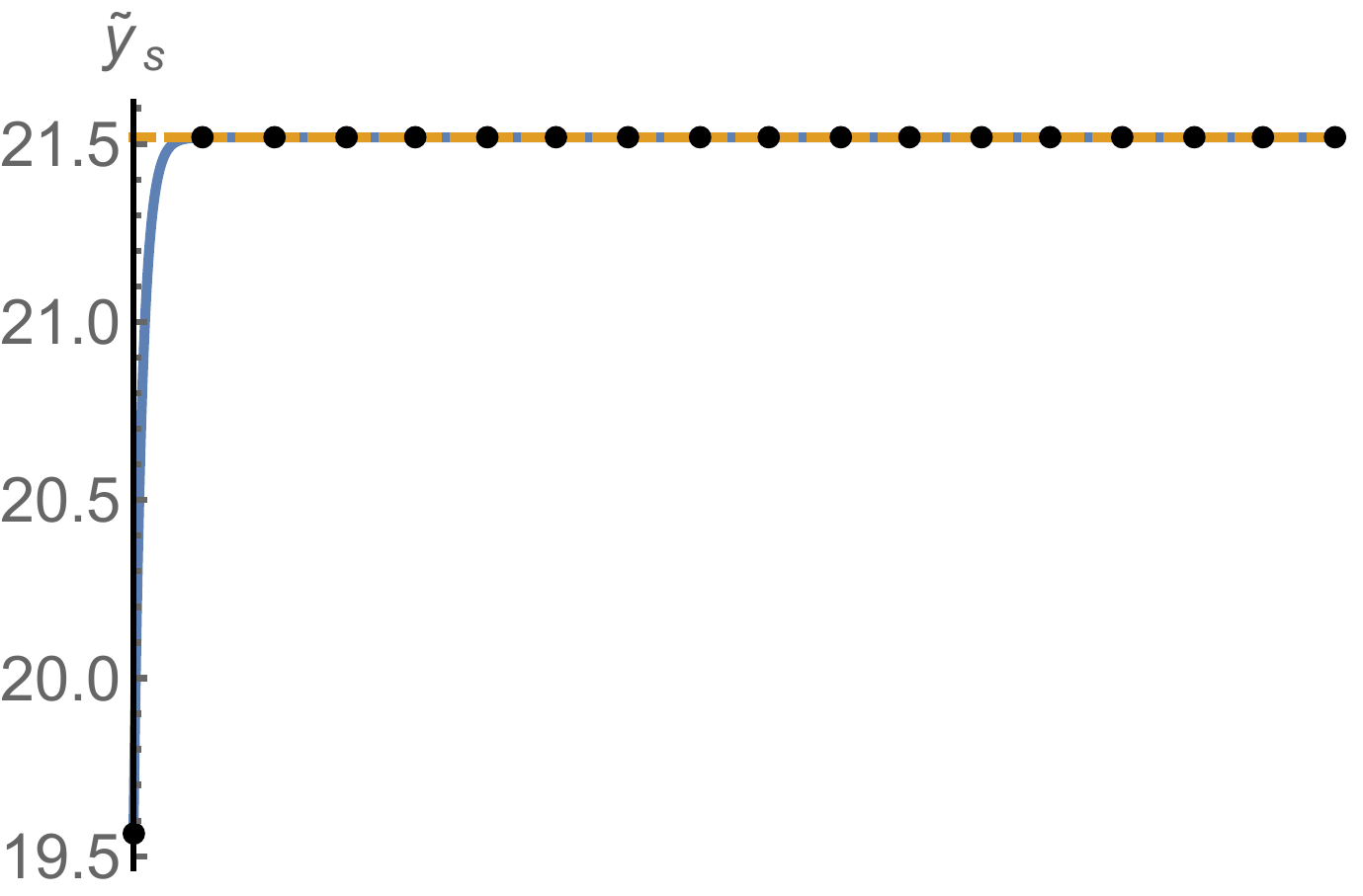}
\par\end{centering}
}\subfloat[]{\begin{centering}
\includegraphics[width=4.8cm]{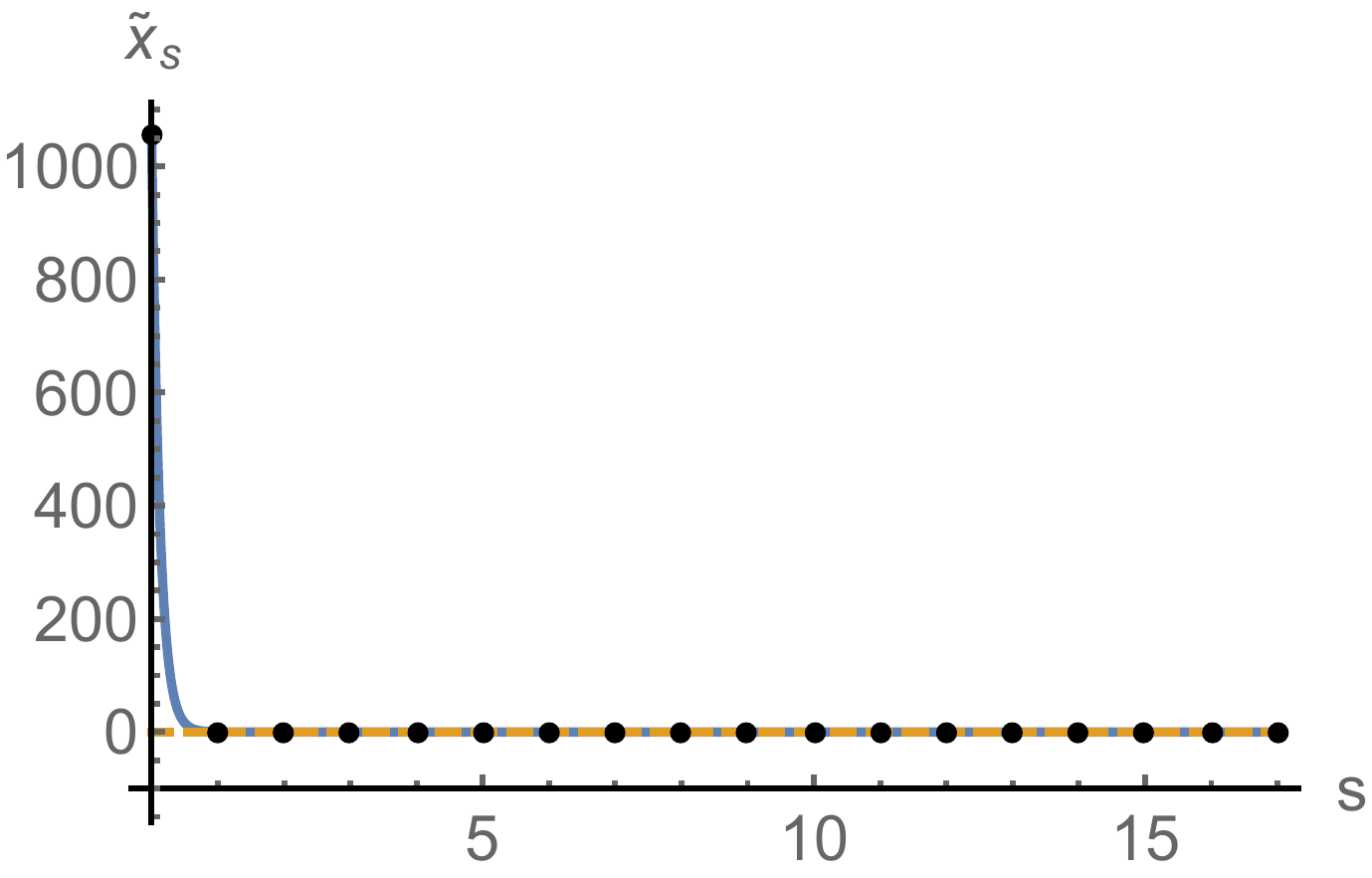}
\par\end{centering}
}\subfloat[]{\begin{centering}
\includegraphics[width=4.8cm]{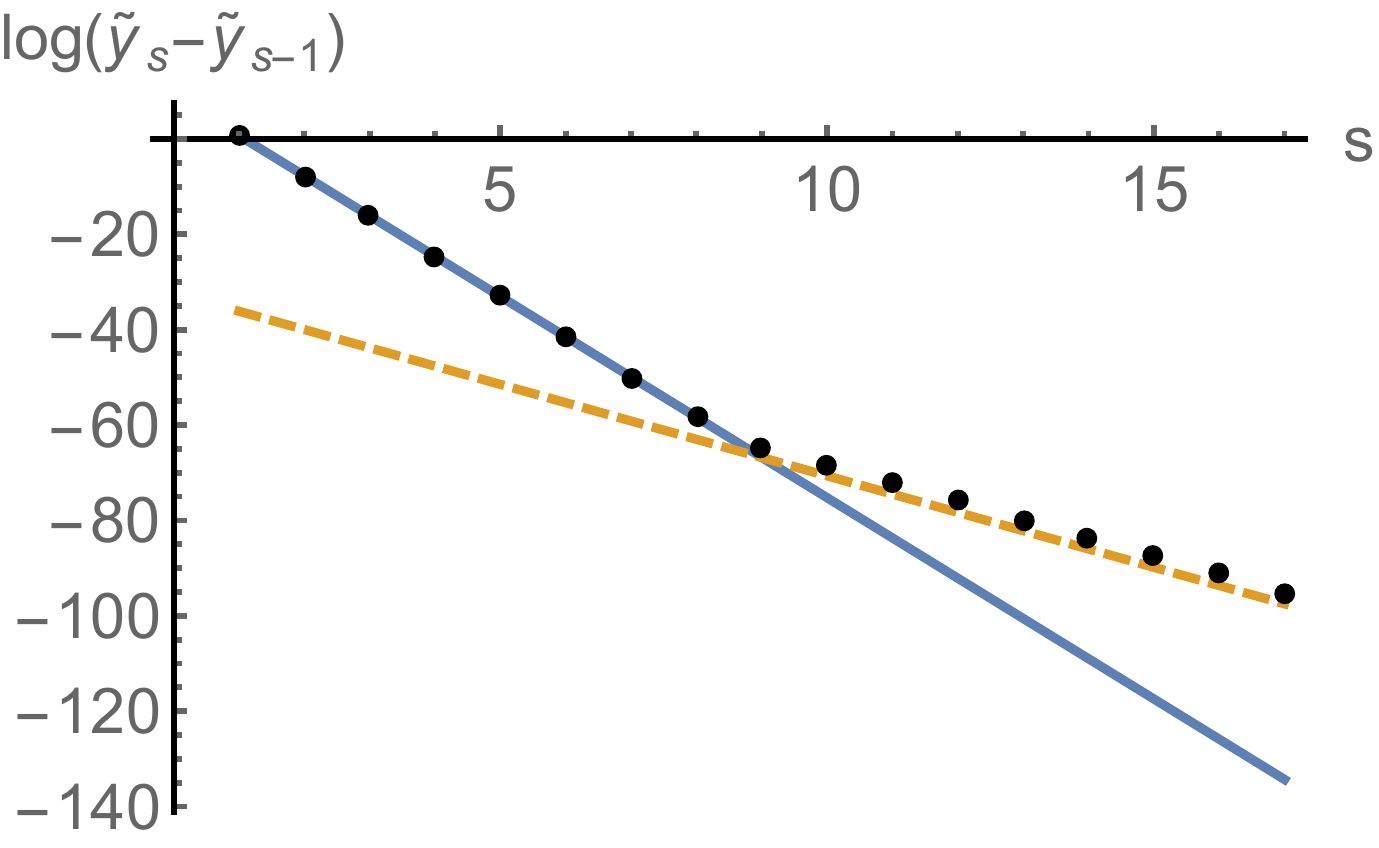}
\par\end{centering}
}
\par\end{centering}
\caption{(a) Exact solution of $\tilde{y}_{s}$ versus approximation $\tilde{y}(s)$.
(b) Exact solution of $\tilde{x}_{s}$ versus approximation $\tilde{x}(s)$.
(c) Exact solution of $\log(\tilde{y}_{s}-\tilde{y}_{s-1})$ versus
approximation $\log(\tilde{y}(s)-\tilde{y}(s-1))$. All settings are
the same as Fig. \ref{fig:xy}. \label{fig:xtyt}}
\end{figure}
It is very similar to solve the other recurrence sequence using the
differential equation approximation. However, from the second equation
in (\ref{eq:75}), $\tilde{y}_{s}$ is a monotonically increasing
function. Hence, the solution to the same differential equation (\ref{eq:79})
should be chosen as
\begin{equation}
\tilde{y}(s)=\begin{cases}
\a^{1/2}\exp\tilde{c}_{1}\tanh(\tilde{c}_{1}s+\tilde{b}_{1}) & s<\left\lfloor k/2\right\rfloor \\
\a^{-1/2}\exp\tilde{c}_{2}\tanh(\tilde{c}_{2}s+\tilde{b}_{2}) & s>\left\lfloor k/2\right\rfloor 
\end{cases}\label{eq:86}
\end{equation}
where the parameters should be determined in the same way as
\begin{align}
\tilde{c}_{1} & =\log(\tilde{y}_{\infty}/\a^{1/2}),\quad\tilde{b}_{1}=\text{arctanh}(\log(\tilde{y}_{0}/\a^{1/2})/\log(\tilde{y}_{\infty}/\a^{1/2}))\\
\tilde{c}_{2} & =\log(\tilde{y}_{\infty}\a^{1/2}),\quad\tilde{b}_{2}=\text{arctanh}\left(\f{\log\a+\tilde{c}_{1}\tanh(\tilde{c}_{1}\left\lfloor k/2\right\rfloor +\tilde{b}_{1})}{\tilde{c}_{2}}\right)-\tilde{c}_{2}\left\lfloor k/2\right\rfloor 
\end{align}
It follows that
\begin{equation}
\tilde{x}(s)=\begin{cases}
\f{\tilde{x}_{0}\cosh^{2}\tilde{b}_{1}}{\cosh^{2}(\tilde{c}_{1}s+\tilde{b}_{1})} & s<\left\lfloor k/2\right\rfloor \\
\f{\tilde{x}_{0}\cosh^{2}\tilde{b}_{1}\cosh^{2}(\tilde{c}_{2}\left\lfloor k/2\right\rfloor +\tilde{b}_{2})}{\cosh^{2}(\tilde{c}_{1}\left\lfloor k/2\right\rfloor +\tilde{b}_{1})\cosh^{2}(\tilde{c}_{2}s+\tilde{b}_{2})} & s>\left\lfloor k/2\right\rfloor 
\end{cases}\label{eq:84-1}
\end{equation}
From Fig. \ref{fig:xtyt}, we see clearly that our approximation works
very well. By our solution of $s=0$, we have 
\begin{align}
x_{0} & =\tilde{x}_{0}=\sec^{2}\w(\b_{l}+\b_{r})/2\label{eq:90}\\
y_{0} & =\tilde{y}_{0}=\cos\w(\b_{l}-\b_{r})/2\sec\w(\b_{l}+\b_{r})/2\label{eq:91}
\end{align}
In terms of $\tilde{x}(s)$ and $\tilde{y}(s)$, we have
\begin{align}
g_{rl}(s)&e^{\s_{rl}^{s}(\tau_{1},\tau_{2})/q}
=  \f{\sgn(g_{rl}(s))}2\left(\w\tilde{\lambda}\mJ^{-1}\tilde{x}(s)^{1/2}\tilde{y}(s)[(\sin\w(\b_{r}-\tau_{1})+\tilde{y}(s)\sin\w\tau_{1})\right. \nn\\
&\left.\times(\sin\w\tau_{2}+\tilde{y}(s)\sin\w(\b_{l}-\tau_{2}))+\tilde{\lambda}\tilde{x}(s)\sin\w\tau_{2}\sin\w(\b_{r}-\tau_{1})]^{-1}\right)^{2/q}
\end{align}

For $\s_{rr}^{s}$ and $\s_{lr}^{s}$, we can simply switch $\b_{l}\leftrightarrow\b_{r}$.
Note that to get $\s_{lr}^{s}$, we can also use symmetry (\ref{eq:62}),
which turns out to be the same as swap $\b_{l}\leftrightarrow\b_{r}$.
This is a consistent check that based on the fact that $\tilde{x}_{s}$
and $\tilde{y}_{s}$ are both invariant under swap $\b_{l}\leftrightarrow\b_{r}$,
which is because initial values $\tilde{x}_{0}$ and $\tilde{y}_{0}$
and recurrence equations all preserve this symmetry.

\begin{figure}
\begin{centering}
\subfloat[]{\begin{centering}
\includegraphics[width=3.7cm]{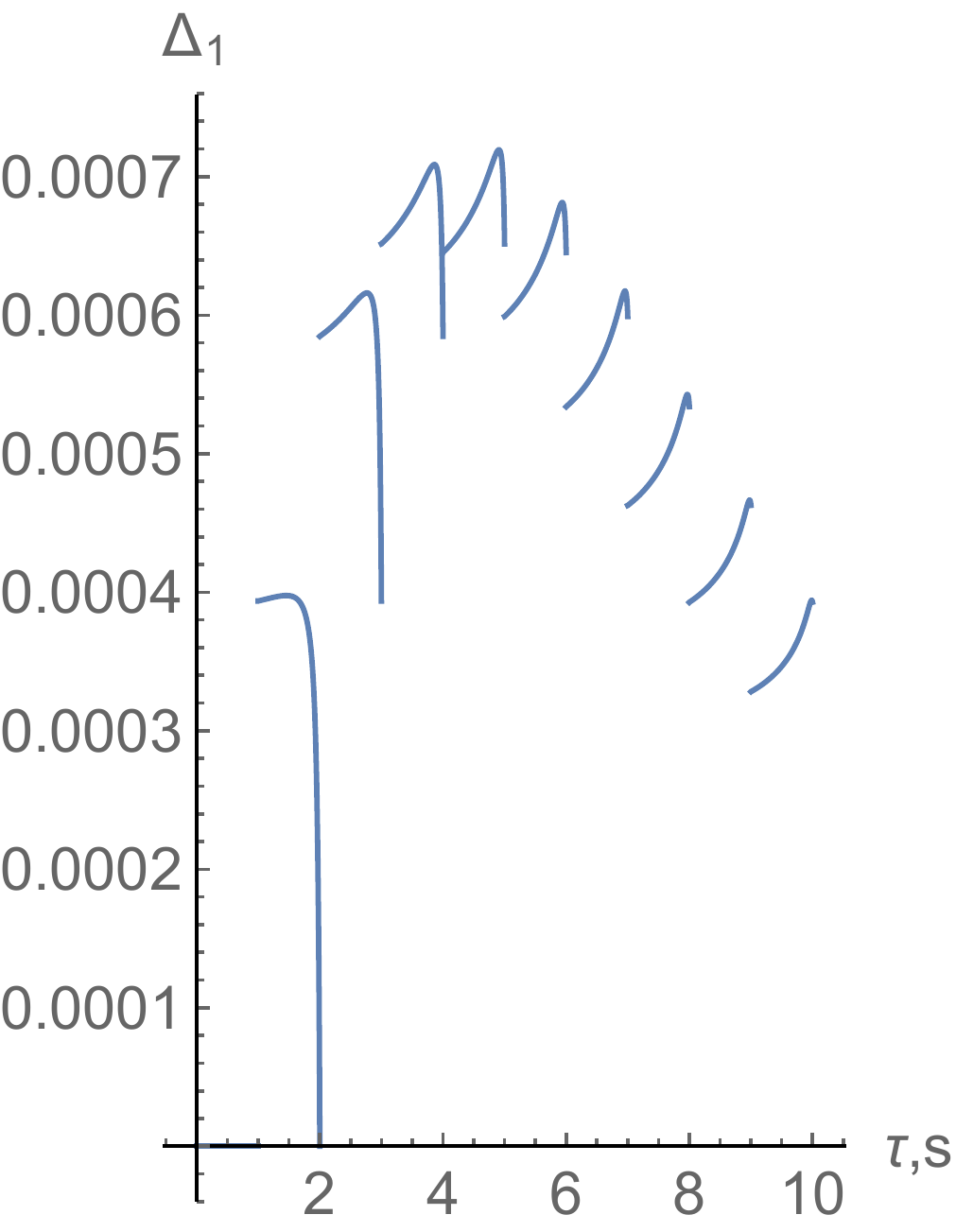}
\par\end{centering}
}\subfloat[]{\begin{centering}
\includegraphics[width=3.7cm]{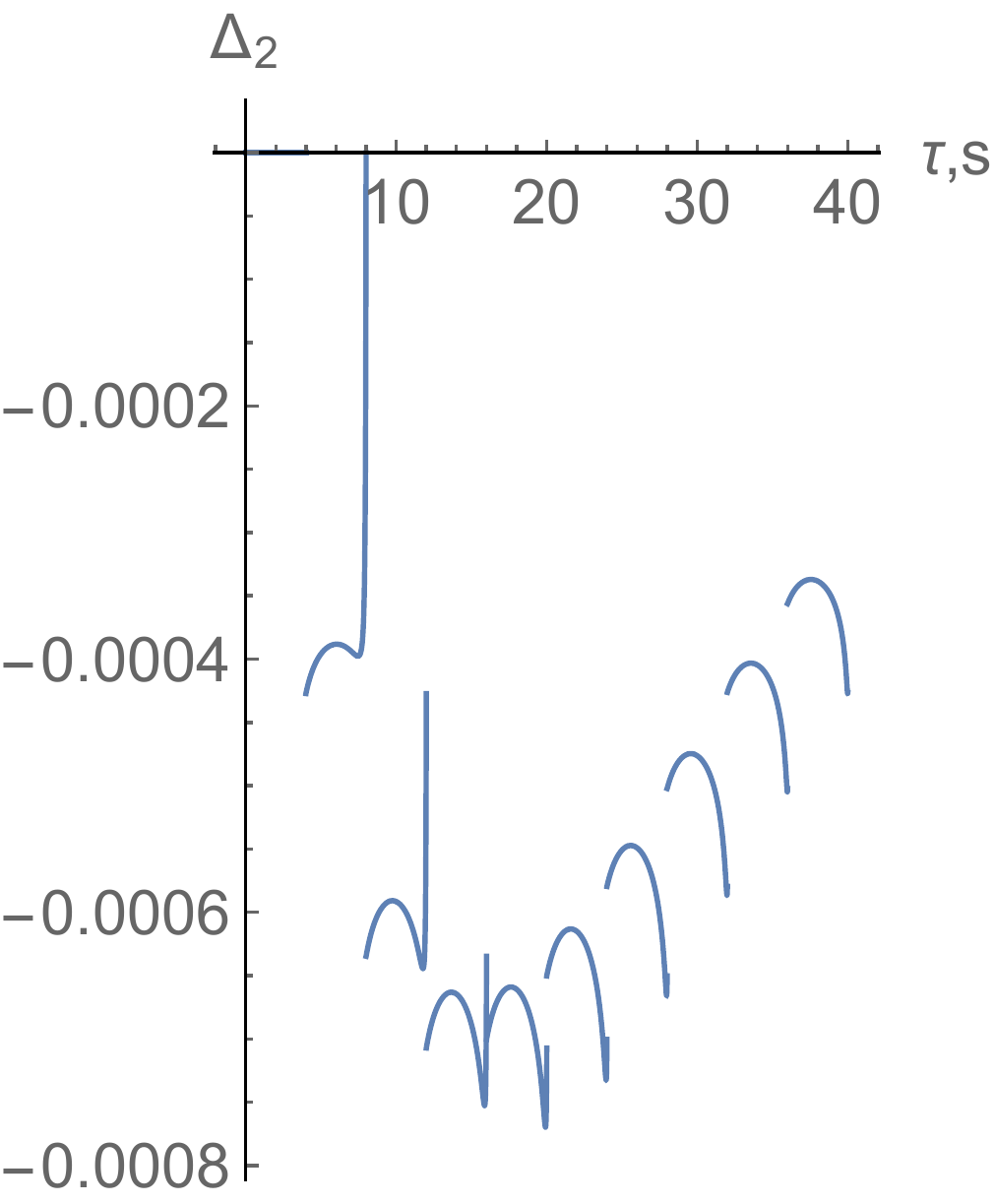}
\par\end{centering}
}\subfloat[]{\begin{centering}
\includegraphics[width=3.7cm]{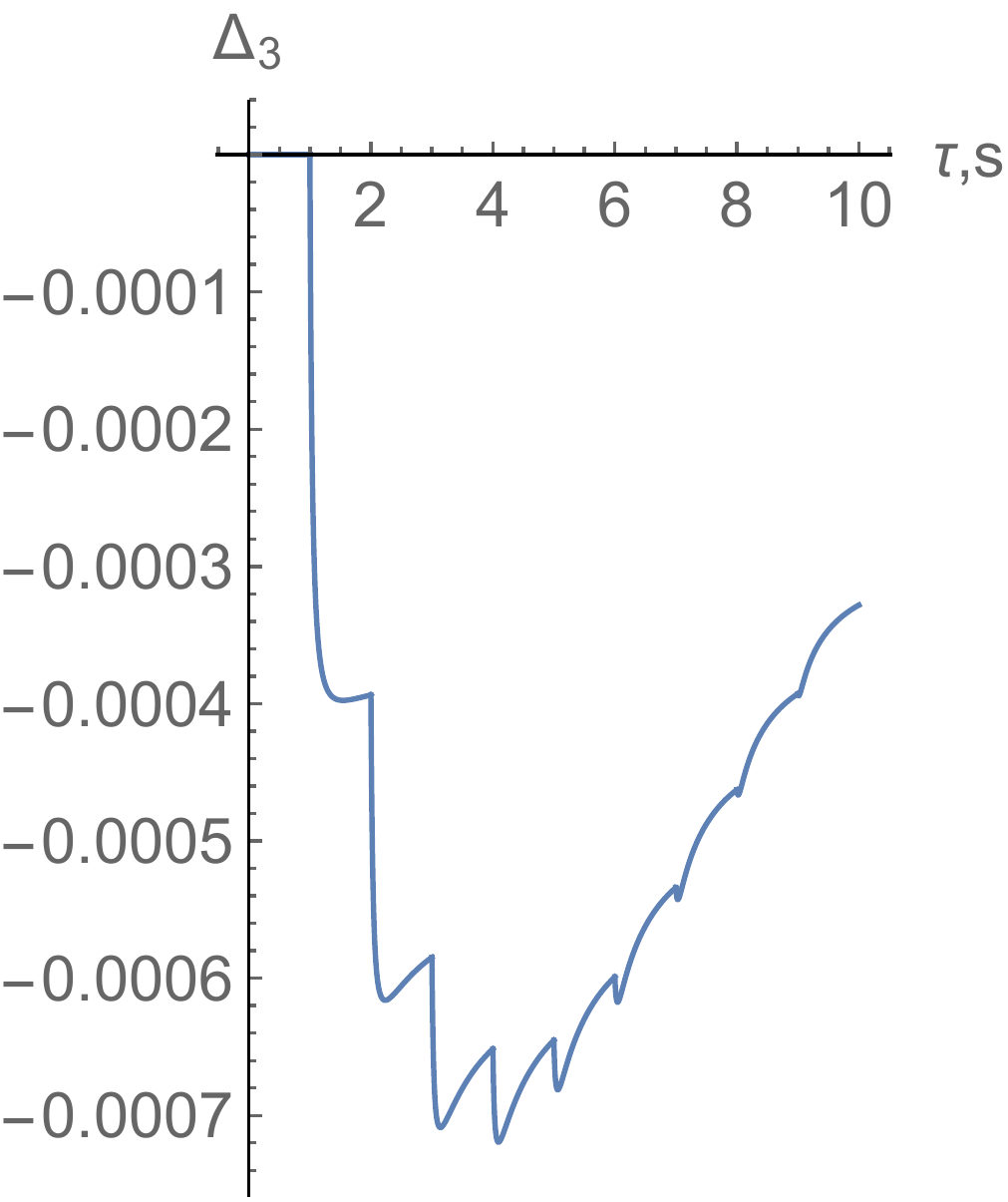}
\par\end{centering}
}\subfloat[]{\begin{centering}
\includegraphics[width=3.7cm]{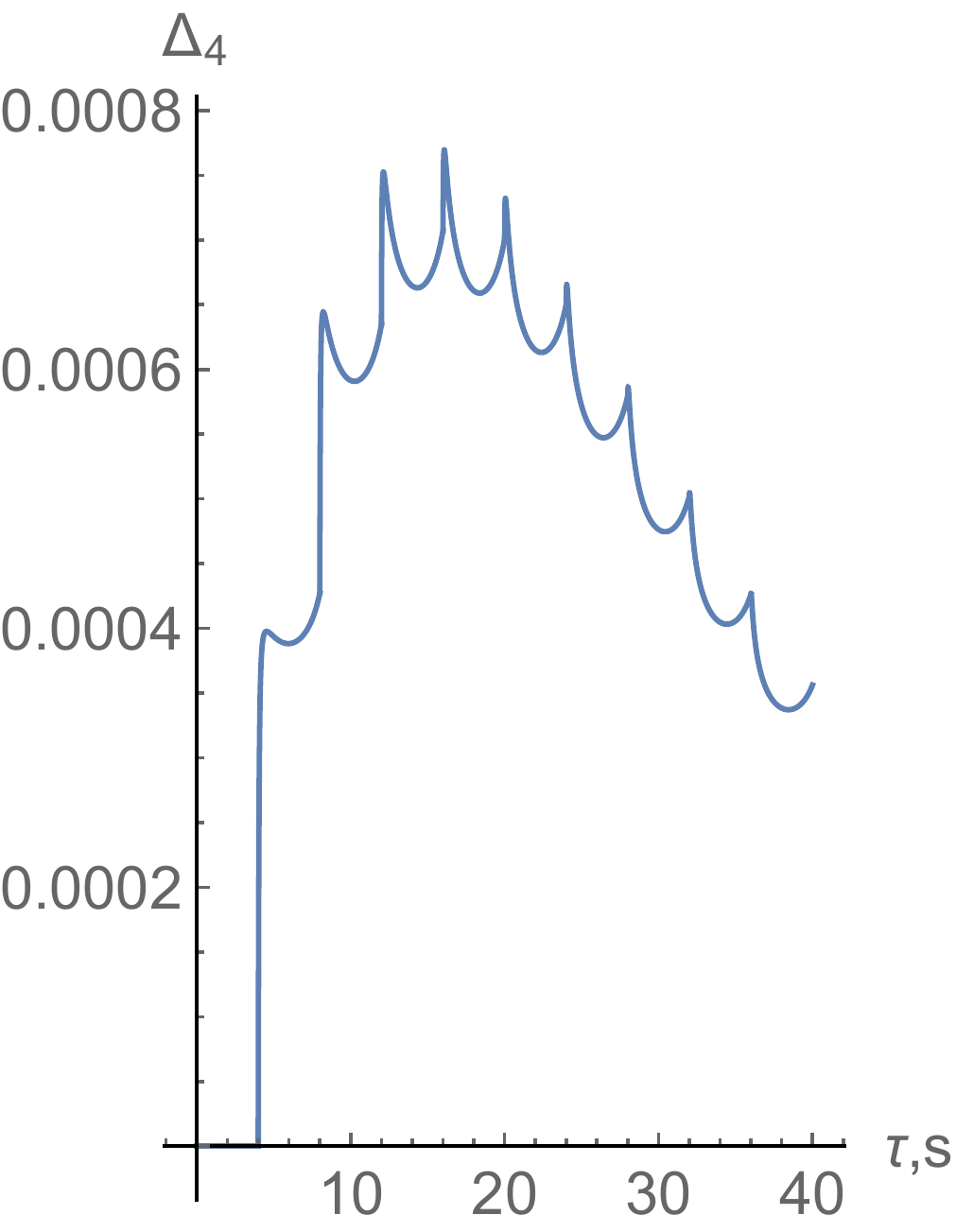}
\par\end{centering}
}
\par\end{centering}
\caption{The errors of twist boundary condition. The horizontal axis is $\tau$
plotted over $[0,\protect\b_{l}]$ for $\protect\D_{1,3}$ and over
$[0,\protect\b_{r}]$ for $\protect\D_{2,4}$ for all $s$ in order,
that is, putting all different $s$ in one plot where $[0,\protect\b_{a}]$
is for $s=0$, $[\protect\b_{a},2\protect\b_{a}]$ is for $s=1$ and
so on. Here the parameters are $\protect\b_{l}=1$, $\protect\b_{r}=4$,
$\protect\mJ=20$, $\protect\a=1/500$, $q=20$ and $k=9$. If we
increase $\protect\b$, namely decrease $\protect\a$, the error will
overall be smaller. We see that the error is much smaller than $1/q=0.05$
for this choice of small $\protect\a$. \label{fig:The-errors-of}}
\end{figure}
Given this solution, we need to check how much the twist boundary
condition in (\ref{eq:21})-(\ref{eq:24}) are violated. Note that
in large $\b$ limit, the factors involving hyperbolic functions become
\begin{align}
\left|\f{\sinh\mu\sinh\f{(k-2(s-1))\mu}2}{\cosh\f{(k-2s)\mu}2}\right| & \app\left|\f{\sinh\mu\cosh\f{(k-2(s-1))\mu}2}{\sinh\f{(k-2s)\mu}2}\right|\ra\begin{cases}
\f 12e^{2\mu} & s\leq\left\lfloor k/2\right\rfloor \\
\f 12e^{\mu} & s=\left\lfloor k/2\right\rfloor +1\\
\f 12 & s>\left\lfloor k/2\right\rfloor +1
\end{cases}\\
\left|\f{\sinh\mu\sinh\f{(k-2s)\mu}2}{\cosh\f{(k-2(s-1))\mu}2}\right| & \app\left|\f{\sinh\mu\cosh\f{(k-2s)\mu}2}{\sinh\f{(k-2(s-1))\mu}2}\right|\ra\begin{cases}
\f 12 & s\leq\left\lfloor k/2\right\rfloor \\
\f 12e^{\mu} & s=\left\lfloor k/2\right\rfloor +1\\
\f 12e^{2\mu} & s>\left\lfloor k/2\right\rfloor +1
\end{cases}
\end{align}
All LHS of (\ref{eq:21})-(\ref{eq:24}) are zero by our ansatz. RHS
are generally nonzero and can be categorized into four types
\begin{align}
\D_{1} & =e^{2\mu}(\exp(\s_{ll}^{s}(\b_{l},\tau)/q)-\exp(\s_{rl}^{s}(\b_{r},\tau)/q))\\
\D_{2} & =e^{2\mu}(\exp(\s_{lr}^{s}(\b_{l},\tau)/q)-\exp(\s_{rr}^{s}(\b_{r},\tau)/q))\\
\D_{3} & =e^{2\mu}(\exp(\s_{lr}^{s}(\tau,0)/q)-\exp(\s_{ll}^{s}(\tau,0)/q))\\
\D_{4} & =e^{2\mu}(\exp(\s_{rr}^{s}(\tau,0)/q)-\exp(\s_{rl}^{s}(\tau,0)/q))
\end{align}
which are upper bound of errors in RHS. In Fig. \ref{fig:The-errors-of},
we plot $\D_{i}$ for all choices of $\tau$ and $s$. In this figure,
we find that when we increase $\b$, equivalently decrease $\a$,
the errors decrease. With the parameters Fig. \ref{fig:The-errors-of},
we see that the errors are typically much smaller than $1/q$. Therefore,
we should trust our solution in large $\b$ limit.

\section{Validity of large $q$ solution} \label{app:b}

Although we find perfect match between our large $q$ solution with
bulk semiclassical computation, we should not expect the solution
well describing black hole physics for arbitrary large $s$. On one
hand, SYK model has distinct long time behavior than semiclassical
gravity, e.g. ramp and plateau in the form factor \cite{Saad:2018bqo} are described
non-pertrubative effects in JT gravity. On the other hand, for a black
hole, the probe will not extend its worldline inside horizon for
infinite proper time because it will eventually hit singularity. However,
it seems neither of these two bounds can be applied our current analysis.
The first type of long time behavior is for boundary time. It is unclear
how that will be related to the proper time of an infalling probe
behind horizon. In particular, from Fig. \ref{fig:7a} and Fig. \ref{fig:7b},
it is clear that after just order one proper time evolution, the spatial
slice of probe already goes beyond two Rindler wedges. The second
type of limitation from singularity unfortunately does not exist in
JT gravity because it has constant curvature everywhere. One could
define the singularity of JT gravity as the curve with large and negative
dilaton value $-\phi_{0}$ understood as dimensional reduction from
higher dimensional near extremal black hole \cite{Harlow:2018tqv}.
However, for large $\phi_{0}$, the singularity is time-like and most
probes are free from hitting it. It was argued in \cite{Jafferis:2020ora}
that the modular flow formula should hold up to scrambling time order
of proper time. This seems to be the only bound for $s$. This bound
is quite high and grants our solution to see the regions way behond
horizon. 

Besides $s$, we still need to check in more details on how other
parameters are bounded for the validity of the large $q$ solution.
These bounds mainly come from the limitation of various approximations
we take in the solution. The first approximation is taking the correlation
function in thermofield double state for $\s_{ab}^{0}$ in (\ref{eq:51})
and (\ref{eq:52}) when $\mu$ is large. To estimate the error,
we need to use the following identity
\begin{align}
e^{-\b V} & \propto\prod_{j=1}^{N}(1-2i\psi_{l}^{j}\psi_{r}^{j}\tanh\f{\mu}2)\propto\prod_{j=1}^{N}\left[\bra{0_{j}}\ket{0_{j}}+2e^{-\mu}\psi_{l}^{j}\bra{0_{j}}\ket{0_{j}}\psi_{l}^{j}\right]\nonumber \\
 & \app\bra 0\ket 0+e^{-\mu}\sum_{j=1}^{N}2\psi_{l}^{j}\bra 0\ket 0\psi_{l}^{j}
\end{align}
where we used $\bra{0_{j}}\ket{0_{j}}=(1-2i\psi_{l}^{j}\psi_{r}^{j})$
up to normalization and assumed $e^{-\mu}\ll1$. Taking this into
$s=0$ correlation function (here we suppress the average over indices
for simplicity)
\begin{equation}
\hat{g}(\tau_{1},\tau_{2})=\f{\Tr\r^{k}\psi_{a}(\tau_{1})\psi_{b}(\tau_{2})}{\Tr\r^{k}}\app\hat{g}_{\tfd}(\tau_{1},\tau_{2})\left(1+\f{e^{-\mu k}\xi^{k}\mF(\tau_{1},\tau_{2})}{1+Ne^{-\mu k}\xi^{k}}\right)+\text{subleading}
\end{equation}
where 
\begin{align}
\xi & \equiv\f 2N\sum_{j}\f{\avg{0|\psi_{l}^{j}e^{-\b_{l}H_{l}-\b_{r}H_{r}}\psi_{l}^{j}|0}}{Z_{\b}}\sim O(1),\quad Z_{\b}\equiv\avg{0|e^{-\b_{l}H_{l}-\b_{r}H_{r}}|0}=\Tr_{\mH_{l}}e^{-\b H_{l}}\\
\mF(\tau_{1},\tau_{2}) & \equiv N\left[\f{\avg{0|\psi_{l}^{j}e^{-\b_{l}H_{l}/2-\b_{r}H_{r}/2}\psi_{a}(\tau_{1})\psi_{b}(\tau_{2})e^{-\b_{l}H_{l}/2-\b_{r}H_{r}/2}\psi_{l}^{j}|0}}{\xi\avg{0|e^{-\b H_{l}/2}\psi_{a}(\tau_{1})\psi_{b}(\tau_{2})e^{-\b H_{l}/2}|0}}-1\right]\sim O(1)
\end{align}
To derive this, we used large $N$ factorization and $SO(N)$ symmetry
of correlators. To gurantee our thermofield double approximation works
for all $k\geq1$, we need to impose
\begin{equation}
e^{-\mu}/N\ll1
\end{equation}
which is obviously satisfied given $e^{-\mu}\ll1$.

The second approximation is assuming $\s_{ab}^{s}$ continuous and
checking if errors $\D_{i}\ll1/q$. All four $\D_{i}$ are in the
same order, and let us check $\D_{1}$ as an example. In large $\mJ$
limit, we can use (\ref{eq:104}) to show that $\tilde{y}_{s}\gg1$
(unless $\d_{l}$ is too close to 0 or $\pi$). Similarly, we can
use recurrence to show that $y_{0}=\tilde{y}_{0}\gg1$, $y_{1}\app y_{0}(1-\a)$,
$y_{k}\app y_{k-1}(1-O(\a(\a y_{0}^{-2})^{s-1}))$ and $\tilde{y}_{s}-y_{s}\sim y_{0}\a$
for $s\geq1$. Besides, $x_{s}$ has same scaling as $\tilde{x}_{s}$
in (\ref{eq:105}) and their difference is $x_{s}-\tilde{x}_{s}\sim O(\a^{2}(\a\tilde{y}_{0}^{-2})^{s-1})$
for $s\geq2$ (and is zero for $s=0,1$). Then, we have
\begin{align}
\D_{1} & \app e^{2\mu}\left[\left(\f{\w\sin\pi\d_{l}x(s)^{1/2}/\mJ}{\sin\w\tau+y(s)\sin\w(\b_{l}-\tau)}\right)^{2/q}-\left(\f{\w\sin\pi\d_{l}\tilde{x}(s)^{1/2}/\mJ}{\sin\w\tau+\tilde{y}(s)\sin\w(\b_{l}-\tau)}\right)^{2/q}\right]\nonumber \\
 & \lesssim e^{2\mu}\left(\f{\a^{1/2}(\a y_{0}^{-2})^{\f{s-1}2}}{(\b\mJ)^{2}}\right)^{2/q}\f{\a}q\nonumber \\
 & \lesssim \f 1qe^{-\mu q}(\b\mJ)^{-4/q}\label{eq:145}
\end{align}
where we assume $\sin\w\tau\sim\sin\w(\b_{l}-\tau)\sim O(1)$ and
in the last line we take $s=1$ to get the upper bound. To guarantee
it being smaller than $1/q$, we need to impose 
\begin{equation}
e^{-\mu q}(\b\mJ)^{-4/q}\ll1\label{eq:146}
\end{equation}

The third approximation is replacing the recurrence sequence with
differential equation. This approximation causes errors for $x_{s}$
and $y_{s}$ (and their tilde version). This error should be smaller
than $x_{s}-\tilde{x}_{s}$ and $y_{s}-\tilde{y}_{s}$. As (\ref{eq:145})
could also be understood as counting for the error of latter type,
we shoul validate this approximation under condition (\ref{eq:146}).

The last approximation is using the sum over image as the solution
to Schwinger-Dyson equation. This error is exponentially small for
$s$ not close to $\left\lfloor k/2\right\rfloor $ as indicated in
Fig. \ref{fig:The-plot-of}. For $s$ close to $\left\lfloor k/2\right\rfloor $,
the image and correlator itself are both exponentially small, there
could exist $O(1)$ relative error though the absolute error is still
exponentially small. It is not easy to analyze the error precisely
there because we do not have a full solution to Schwinger-Dyson equation.
However, we could understand this error as putting some restriction
on our analytic continuation of $s$. In other words, we should require
$\Im W_{2}\ll\Im W_{1}$ in (\ref{eq:102}) for some range of $s$.
In large $\mJ$ limit, we have
\begin{equation}
\tilde{c}_{2}\app\log\f{\b\mJ\a^{1/2}}{\pi}\sin\pi\d_{l},\quad\tilde{b}_{2}\app\f 12\log(\tilde{c}_{2}/\a)
\end{equation}
Here we see a competition between $\b\mJ$ and $\a$ in $\tilde{c}_{2}$
and $\tilde{b}_{2}$ will have $\pm i\pi/2$ imaginary part if $\tilde{c}_{2}<0$.
Nevertheless, we could require $|\Re(\tilde{c}_{2}+\tilde{b}_{2})|\gg0$
for simplicity. This leads to
\begin{equation}
|\log\b\mJ+\f 12\Re\log\tilde{c}_{2}|\gg0\implies\begin{cases}
\a^{1/2}\b\mJ\gg\exp(\b^{-2}\mJ^{-2})\text{ or }\a^{1/2}\b\mJ\ll\exp(-\b^{-2}\mJ^{-2}) & \text{(a)}\\
\left|\f{\b\mJ\a^{1/2}}{\pi}\sin\pi\d_{l}-1\right|\ll\b^{-2}\mJ^{-2} & \text{(b)}
\end{cases}
\end{equation}
where case (a) means overwhelming large $\b\mJ$ or $1/\a$
leads to large $|\tilde{c}_{2}|$, and case (b) means $\tilde{c}_{2}$
is very close to zero. For case (a), $\tilde{y}(1-is)$ is again small
oscillating function around its average value $\tilde{y}(1)$. Using
a similar approximation towards (\ref{eq:107}), we have
\begin{equation}
W_{2}(s,t)\app\left(\f{(2\pi\sin\pi\d_{l}/2)/(\b\mJ)}{Y(s)\sin\w(\b_{l}/2+it)+Y(s)^{-1}\sin\w(\b_{l}/2-it)}\right)^{2/q}
\end{equation}
where
\begin{equation}
Y(s)=\f{(1+\a)(\cosh(\tilde{c}_{2}+\tilde{b}_{2})\cos\tilde{c}_{1}s+i\sinh(\tilde{c}_{2}+\tilde{b}_{2})\sin\tilde{c}_{1}s)}{\cosh\tilde{b}_{2}}
\end{equation}
As $|\tilde{c}_{2}|$ is very large, this leads to 
\begin{equation}
|Y(s)|\sim e^{|\tilde{c}_{2}|}\implies|W_{2}/W_{1}|\sim e^{-2|\tilde{c}_{2}|/q}
\end{equation}
For small image contribution, we need
\begin{equation}
(\b\mJ\a^{1/2})^{\mp2/q}\ll1\label{eq:152}
\end{equation}
where minus sign is for $\b\mJ\a^{1/2}\gg1$ and plus sign is
for $\b\mJ\a^{1/2}\ll1$. For case (b), we have $\tilde{y}(1-is)\app1$
and $\tilde{x}(1-is)\app\tilde{x}_{0}$ for all $s\ll1/|\tilde{c}_{2}|\ra\infty$,
this leads to 
\begin{equation}
W_{2}(s,t)\sim1/x_{0}^{2/q}\sim(\b^{2}\mJ^{2})^{-2/q}\implies|W_{2}/W_{1}|\sim(\b\mJ)^{-2/q}
\end{equation}
For small image contribution, we need
\begin{equation}
(\b\mJ)^{-2/q}\ll1
\end{equation}

Summarizing above analysis, we should expect our solution valid generally
for $e^{-\mu}\ll1$ and $(\b\mJ)^{-1/q}\ll1$. If the $e^{-\mu q/2}$
and $1/\b\mJ$ are two distinct scales, we require the distinct
large enough as (\ref{eq:152}). Otherwise, we require them to be
in very close scales as $\b\mJ e^{-\mu q/2}\ra\pi/\sin\pi\d_{l}$.
For the case we are mostly interested in, we can first take large
$\mu$ limit and then take large $\b\mJ$, which is in validity
of our solution.

\section{Euclidean wormhole $SL(2,R)$ charge from large $q$ SYK solution} \label{app:d}

The essence of $M$ in gravitational computation is the magnitude
of $SL(2,R)$ charge carried by insertion of $\r_{0}$. Therefore, we
should first find a way to define $SL(2,R)$ charge for a given solution
\eqref{eq:35} on the ``necklace" diagram. In the following,
we will only focus on $\s_{rl}^{s}$, whose solution is copied here
\begin{equation}
e^{\s_{rl}^{s}(\tau_{1},\tau_{2})}=\f{h_{s}'(\tau_{1})f'(\tau_{2})}{\mJ\tilde{\mJ}_{s}(1-h_{s}(\tau_{1})f(\tau_{2}))^{2}}
\end{equation}
Let us first forget about all conditions that we impose to fix these
functions as in Section \ref{sec:approxsoln}. After
fixing $f$, we could restrict ourselves to the subspace of solutions
to Liouville equation in which all $h_{s}$ are related to each other
by an $SL(2,R)$ transformation. This subspace is isomorphic to the
group manifold of $SL(2,R)$. In this sense, our solution for each
$s$ given in Section \ref{sec:approxsoln} is a point
in this subpace. Finding a quantity to characterize the effect of
insertion $\r_{0}$ is equivalent to measuring the ``distance''
between $s$-th and $(s+1)$-th points.

Such ``distance'' has a natural constraint that if the translation
$\tau_{1}\ra\tau_{1}+c$ for a constant $c$ is an $SL(2,R)$ transformation
of $h_{s}$, we should count it as no ``distance'' away from original
solution. This corresponds to the $SL(2,R)$ charge defined in \eqref{sl2rcharge}
for circular boundary particle trjectory in EAdS$_{2}$ being invariant
under translation in $\t$. In other words, we are counting the effect
of $\r_{0}$ relative to the time translation generated by SYK Hamiltonian.

The $PSL(2,R)$ group has a 3-dimensional faithful representation
by acting $SL(2,R)$ on its $sl(2,R)$ algebra by conjugation, where
for a given $SL(2,R)$ transformation
\begin{equation}
h_{s}\ra\f{ah_{s}+b}{ch_{s}+d},\quad ad-bc=1
\end{equation}
we represent it as
\begin{equation}
Q\equiv\begin{pmatrix}V_{0} & V_{+}\\
V_{-} & -V_{0}
\end{pmatrix}\ra\begin{pmatrix}a & b\\
c & d
\end{pmatrix}\begin{pmatrix}V_{0} & V_{+}\\
V_{-} & -V_{0}
\end{pmatrix}\begin{pmatrix}a & b\\
c & d
\end{pmatrix}^{-1},\quad V_{\pm,0}\in\R\label{eq:app4}
\end{equation}
where $V_{0,\pm}$ parameterize the representation space. Indeed,
our transformations of solution from $s$ to $s+1$ are all in the
subgroup $PSL(2,R)$ because we will keep the direction of time unflipped.
As $Q$ are elements of the representation space, it is natural to
define it as the charge for each solution and use it to measure ``distance''.
For any two charges $Q_{1}$ and $Q_{2}$, the inner product is defined
as 
\begin{equation}
(Q_{1},Q_{2})\equiv\Tr Q_{1}Q_{2}
\end{equation}
and the norm of $Q_{1}-Q_{2}$ is their ``distance''. This ``distance''
coincides with the charge of $\r_{0}$ (namely $M$) if $Q_{1,2}$
are charges of $s$-th and $(s+1)$-th solution respectively assuming
charge conservation.

For $s=0$, we have $h_{0}(\tau)=\tan\w(\tau+\tilde{\g}_{0})$, whose
time translation acts as
\begin{equation}
h_{0}(\tau+\tau_{0})=\f{\cos\w\tau_{0}h_{0}(\tau)+\sin\w\tau_{0}}{-\sin\w\tau_{0}h_{0}(\tau)+\cos\w\tau_{0}}
\end{equation}
Invariance of $Q=Q_{0}$ under this $SL(2,R)$ transformation solves
$V_{0,\pm}$ in (\ref{eq:app4}) as
\begin{equation}
Q_{0}:\quad V_{0}=0,\quad V_{+}=-V_{-}=\kappa/\b
\end{equation}
Here $1/\b$ is due to dimensional analysis that $Q_{0}$ should match
with mass $M$. $\kappa$ is an order one number that does not depend
on $\mu$. This is important because $s=0$ solution should not know
anything about $\r_{0}$. Starting with $Q_{0}$, we can represent
the space of $PSL(2,R)/U(1)$ by conjugation (\ref{eq:app4}) of $Q_{0}$,
where $U(1)$ counts for the constraint from the time translation
as we proposed above.

Moving to a finite $s$ solution leads to $SL(2,R)$ matrix
\begin{equation}
R_{s}=\begin{pmatrix}\tilde{v}_{s}-\tilde{u}_{s}\tan\tilde{\g}_{s0} & \tilde{u}_{s}+\tilde{v}_{s}\tan\tilde{\g}_{s0}\\
-\tan\tilde{\g}_{s0} & 1
\end{pmatrix}\f{\cos\tilde{\g}_{s0}}{\sqrt{\tilde{v}_{s}}}
\end{equation}
where $\tilde{\g}_{s0}\equiv \tilde{\g}_s -\tilde{\g}_0$ and which conjugating $Q_{0}$ leads to
\begin{equation}
Q_{s}=\kappa/\b\begin{pmatrix}-\tilde{u}_{s}/\tilde{v}_{s} & \tilde{u}_{s}^{2}/\tilde{v}_{s}+\tilde{v}_{s}\\
-1/\tilde{v}_{s} & \tilde{u}_{s}/\tilde{v}_{s}
\end{pmatrix}
\end{equation}
Using \eqref{eq:44} and \eqref{eq:B.2},
we can write $Q_{s}$ in terms of $\tilde{x}_{s}$ and $\tilde{y}_{s}$.
Further using recurrence \eqref{eq:75}, we can represent
$Q_{s}$ in terms of $\tilde{x}_{s-1}$ and $\tilde{y}_{s-1}$. The
norm square of $Q_{s+1}-Q_{s}$ is
\begin{align}
M_{s}^{2}= & \Tr(Q_{s+1}-Q_{s})(Q_{s+1}-Q_{s})\nonumber \\
= & \f{2\kappa^{2}}{\alpha\b^{2}\tilde{y}_{s}^{2}}\left[2\alpha\tilde{x}_{s}\left(\alpha+\tilde{y}_{s}^{2}\right)\sin\text{\ensuremath{\w\b_{l}}}\csc\text{\ensuremath{\w\b_{r}}} +\left(\alpha^{2}+\tilde{y}_{s}^{4}+(\alpha(\alpha+4)+1)\tilde{y}_{s}^{2}\right)\csc^{2}\text{\ensuremath{\w\b_{r}}} \right.\nonumber \\
 & \left.+\alpha\left(\alpha\tilde{x}_{s}^{2}\sin^{2}\text{\ensuremath{\w\b_{l}}}-4\tilde{y}_{s}^{2}\right)-2(\alpha+1)\tilde{y}_{s}\cot\text{\ensuremath{\w\b_{r}}}\left(\alpha\tilde{x}_{s}\sin\text{\ensuremath{\w\b_{l}}}+\left(\alpha+\tilde{y}_{s}^{2}\right)\csc\text{\ensuremath{\w\b_{r}}}\right)\right]\label{eq:recid}
\end{align}
where we take large $\mu$ to set all $\a_{s}$ equal to $\a$. One
can easily show that this is indeed an exact identity of recurrence
\eqref{eq:75} if we set $\a_{s}=\a$, which means
$M_{s}^{2}$ is a constant for all $s$. This exactly corresponds
to the gravitational computation in Section \ref{sec:eucworm}
where the magnitude of $SL(2,R)$ charges of all $\r_{0}$ insertion
are the same and the $SL(2,R)$ transformation of boundary circular
trajectory is just power of $B(x,y)$. In particular, taking $s=0$
leads to
\begin{equation}
M_{s}^{2}=M_{0}^{2}=\f{2\kappa^{2}}{\b^{2}}\left(\f{(1+\a)^{2}}{\a\cos^{2}\f{\w\b}2}-4\right)\ra\f{2\kappa^{2}\mJ^{2}}{\pi^{2}\a}\label{eq:recM0}
\end{equation}
where in the last step we take large $\b\mJ$ and small $\a$. This
corresponds to the norm of charge carried by $\r_{0}$. To match with
\eqref{eq:Mmatch}, we simply choose $\kappa=\pi/\sqrt{2}$. 

Another immediate application of recurrence identity is to compute
$\tilde{y}_{\infty}$. Given $\tilde{x}_{\infty}=0$, using (\ref{eq:recM0})
for $s\ra\infty$ leads to 
\begin{align}
\tilde{y}_{\infty}= &\frac{1}{4}\sec\w\b/2\Big[2(\alpha+1)\cos\w(\b_{l}-\b_{r})/2\nn\\
& +\sqrt{2\left((\alpha-1)^{2}+(\alpha+1)^{2}\cos\w(\b_{l}-\b_{r})-4\alpha\cos\w\b\right)}\Big] \label{eq:yinfty}
\end{align}
Expanding in small $\a$ and large $\b\mJ$ limit, we see at leading
order
\begin{equation}
\tilde{y}_{\infty}\app\tilde{y}_{0}(1+\a)\app\tilde{y}_{1}
\end{equation}
which verifies our approximation \eqref{eq:104}. Using similar method, we can find another recurrence identity for
$x_{s}$ and $y_{s}$, from which we can derive
\begin{align}
y_{\infty}= & \frac{1}{2}\sec\w\b/2\left[\cos\left(\frac{\w(\b_{l}-\b_{r})}{2}\right)+\alpha\cos\left(\frac{\w(3\b_{l}+\b_{r})}{2}\right) \right.\nonumber \\
 & \left.+\sqrt{(1-\alpha )^2-\left(\sin \left(\frac{\w(\b_l-\b_r)}{2}\right)+\alpha  \sin \left(\frac{\w(3  \b_l+ \b_r)}{2} \right)\right)^2}\right] \label{eq:yinfty2}
\end{align}

\section{Bulk phase transition in large $q$ SYK} \label{app:e}

There is a very simple estimation for the bulk phase transition by changing parameter $\mu$ in large $q$ SYK model. For disconnected phase, the correlation function between SYK$_l$ and SYK$_r$ scales as $N^{-(q-1)}$ because it can only be built by classical correlation of random coupling $J$ between left and right \cite{Kourkoulou:2017zaj}. It follows that the contribution from insertion of probe scales as $\mu N \avg{\psi_l \psi_r}\sim\mu N^{-(q-2)}$ which vanishes in large $N$ limit (for $q>2$). At nonlinear orders, the insertion of probe contribute by the correlations within each SYK system. Nevertheless, we can treat the partition function of $k$ replica as product of the partition function of two SYK models with inverse temperature $k\b_l$ and $k\b_r$ respectively plus $\mu^2$ and higher order perturbation.  

For each SYK model with temperature $\b$, the large $q$ effective action is derived in \cite{Choi:2020tdj} 
\be
S_\eff=\f N{4q^{2}}\int d\tau_{1}d\tau_{2}\left[\f 14\del_{1}\s(\tau_{1},\tau_{2})\del_{2}\s(\tau_{1},\tau_{2})-\mJ^{2}e^{\s(\tau_{1},\tau_{2})}\right]\label{eq:4.29}
\ee
where the correlation function is in the form of $G(\tau_1,\tau_2)=\f 1 2 \sgn(\tau_{12})e^{\s(\tau_1,\tau_2 )/q}$. The equation of motion of \eqref{eq:4.29} is Liouville equation and its equilibrium solution is
\be
e^{\s(\tau_1,\tau_2)}=\f{\w^{2}}{\mJ^{2}\cos^{2}\w(|\tau_{12}|-k\b/2)}
\ee
where $\w$ is defined by $\w=\mJ \cos k\b/2$. Taking this solution back to \eqref{eq:4.29}, we get the on-shell action of disconnected phase to be \cite{Sarosi:2017ykf}
\begin{align}
S_\eff(k\b)&=\f N{2q^{2}}\int_{\tau_{1}>\tau_{2}}d\tau_{1}d\tau_{2}\w^{2}(1-2\sec^{2}(\w\tau_{12}-\w k\b/2))\nn\\
&=\f N{4q^{2}}k\b\w\left(k\b\w-4\tan\f{\w k\b}2\right)\ra -N \f {k\b \mJ}{q^2}
\end{align}
where in the last step we take large $\b$ limit for simplicity. To count for the correct partition function, we also need to include a constant extremal entropy $S_0=-\f N 2 \log 2$. This can be seen from high temperature limit $\beta\ra 0$ where partition function should count the total dimension of Hilbert space. For two SYK models, the total partition function is
\be
Z_{disconn.}(k\b_l,k\b_r)\app e^{-2S_0-S_\eff (k\b)},~\b\equiv \b_l+\b_r
\ee

The $\mu$ dependence at quadratic order is derived from expanding $\r_0$ as 
\be
\r_0=\cosh^N\f\mu 2 \prod_{j=1}^N\left(1-2i\psi^j_l\psi^j_r\tanh\f \mu 2\right)
\ee
and contracting $k$ insertions of $1-2i\psi^j_l\psi^j_r\tanh\f \mu 2$ within each SYK model respectively for each $j$. Given insertions are located at equal spacing of $\b_{l,r}$ on the thermal circle of circumstance $k\b_{l,r}$, we just need to consider the nearest contractions in large $\b$ limit. Due to $SO(N)$ symmetry and assuming large $N$ factorization, we can derive the following contribution to on-shell action from the nearest contractions 
\be
e^{-\d S_{\eff}/N}= \left(\f{1-\sqrt{1+4 x}}{2}\right)^k+\left(\f{1+\sqrt{1+4 x}}{2}\right)^k,~~x\equiv 4 \tanh^2 \f \mu 2 G_l(\b_l)G_r(\b_r)
\ee
where $G_a(\b_a)=\avg{\psi_a(\b_a)\psi_a}_{k\b_a}\in[0,1/2]$ are correlation functions of Majorana fermions with $\b_a$ spacing on the thermal circle with circumstance of $k\b_a$. Note that $x\in[0,1]$ for $\mu\in\R$ and exponentially suppressed in large $\b_{l,r}$ limit. It turns out that $\d S_\eff$ decreases monotonically until a finite value for increasing $\mu$.\footnote{In this computation, we ignore the backreaction of $\r_0$ to the background SYK solution on the thermal circle with circumstance of $k\b_{l,r}$ even in some large $\mu$ case. But we should expect this backreaction does not affect our result qualitatively.}

\begin{figure}
\begin{centering}
\includegraphics[width=8cm]{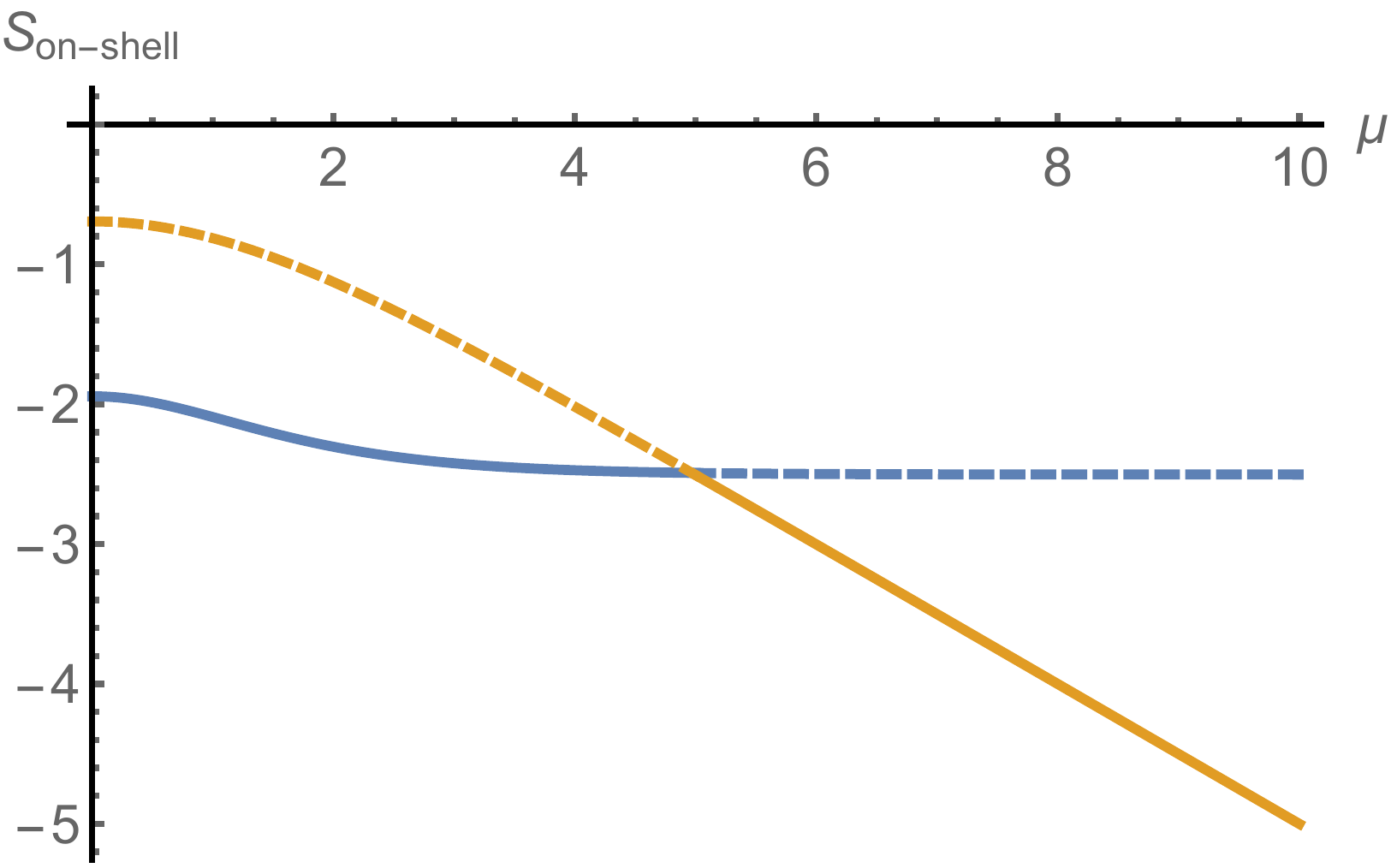}
\par\end{centering}
\caption{The comparison of on-shell action between disconnected phase (blue) and connected phase (yellow) as we increase $\mu$. At some finite $\mu=\mu_{cr}$, the dominant phase changes from the disconnected to the connected. The numbers in this plot are just for illustrative purpose. \label{fig:phasetras}}
\end{figure}

On the other hand, for connected phase where $\mu$ is large, we can roughly ignore all SYK contributions but only keep the one coming from insertion of probe. This is equivalent to evaluating \be
Z_{conn.}(k)\app\Tr \prod_{j=1} ^N\exp\left(-i\mu k \psi_l ^j\psi_r^j\right)=\left(2\cosh\f {\mu k}{2}\right)^N \app e^{-N(-\mu k/2)}
\ee
where in the last step we take large $\mu$ limit. We can regard $-\mu k/2$ as the on-shell action for connected phase. It is important that this action does not include a constant extremal entropy term. Indeed, from JT gravity point of view, this reflects the fact that disconnected and connected phase have different contributions of topological term proportional to $S_0$. It is clear that when $-\mu k N/2 > 2S_0+S_\eff(k\b)+\d S_\eff$, the dominant phase will be disconnected and vice versa. If we ignore $\d S$, the critical value of $\mu$ for phase transition is $\mu_{cr}\sim 2\b\mJ/q^2$ in large $\b \mJ$ limit. See Fig. \ref{fig:phasetras} for an illustration. In this paper, we basically consider the regime $\mu> \mu_{cr}$ such that connected phase dominates.

\bibliographystyle{JHEP.bst}
\bibliography{main.bib}

\providecommand{\href}[2]{#2}\begingroup\raggedright\begin{thebibliography}{10}

\bibitem{Jafferis:2020ora}
D.L.~Jafferis and L.~Lamprou, \emph{{Inside the Hologram: Reconstructing the
  bulk observer's experience}},
  \href{https://arxiv.org/abs/2009.04476}{{\ttfamily 2009.04476}}.

\bibitem{Yoshida:2019qqw}
B.~Yoshida, \emph{{Firewalls vs. Scrambling}},
  \href{https://doi.org/10.1007/JHEP10(2019)132}{\emph{JHEP} {\bfseries 10}
  (2019) 132} [\href{https://arxiv.org/abs/1902.09763}{{\ttfamily
  1902.09763}}].

\bibitem{Yoshida:2019kyp}
B.~Yoshida, \emph{{Observer-dependent black hole interior from operator
  collision}}, \href{https://doi.org/10.1103/PhysRevD.103.046004}{\emph{Phys.
  Rev. D} {\bfseries 103} (2021) 046004}
  [\href{https://arxiv.org/abs/1910.11346}{{\ttfamily 1910.11346}}].

\bibitem{LamprouJafferisdeBoer}
J.~de~Boer, D.L.~Jafferis and L.~Lamprou, \emph{{Inside holographic black
  holes: Melting the frozen vacuum}},  \href{https://arxiv.org/abs/to
  appear}{{\ttfamily to appear}}.

\bibitem{Sachdev:2010um}
S.~Sachdev, \emph{{Holographic metals and the fractionalized Fermi liquid}},
  \href{https://doi.org/10.1103/PhysRevLett.105.151602}{\emph{Phys. Rev. Lett.}
  {\bfseries 105} (2010) 151602}
  [\href{https://arxiv.org/abs/1006.3794}{{\ttfamily 1006.3794}}].

\bibitem{sachdev1993gapless}
S.~Sachdev and J.~Ye, \emph{Gapless spin-fluid ground state in a random quantum
  heisenberg magnet},
  \href{https://doi.org/10.1103/physrevlett.70.3339}{\emph{Physical Review
  Letters} {\bfseries 70} (1993) 3339}.

\bibitem{kitaev2015simple}
A.~Kitaev, \emph{A simple model of quantum holography},  in \emph{KITP strings
  seminar and Entanglement}, vol.~12, 2015.

\bibitem{maldacena2016remarks}
J.~Maldacena and D.~Stanford, \emph{{Remarks on the Sachdev-Ye-Kitaev model}},
  \href{https://doi.org/10.1103/PhysRevD.94.106002}{\emph{Phys. Rev. D}
  {\bfseries 94} (2016) 106002}
  [\href{https://arxiv.org/abs/1604.07818}{{\ttfamily 1604.07818}}].

\bibitem{Stanford:2020wkf}
D.~Stanford, \emph{{More quantum noise from wormholes}},
  \href{https://arxiv.org/abs/2008.08570}{{\ttfamily 2008.08570}}.

\bibitem{maldacena2013cool}
J.~Maldacena and L.~Susskind, \emph{{Cool horizons for entangled black holes}},
  \href{https://doi.org/10.1002/prop.201300020}{\emph{Fortsch. Phys.}
  {\bfseries 61} (2013) 781} [\href{https://arxiv.org/abs/1306.0533}{{\ttfamily
  1306.0533}}].

\bibitem{Maldacena:2017axo}
J.~Maldacena, D.~Stanford and Z.~Yang, \emph{{Diving into traversable
  wormholes}}, \href{https://doi.org/10.1002/prop.201700034}{\emph{Fortsch.
  Phys.} {\bfseries 65} (2017) 1700034}
  [\href{https://arxiv.org/abs/1704.05333}{{\ttfamily 1704.05333}}].

\bibitem{Roberts:2018mnp}
D.A.~Roberts, D.~Stanford and A.~Streicher, \emph{{Operator growth in the SYK
  model}}, \href{https://doi.org/10.1007/JHEP06(2018)122}{\emph{JHEP}
  {\bfseries 06} (2018) 122}
  [\href{https://arxiv.org/abs/1802.02633}{{\ttfamily 1802.02633}}].

\bibitem{Qi:2018bje}
X.-L.~Qi and A.~Streicher, \emph{{Quantum Epidemiology: Operator Growth,
  Thermal Effects, and SYK}},
  \href{https://doi.org/10.1007/JHEP08(2019)012}{\emph{JHEP} {\bfseries 08}
  (2019) 012} [\href{https://arxiv.org/abs/1810.11958}{{\ttfamily
  1810.11958}}].

\bibitem{Nezami:2021yaq}
S.~Nezami, H.W.~Lin, A.R.~Brown, H.~Gharibyan, S.~Leichenauer, G.~Salton
  et~al., \emph{{Quantum Gravity in the Lab: Teleportation by Size and
  Traversable Wormholes, Part II}},
  \href{https://arxiv.org/abs/2102.01064}{{\ttfamily 2102.01064}}.

\bibitem{Haehl:2021emt}
F.M.~Haehl and Y.~Zhao, \emph{{Size and momentum of an infalling particle in
  the black hole interior}},
  \href{https://doi.org/10.1007/JHEP06(2021)056}{\emph{JHEP} {\bfseries 06}
  (2021) 056} [\href{https://arxiv.org/abs/2102.05697}{{\ttfamily
  2102.05697}}].

\bibitem{Jian:2020qpp}
S.-K.~Jian, B.~Swingle and Z.-Y.~Xian, \emph{{Complexity growth of operators in
  the SYK model and in JT gravity}},
  \href{https://doi.org/10.1007/JHEP03(2021)014}{\emph{JHEP} {\bfseries 03}
  (2021) 014} [\href{https://arxiv.org/abs/2008.12274}{{\ttfamily
  2008.12274}}].

\bibitem{Lensky:2020ubw}
Y.D.~Lensky, X.-L.~Qi and P.~Zhang, \emph{{Size of bulk fermions in the SYK
  model}}, \href{https://doi.org/10.1007/JHEP10(2020)053}{\emph{JHEP}
  {\bfseries 10} (2020) 053}
  [\href{https://arxiv.org/abs/2002.01961}{{\ttfamily 2002.01961}}].

\bibitem{Gao:2019nyj}
P.~Gao and D.L.~Jafferis, \emph{{A traversable wormhole teleportation protocol
  in the SYK model}},
  \href{https://doi.org/10.1007/JHEP07(2021)097}{\emph{JHEP} {\bfseries 07}
  (2021) 097} [\href{https://arxiv.org/abs/1911.07416}{{\ttfamily
  1911.07416}}].

\bibitem{Lucas:2018wsc}
A.~Lucas, \emph{{Operator size at finite temperature and Planckian bounds on
  quantum dynamics}},
  \href{https://doi.org/10.1103/PhysRevLett.122.216601}{\emph{Phys. Rev. Lett.}
  {\bfseries 122} (2019) 216601}
  [\href{https://arxiv.org/abs/1809.07769}{{\ttfamily 1809.07769}}].

\bibitem{Schuster:2021uvg}
T.~Schuster, B.~Kobrin, P.~Gao, I.~Cong, E.T.~Khabiboulline, N.M.~Linke et~al.,
  \emph{{Many-body quantum teleportation via operator spreading in the
  traversable wormhole protocol}},
  \href{https://arxiv.org/abs/2102.00010}{{\ttfamily 2102.00010}}.

\bibitem{Maldacena:2018lmt}
J.~Maldacena and X.-L.~Qi, \emph{{Eternal traversable wormhole}},
  \href{https://arxiv.org/abs/1804.00491}{{\ttfamily 1804.00491}}.

\bibitem{Maldacena:2016upp}
J.~Maldacena, D.~Stanford and Z.~Yang, \emph{{Conformal symmetry and its
  breaking in two dimensional Nearly Anti-de-Sitter space}},
  \href{https://doi.org/10.1093/ptep/ptw124}{\emph{PTEP} {\bfseries 2016}
  (2016) 12C104} [\href{https://arxiv.org/abs/1606.01857}{{\ttfamily
  1606.01857}}].

\bibitem{Gao:2021uro}
P.~Gao, D.L.~Jafferis and D.K.~Kolchmeyer, \emph{{An effective matrix model for
  dynamical end of the world branes in Jackiw-Teitelboim gravity}},
  \href{https://arxiv.org/abs/2104.01184}{{\ttfamily 2104.01184}}.

\bibitem{Saad:2019lba}
P.~Saad, S.H.~Shenker and D.~Stanford, \emph{{JT gravity as a matrix
  integral}},  \href{https://arxiv.org/abs/1903.11115}{{\ttfamily 1903.11115}}.

\bibitem{Engelhardt:2020qpv}
N.~Engelhardt, S.~Fischetti and A.~Maloney, \emph{{Free energy from replica
  wormholes}}, \href{https://doi.org/10.1103/PhysRevD.103.046021}{\emph{Phys.
  Rev. D} {\bfseries 103} (2021) 046021}
  [\href{https://arxiv.org/abs/2007.07444}{{\ttfamily 2007.07444}}].

\bibitem{Saad:2018bqo}
P.~Saad, S.H.~Shenker and D.~Stanford, \emph{{A semiclassical ramp in SYK and
  in gravity}},  \href{https://arxiv.org/abs/1806.06840}{{\ttfamily
  1806.06840}}.

\bibitem{Hamilton:2005ju}
A.~Hamilton, D.N.~Kabat, G.~Lifschytz and D.A.~Lowe, \emph{{Local bulk
  operators in AdS/CFT: A Boundary view of horizons and locality}},
  \href{https://doi.org/10.1103/PhysRevD.73.086003}{\emph{Phys. Rev. D}
  {\bfseries 73} (2006) 086003}
  [\href{https://arxiv.org/abs/hep-th/0506118}{{\ttfamily hep-th/0506118}}].

\bibitem{Hamilton:2006az}
A.~Hamilton, D.N.~Kabat, G.~Lifschytz and D.A.~Lowe, \emph{{Holographic
  representation of local bulk operators}},
  \href{https://doi.org/10.1103/PhysRevD.74.066009}{\emph{Phys. Rev. D}
  {\bfseries 74} (2006) 066009}
  [\href{https://arxiv.org/abs/hep-th/0606141}{{\ttfamily hep-th/0606141}}].

\bibitem{Hamilton:2006fh}
A.~Hamilton, D.N.~Kabat, G.~Lifschytz and D.A.~Lowe, \emph{{Local bulk
  operators in AdS/CFT: A Holographic description of the black hole interior}},
  \href{https://doi.org/10.1103/PhysRevD.75.106001}{\emph{Phys. Rev. D}
  {\bfseries 75} (2007) 106001}
  [\href{https://arxiv.org/abs/hep-th/0612053}{{\ttfamily hep-th/0612053}}].

\bibitem{Gao:2016bin}
P.~Gao, D.L.~Jafferis and A.C.~Wall, \emph{{Traversable Wormholes via a Double
  Trace Deformation}},
  \href{https://doi.org/10.1007/JHEP12(2017)151}{\emph{JHEP} {\bfseries 12}
  (2017) 151} [\href{https://arxiv.org/abs/1608.05687}{{\ttfamily
  1608.05687}}].

\bibitem{Gao:2018yzk}
P.~Gao and H.~Liu, \emph{{Regenesis and quantum traversable wormholes}},
  \href{https://doi.org/10.1007/JHEP10(2019)048}{\emph{JHEP} {\bfseries 10}
  (2019) 048} [\href{https://arxiv.org/abs/1810.01444}{{\ttfamily
  1810.01444}}].

\bibitem{Lin:2019qwu}
H.W.~Lin, J.~Maldacena and Y.~Zhao, \emph{{Symmetries Near the Horizon}},
  \href{https://doi.org/10.1007/JHEP08(2019)049}{\emph{JHEP} {\bfseries 08}
  (2019) 049} [\href{https://arxiv.org/abs/1904.12820}{{\ttfamily
  1904.12820}}].

\bibitem{Chandrasekaran:2021tkb}
V.~Chandrasekaran, T.~Faulkner and A.~Levine, \emph{{Scattering strings off
  quantum extremal surfaces}},
  \href{https://arxiv.org/abs/2108.01093}{{\ttfamily 2108.01093}}.

\bibitem{Kourkoulou:2017zaj}
I.~Kourkoulou and J.~Maldacena, \emph{{Pure states in the SYK model and
  nearly-$AdS_2$ gravity}},  \href{https://arxiv.org/abs/1707.02325}{{\ttfamily
  1707.02325}}.

\bibitem{DeBoer:2019yoe}
J.~De~Boer, R.~Van~Breukelen, S.F.~Lokhande, K.~Papadodimas and E.~Verlinde,
  \emph{{Probing typical black hole microstates}},
  \href{https://doi.org/10.1007/JHEP01(2020)062}{\emph{JHEP} {\bfseries 01}
  (2020) 062} [\href{https://arxiv.org/abs/1901.08527}{{\ttfamily
  1901.08527}}].

\bibitem{Leutheusser:2021qhd}
S.~Leutheusser and H.~Liu, \emph{{Causal connectability between quantum systems
  and the black hole interior in holographic duality}},
  \href{https://arxiv.org/abs/2110.05497}{{\ttfamily 2110.05497}}.

\bibitem{Bouland:2019pvu}
A.~Bouland, B.~Fefferman and U.~Vazirani, \emph{{Computational
  pseudorandomness, the wormhole growth paradox, and constraints on the AdS/CFT
  duality}},  \href{https://arxiv.org/abs/1910.14646}{{\ttfamily 1910.14646}}.

\bibitem{Blake:2021wqj}
M.~Blake and H.~Liu, \emph{{On systems of maximal quantum chaos}},
  \href{https://doi.org/10.1007/JHEP05(2021)229}{\emph{JHEP} {\bfseries 05}
  (2021) 229} [\href{https://arxiv.org/abs/2102.11294}{{\ttfamily
  2102.11294}}].

\bibitem{Blake:2017ris}
M.~Blake, H.~Lee and H.~Liu, \emph{{A quantum hydrodynamical description for
  scrambling and many-body chaos}},
  \href{https://doi.org/10.1007/JHEP10(2018)127}{\emph{JHEP} {\bfseries 10}
  (2018) 127} [\href{https://arxiv.org/abs/1801.00010}{{\ttfamily
  1801.00010}}].

\bibitem{DAlessio:2015qtq}
L.~D'Alessio, Y.~Kafri, A.~Polkovnikov and M.~Rigol, \emph{{From quantum chaos
  and eigenstate thermalization to statistical mechanics and thermodynamics}},
  \href{https://doi.org/10.1080/00018732.2016.1198134}{\emph{Adv. Phys.}
  {\bfseries 65} (2016) 239}
  [\href{https://arxiv.org/abs/1509.06411}{{\ttfamily 1509.06411}}].

\bibitem{Collier:2019weq}
S.~Collier, A.~Maloney, H.~Maxfield and I.~Tsiares, \emph{{Universal dynamics
  of heavy operators in CFT$_{2}$}},
  \href{https://doi.org/10.1007/JHEP07(2020)074}{\emph{JHEP} {\bfseries 07}
  (2020) 074} [\href{https://arxiv.org/abs/1912.00222}{{\ttfamily
  1912.00222}}].

\bibitem{Belin:2020hea}
A.~Belin and J.~de~Boer, \emph{{Random statistics of OPE coefficients and
  Euclidean wormholes}},
  \href{https://doi.org/10.1088/1361-6382/ac1082}{\emph{Class. Quant. Grav.}
  {\bfseries 38} (2021) 164001}
  [\href{https://arxiv.org/abs/2006.05499}{{\ttfamily 2006.05499}}].

\bibitem{Belin:2021ryy}
A.~Belin, J.~de~Boer and D.~Liska, \emph{{Non-Gaussianities in the Statistical
  Distribution of Heavy OPE Coefficients and Wormholes}},
  \href{https://arxiv.org/abs/2110.14649}{{\ttfamily 2110.14649}}.

\bibitem{Foini:2018sdb}
L.~Foini and J.~Kurchan, \emph{{Eigenstate thermalization hypothesis and out of
  time order correlators}},
  \href{https://doi.org/10.1103/PhysRevE.99.042139}{\emph{Phys. Rev. E}
  {\bfseries 99} (2019) 042139}
  [\href{https://arxiv.org/abs/1803.10658}{{\ttfamily 1803.10658}}].

\bibitem{Haehl:2021tft}
F.M.~Haehl, A.~Streicher and Y.~Zhao, \emph{{Six-point functions and collisions
  in the black hole interior}},
  \href{https://doi.org/10.1007/JHEP08(2021)134}{\emph{JHEP} {\bfseries 08}
  (2021) 134} [\href{https://arxiv.org/abs/2105.12755}{{\ttfamily
  2105.12755}}].

\bibitem{Shenker:2013pqa}
S.H.~Shenker and D.~Stanford, \emph{{Black holes and the butterfly effect}},
  \href{https://doi.org/10.1007/JHEP03(2014)067}{\emph{JHEP} {\bfseries 03}
  (2014) 067} [\href{https://arxiv.org/abs/1306.0622}{{\ttfamily 1306.0622}}].

\bibitem{Goel:2018ubv}
A.~Goel, H.T.~Lam, G.J.~Turiaci and H.~Verlinde, \emph{{Expanding the Black
  Hole Interior: Partially Entangled Thermal States in SYK}},
  \href{https://doi.org/10.1007/JHEP02(2019)156}{\emph{JHEP} {\bfseries 02}
  (2019) 156} [\href{https://arxiv.org/abs/1807.03916}{{\ttfamily
  1807.03916}}].

\bibitem{Brown:2019rox}
A.R.~Brown, H.~Gharibyan, G.~Penington and L.~Susskind, \emph{{The
  Python\textquoteright{}s Lunch: geometric obstructions to decoding Hawking
  radiation}}, \href{https://doi.org/10.1007/JHEP08(2020)121}{\emph{JHEP}
  {\bfseries 08} (2020) 121}
  [\href{https://arxiv.org/abs/1912.00228}{{\ttfamily 1912.00228}}].

\bibitem{Harlow:2018tqv}
D.~Harlow and D.~Jafferis, \emph{{The Factorization Problem in
  Jackiw-Teitelboim Gravity}},
  \href{https://doi.org/10.1007/JHEP02(2020)177}{\emph{JHEP} {\bfseries 02}
  (2020) 177} [\href{https://arxiv.org/abs/1804.01081}{{\ttfamily
  1804.01081}}].

\bibitem{Choi:2020tdj}
C.~Choi, M.~Mezei and G.~S\'arosi, \emph{{Pole skipping away from maximal
  chaos}},  \href{https://arxiv.org/abs/2010.08558}{{\ttfamily 2010.08558}}.

\bibitem{Sarosi:2017ykf}
G.~S\'arosi, \emph{{AdS$_{2}$ holography and the SYK model}},
  \href{https://doi.org/10.22323/1.323.0001}{\emph{PoS} {\bfseries Modave2017}
  (2018) 001} [\href{https://arxiv.org/abs/1711.08482}{{\ttfamily
  1711.08482}}].

\end{thebibliography}\endgroup

\end{document}